\documentclass[11pt]{article}
\usepackage{amsfonts}
\usepackage{amssymb}
\usepackage{amsmath,amssymb}
\usepackage{bbm}

\def\slasha#1{\setbox0=\hbox{$#1$}#1\hskip-\wd0\hbox to\wd0{\hss\sl/\/\hss}}

\def\periodb#1{\setbox0=\hbox{$#1$}#1\hskip-\wd0\hbox to\wd0{-}}

\newcommand{\pZ}{p}    % points in Twistor space

\newcommand{\CN}{\mathcal{N}}    % number of supersymmetries
\newcommand{\CA}{\mathcal{A}}    % super gauge potential
\newcommand{\CO}{\mathcal{O}}    % bundle classification
\newcommand{\CV}{\mathcal{V}}    % Variety
\newcommand{\CL}{\mathcal{L}}    % Lagrangian
\newcommand{\CU}{\mathcal{U}}    % patches
\newcommand{\CF}{\mathcal{F}}    % f-theory
\newcommand{\CM}{\mathcal{M}}    % m-theory, manifold
\newcommand{\CP}{\mathcal{P}}    % twistor space
\newcommand{\CT}{\mathcal{T}}    % real twistor space
\newcommand{\CZ}{\mathcal{Z}}    % CY
\newcommand{\CE}{\mathcal{E}}    % complex vector bundle
\newcommand{\CR}{\mathcal{R}}    % REAL superspace
\newcommand{\FR}{\mathbbm{R}}     % field of real numbers
\newcommand{\FC}{\mathbbm{C}}     % field of complex numbers
\newcommand{\CPP}{{\mathbbm{C}P}}    % complex projective plane
\newcommand{\RPS}{{\mathbbm{R}P}}    % complex projective plane
\newcommand{\RZ}{\mathbbm{Z}}     % ring of integers
\newcommand{\dd}{\mathrm{d}}     % total differential
\newcommand{\dpar}{\partial}     % partial differential
\newcommand{\dparb}{{\bar{\partial}}}     % partial differential with bar
\newcommand{\embd}{{\hookrightarrow}}     % embedded
\newcommand{\diag}{{\mathrm{diag}}}     % embedded
\newcommand{\cb}{\,,\,}             % Coordinate comma
\newcommand{\cm}{{}\hspace{-0.05cm}{}}             % Other spacing
\newcommand{\de}{\mathrm{e}}     % Euler's number
\newcommand{\di}{\mathrm{i}}     % imaginary unit
\newcommand{\bz}{{\bar{z}}}     % bar \"{u}ber z
\newcommand{\bl}{{\bar{\lambda}}}     % bar \"{u}ber j
\newcommand{\bs}{{\bar{\zeta}}}     % bar \"{u}ber j
\newcommand{\ald}{{\dot{\alpha}}}     % bar \"{u}ber j
\newcommand{\bed}{{\dot{\beta}}}     % bar \"{u}ber j
\newcommand{\eps}{{\varepsilon}}     % bar \"{u}ber j
\newcommand{\der}[1]{\frac{\dpar}{\dpar #1}}   % partielle ableitung
\newcommand{\tr}{\,\mathrm{tr}\,}     % trace
\newcommand{\sSU}{\mathrm{SU}}     % Euler's number
\newcommand{\sSO}{\mathrm{SO}}     % Euler's number
\newcommand{\sGL}{\mathrm{GL}}     % Euler's number

\makeatletter
\renewcommand*\l@section{\@dottedtocline{1}{1.5em}{4em}}%1.5em% %2.3em%
\@addtoreset{equation}{section}
\makeatother
\renewcommand{\thesection}{\arabic{section}.}
\renewcommand{\theequation}{\thesection\arabic{equation}}

\begin{document}
\begin{titlepage}
\setcounter{page}{0}
\begin{flushright}
hep--th/0405123\\
ITP--UH--11/04
\end{flushright}
\vskip 1.5cm
\begin{center}
{\LARGE \bf On Supertwistors, the Penrose-Ward Transform}
\vskip 0.5cm
{\LARGE \bf and $\CN$=4 super Yang-Mills Theory}
\vskip 1.5cm
\renewcommand{\thefootnote}{\fnsymbol{footnote}}
{\Large Alexander D. Popov\footnote{On leave from Bogoliubov
Laboratory of Theoretical Physics, JINR, Dubna, Russia.} and
Christian S\"amann} \setcounter{footnote}{0}
\renewcommand{\thefootnote}{\arabic{thefootnote}}
\vskip 1cm
{\em Institut f\"{u}r Theoretische Physik\\
Universit\"{a}t Hannover\\
Appelstra{\ss}e 2, 30167 Hannover, Germany}\\[5mm]
{Email: popov, saemann@itp.uni-hannover.de}
\vskip 1.1cm
\end{center}
\begin{center}
{\bf Abstract}
\end{center}
\begin{quote}
It was recently shown by Witten that B-type open topological
string theory with the supertwistor space $\CPP^{3|4}$ as a target
space is equivalent to holomorphic Chern-Simons (hCS) theory on
the same space. This hCS theory in turn is equivalent to self-dual
$\CN{=}4$ super Yang-Mills (SYM) theory in four dimensions. We
review the supertwistor description of self-dual and
anti-self-dual $\CN$-extended SYM theory as the integrability of super
Yang-Mills fields on complex $(2|\CN)$-dimensional superplanes and
demonstrate the equivalence of this description to Witten's
formulation. The equivalence of the field equations of hCS theory on
an open subset of $\CPP^{3|\CN}$ to the field equations of
self-dual $\CN$-extended SYM theory in four dimensions is made
explicit. Furthermore, we extend the picture to the full $\CN{=}4$
SYM theory and, by using the known supertwistor description of
this case, we show that the corresponding constraint equations are
(gauge) equivalent to the field equations of hCS theory on a
quadric in $\CPP^{3|3}\times \CPP^{3|3}$. \vskip 5mm
\end{quote}
\end{titlepage}
\newpage
\setcounter{page}{1}
\tableofcontents
\section{Introduction}
Let $\CZ$ be a complex three-dimensional Calabi-Yau (CY) manifold, $E$
a rank $n$ complex vector bundle over $\CZ$ and $A$ a connection
one-form on $E$. Consider the action \cite{Witten:1992fb}
\begin{equation}\label{hCSaction}
S_{hCS}=\int_\CZ\Omega_0\wedge \tr\left(A^{0,1}\wedge\bar{\dpar}A^{0,1}
+\frac{2}{3}A^{0,1}\wedge A^{0,1}\wedge A^{0,1}\right),
\end{equation}
where $\Omega_0$ is the nowhere vanishing holomorphic $(3,0)$-form
on $\CZ$ and $A^{0,1}$ is the $(0,1)$-component of the connection
one-form $A$. Witten has obtained \eqref{hCSaction} as the full
target space action of the open topological B-model on a complex
three-dimensional target space, on which the CY restriction arises
from $N{=}2$ supersymmetry of the corresponding topological sigma
model and an anomaly cancellation condition. This holomorphic
Chern-Simons (hCS) theory \eqref{hCSaction} describes inequivalent
complex structures on the bundle $E\rightarrow \CZ$.

In a beautiful recent paper \cite{Witten:2003nn}, Witten observed
that the above-mentioned severe CY restriction can be relaxed by
considering a topological B-model whose target space is a
Calabi-Yau {\em super\-mani\-fold}\hspace{0.5mm}\footnote{For a
definition, see appendix B.}. Here, the fermionic dimensions will
also make a contribution to the first Chern class, and this yields
more freedom in the choice of the bosonic dimensions to have an
overall vanishing first Chern class. In particular, an extension
\begin{equation}\label{ShCSaction}
\hat{S}_{hCS}=\int_\mathcal{Y}\Omega\wedge
\tr\left(\hat{\CA}^{0,1}\wedge\bar{\dpar}\hat{\CA}^{0,1}
+\frac{2}{3}\hat{\CA}^{0,1}\wedge \hat{\CA}^{0,1}\wedge
\hat{\CA}^{0,1}\right)
\end{equation}
of the action \eqref{hCSaction} to the supertwistor space
$\CPP^{3|4}$ was considered. Here $\mathcal{Y}$ is a subspace of
$\CPP^{3|4}$ parametrized by three complex bosonic coordinates
together with their complex conjugate and four (holomorphic)
fermionic coordinates, $\Omega$ is a holomorphic measure for
bosonic and fermionic coordinates, and $\hat{\CA}^{0,1}$ is the
$(0,1)$-component of a connection one-form $\hat{\CA}$ on a rank
$n$ complex vector bundle $\CE$ over $\CPP^{3|4}$ depending on
both the bosonic and fermionic coordinates. It was shown
\cite{Witten:2003nn} that there is a bijection between the moduli
spaces of holomorphic Chern-Simons theory \eqref{ShCSaction} on
the supermanifold $\CPP^{3|4}\backslash\CPP^{1|4}$ and of
self-dual $\CN{=}4$ super Yang-Mills (SYM) theory on the space
$\FR^4$ with a metric of signature $(\cm+\cm+\cm+\cm+\cm)$ or
$(\cm-\cm-\cm+\cm+\cm)$, depending on the reality conditions
imposed on the supertwistor space (for related works see
\cite{Berkovits:2004hg}-\cite{Siegel:2004dj}). It was also
demonstrated that the above twistor description allows one to
recover Yang-Mills scattering amplitudes, in particular maximally
helicity violating (MHV) ones\footnote{See \cite{Nair:bq} for an
earlier discussion of this point.}, and to clarify the
holomorphicity properties\footnote{The unexpected holomorphicity
properties of the so-called maximally helicity violating
amplitudes provided the original motivation to study the space
$\CP^{3|4}:=\CPP^{3|4}\backslash\CPP^{1|4}$. Holomorphicity is an
earmark of the topological B-model, and there is a six-dimensional
description of self-dual SYM theory via the twistor
correspondence. Thus one had to find a space which is CY and a
complex (super)twistor space of $\FR^4$ at the same time:
$\CP^{3|4}$.} of these amplitudes\footnote{For a preliminary
consideration of gravity amplitudes in this context, see
\cite{Giombi:2004ix}.} and identities appearing in this context
\cite{Roiban:2004vt}-\cite{Gukov:2004ei}.

Note that Witten described the correspondence between hCS theory
on $\CPP^{3|4}$ and anti-self-dual $\CN{=}4$ SYM theory by analyzing
the sheaf cohomology interpretation of the {\em linearized} field
equations on the supertwistor space. The main purpose of this
paper is to give a more detailed and explicit description of this
correspondence for $0\leq\CN\leq 4$ beyond the linearized
level.\footnote{Still, it should be stressed that the supertwistor
space is CY only for $\CN{=}4$.} We will also discuss the supertwistor
description of the full $\CN{=}4$ SYM theory along the lines
proposed in \cite{Witten:2003nn}. In fact, we shall
consider several special cases of the following general situation.
Suppose we are given complex (super)manifolds $X,Y,Z$ and a double
fibration
\begin{equation}\label{gnldblfibration}
\begin{picture}(50,40)
\put(0.0,0.0){\makebox(0,0)[c]{$Z$}}
%\put(32.0,0.1){\makebox(0,0)[c]{$\Leftrightarrow$}}
\put(64.0,0.0){\makebox(0,0)[c]{$X$}}
\put(34.0,33.0){\makebox(0,0)[c]{$Y$}}
\put(7.0,18.0){\makebox(0,0)[c]{$\pi_2$}}
\put(55.0,18.0){\makebox(0,0)[c]{$\pi_1$}}
\put(25.0,25.0){\vector(-1,-1){18}}
\put(37.0,25.0){\vector(1,-1){18}}
\end{picture}
\end{equation}
with surjective holomorphic projections $\pi_1$ and $\pi_2$. Then
we have a {\em correspondence} between $Z$ and $X$, i.e.\ between
points in one space and subspaces of the other one:
\begin{equation}\label{correspond1}\begin{split}
\mbox{points $z$ in $Z$}&~~\leftrightarrow~~\mbox{subspaces
$\pi_1(\pi_2^{-1}(z))$ in $X$}~,\\
\mbox{subspaces $\pi_2(\pi_1^{-1}(x))$ in
$Z$}&~~\leftrightarrow~~\mbox{points $x$ in $X$}~.\end{split}
\end{equation}
Using the correspondence \eqref{correspond1}, one can transfer
data given on $Z$ to data on $X$ (and vice versa). One may
take some analytic objects $h$ on $Z$ (Dolbeault cohomology
classes, holomorphic vector bundles, etc.) and transform them to
objects $f$ on $X$ which will be constrained by some differential
equations, since the pull-back of $h$ to $Y$ has to be constant
along the fibres of $\pi_2$. The map $\CP\mathcal{W}:h\mapsto
f$ is called the {\em Penrose-Ward transform}. Of course, one can
also consider the inverse map $\CP\mathcal{W}^{-1}:f\mapsto h$.
If there is an anti-linear involution $\tau$ (real structure)
on $X$, then the set of fixed points of $\tau$ forms a real
subspace $X_\tau$ of $X$. In the real setup, the double fibration
often simplifies to the nonholomorphic fibration
\begin{equation}
\pi:~Z~\rightarrow X_\tau~.
\end{equation}
This happens when $\pi_1^{-1}(X_\tau)\cong Z$ and therefore
$\pi_2$ becomes a bijection. The correspondence
\eqref{correspond1} is preserved in this case.

In particular, if field theories are given on the spaces $Z$ and
$X$, then a correspondence of the type \eqref{correspond1} between
both spaces can be lifted to a correspondence between solutions to
the field equations on $Z$ and solutions to those on $X$. In
general, this correspondence will not be one-to-one, since fields
are usually defined only up to some (gauge) equivalence. However,
one can often establish a one-to-one correspondence between
elements of the {\em moduli spaces} of theories on $Z$ and those
on $X$. This will be specialized later in concrete examples. Our
considerations in this paper are based on the results of many
authors (see e.g.\ \cite{Penrose:in}-\cite{Ivanova:2000af} and
references therein). More details on twistor theory and the
Penrose-Ward transform can be found in the books
\cite{Penrose:ca}-\cite{Mason:rf}.

\section{Twistor geometry}\label{secTwistorspace}

{\bf Local coordinates.} We start from the complex projective
space $\CPP^3$ (the {\em twistor space}) with homogeneous
coordinates $(\omega^\alpha,\lambda_\ald)$ subject to the
equivalence relation
$(\omega^\alpha,\lambda_\ald)\sim(t\omega^\alpha,t\lambda_\ald)$
for all nonzero complex $t$, $\alpha=1,2$ and
$\ald=\dot{1},\dot{2}$. Consider now the space
$\CP^3=\CPP^3\backslash\CPP^1$ in which $(\lambda_\ald)\neq 0$.
This space can be covered by two patches $\CU_+$
($\lambda_{\dot{1}}\neq 0$) and $\CU_-$ ($\lambda_{\dot{2}}\neq
0$) with coordinates
\begin{equation}\label{coords}
z_+^\alpha=\frac{\omega^\alpha}{\lambda_{\dot{1}}}~,~~~
z_+^3=\frac{\lambda_{\dot{2}}}{\lambda_{\dot{1}}}=:\lambda_+~~~
\mbox{on}~~\CU_+~~~ \mbox{and} ~~~
z_-^\alpha=\frac{\omega^\alpha}{\lambda_{\dot{2}}}~,~~~
z_-^3=\frac{\lambda_{\dot{1}}}{\lambda_{\dot{2}}}=:\lambda_-~~~
\mbox{on}~~\CU_-~,
\end{equation}
related by
\begin{equation}\label{coordstrafo}
z_+^\alpha=z_+^3z_-^\alpha~~~\mbox{ and }~~~z_+^3=\frac{1}{z_-^3}
\end{equation}
on the overlap $\CU_+\cap\CU_-$.

{}From \eqref{coords} and \eqref{coordstrafo}, it is obvious that
$\CP^3=\CU_+\cup\,\CU_-$ coincides with the total space of the
rank 2 holomorphic vector bundle\footnote{The holomorphic line
bundle $\CO(n)$ over $\CPP^1$ has the transition function
$\lambda_+^n$ and the first Chern number $n$. See appendix B for
more details.} $\CP^3=\CO(1)\oplus\CO(1)$ over $\CPP^1$,
\begin{equation}\label{bundle}
\CP^3\rightarrow\CPP^1~,
\end{equation}
where the Riemann sphere $\CPP^1$ is covered by two patches
\begin{equation}
U_+=\CU_+\cap\CPP^1~~~\mbox{and}~~~U_-=\CU_-\cap\CPP^1
\end{equation}
with coordinates $\lambda_+$ on $U_+$ and $\lambda_-$ on $U_-$,
$\CPP^1=U_+\cup U_-$.

Special cases of the correspondence \eqref{gnldblfibration},
\eqref{correspond1} can be established between the twistor space
$Z=\CPP^3$ and the Gra{\ss}mann manifold $X=G_{2,4}(\FC)$
\cite{Penrose:ca} as well as the space
$Z=\CP^3=\CPP^3\backslash\CPP^1$ (which we also call twistor
space) and $X=\FC^4\subset G_{2,4}(\FC)$. In the following, we
will focus on the geometry of the latter correspondence.

\smallskip
\noindent{\bf Moduli space of curves.} Holomorphic sections of the
complex vector bundle \eqref{bundle} are rational curves $\CPP^1_x
\embd \CP^3$ defined by the equations
\begin{equation}\label{sections}
z_+^\alpha=x^{\alpha \dot{1}}+\lambda_+ x^{\alpha \dot{2}}~~
\mbox{for}~~\lambda_+\in U_+~~~\mbox{ and }~~~
z_-^\alpha=\lambda_-x^{\alpha\dot{1}}+x^{\alpha\dot{2}}~~
\mbox{for}~~\lambda_-\in U_-
\end{equation}
and parametrized by moduli $x=(x^{\alpha\ald})\in\FC^4$.
Introducing
\begin{equation}
\left(\lambda_\ald^+\right):=\left(\begin{array}{c}1\\\lambda_+
\end{array}\right)~~~\mbox{and}~~~
\left(\lambda_\ald^-\right):=\left(\begin{array}{c}\lambda_-\\1
\end{array}\right),
\end{equation}
we can rewrite \eqref{sections} as
\begin{equation}\label{sections2}
z_\pm^\alpha=x^{\alpha\dot{\alpha}}\lambda_\ald^\pm~.
\end{equation}
These equations allow us to introduce a double
fibration\footnote{Note that we use the same notation $\pi_1$ and
$\pi_2$ for projections in completely different diagrams
throughout the paper.}
\begin{equation}\label{dblfibration}
\begin{picture}(50,40)
\put(0.0,0.0){\makebox(0,0)[c]{$\CP^3$}}
%\put(32.0,0.1){\makebox(0,0)[c]{$\Leftrightarrow$}}
\put(64.0,0.0){\makebox(0,0)[c]{$\FC^4$}}
\put(34.0,33.0){\makebox(0,0)[c]{$\CF^5$}}
\put(7.0,18.0){\makebox(0,0)[c]{$\pi_2$}}
\put(55.0,18.0){\makebox(0,0)[c]{$\pi_1$}}
\put(25.0,25.0){\vector(-1,-1){18}}
\put(37.0,25.0){\vector(1,-1){18}}
\end{picture}
\end{equation}
where $\CF^5:=\FC^4\times\CPP^1$. {}From this diagram, one
observes that a point $x=(x^{\alpha\ald})\in\FC^4$ corresponds
to the projective line $\CPP^1_x=\pi_2(\pi_1^{-1}(x))$ in $\CP^3$
given by solutions of \eqref{sections2} for fixed $x$, and a
point
$\pZ=(z^\alpha_\pm,\lambda_\ald^\pm)\in\CP^3$ corresponds\footnote{In
the literature, $\CP^3$ is often called the {\em dual} twistor
space, while the space of totally null self-dual 2-planes
($\alpha$-planes) is called the twistor space.} to a totally null
anti-self-dual 2-plane ($\beta$-plane) $\pi_1(\pi_2^{-1}(\pZ))$ in
$\FC^4$ defined by solutions of \eqref{sections2} for fixed
$(z^\alpha_\pm,\lambda_\ald^\pm)$. The double fibration
\eqref{dblfibration} and its $\RZ_2$-graded\footnote{As usual in
theoretical physics, we will use the prefix ``super'' instead of
$\RZ_2$-graded, and do {\em not} imply supersymmetry by that
term.} generalizations play the central r\^ole in the twistor
correspondence.

\smallskip
\noindent{\bf Real structures.} Recall that a real structure on a
complex manifold $M$ is defined as an antiholomorphic involution
$\tau:M \rightarrow M$. {}To introduce real structures on $\CP^3$,
let us consider three anti-linear transformations of dotted and
undotted (commuting) spinors,
\begin{align}\label{trafo1}
(\omega^\alpha)\mapsto\tau_\eps(\omega^\alpha)=&\left(\begin{array}{cc}
0 & \eps\\1 & 0\end{array}\right)\left(\begin{array}{c}
\bar{\omega}^1\\\bar{\omega}^2\end{array}\right)=\left(\begin{array}{c}
\eps\bar{\omega}^2\\\bar{\omega}^1\end{array}\right)=:(\hat{\omega}^\alpha)~,
\\[0.2cm]\label{trafo2}
(\lambda_\ald)\mapsto\tau_\eps(\lambda_\ald)=&\left(\begin{array}{cc}
0 & \eps\\1 & 0\end{array}\right)\left(\begin{array}{c}
\bar{\lambda}_{\dot{1}}\\\bar{\lambda}_{\dot{2}}\end{array}\right)=
\left(\begin{array}{c}
\eps\bar{\lambda}_{\dot{2}}\\\bar{\lambda}_{\dot{1}}\end{array}\right)=:
(\hat{\lambda}_\ald)~,\\[0.2cm]\label{trafo3}
&\tau_0(\omega^\alpha)=(\bar{\omega}^\alpha)~,~~~\tau_0(\lambda_\ald)=(\bl_\ald)~,
\end{align}
where $\eps=\pm 1$. For later use, we define furthermore
$(\hat{\lambda}^\ald):=\tau(\lambda^\ald)$, i.e.\ indices are
raised as $\lambda^\ald=\eps^{\ald\bed}\lambda_\bed$ with
$\eps^{\dot{1}\dot{2}}=-\eps^{\dot{2}\dot{1}}=-1$ before the
action of $\tau$. The transformations
\eqref{trafo1}-\eqref{trafo3} define {\em three} real structures
on $\CP^3$ which in the coordinates \eqref{coords} are given by
the formul\ae
\begin{equation}\label{three}
\tau_\eps(z_+^1,z_+^2,\lambda_+)=\left(\frac{\bz_+^2}{\bl_+},
\frac{\eps\bz_+^1}{\bl_+},\frac{\eps}{\bl_+}\right),~~~
\tau_\eps(z_-^1,z_-^2,\lambda_-)=\left(\frac{\eps\bz_-^2}{\bl_-}
,\frac{\bz_-^1}{\bl_-},\frac{\eps}{\bl_-}\right),
\end{equation}
\begin{equation}
\tau_0(z_\pm^1,z_\pm^2,\lambda_\pm)=(\bz_\pm^1,\bz^2_\pm,\bl_\pm)~.
\end{equation}
It is obvious that the involution $\tau_{-1}$ has no fixed points
but does leave invariant projective lines joining $\pZ$ and
$\tau_{-1}(\pZ)$ for any $\pZ\in\CP^3$. On the other hand, the
involutions $\tau_1$ and $\tau_0$ have fixed points which form a
three-dimensional real manifold
\begin{equation}\label{realtwistor}
\CT^3=\FR P^3\backslash\FR P^1
\end{equation}
fibred over $S^1\cong\FR P^1\subset \CPP^1$. The space
$\CT^3\subset\CP^3$ is called {\em real} twistor space. For the real
structure $\tau_1$, this space is described by the coordinates
$(z_\pm^1,\,\de^{\di\chi}\,\bz_\pm^1,\,\de^{\di\chi})$ with
$0\leq\chi<2\pi$, and for the real structure $\tau_0$, the
coordinates $(z_\pm^1,\,z_\pm^2,\,\lambda_\pm)$ are real. These two
descriptions are equivalent.

We shall concentrate on the real structures $\tau_{\pm 1}$ since
all formul\ae{} for these two cases can be written in a unified
form using $\eps=\pm 1$. For instance, an extension of the
involution $\tau_\eps$ to any function $f(x,\lambda_+)$ is defined
as
\begin{equation}\label{involution}
\tau_\eps(f(x,\lambda_+)):=\overline{f(\tau_\eps(x,\lambda_+))}:=
\overline{f\left(\tau_\eps(x),\frac{\eps}{\bl_+}\right)}~,
\end{equation}
where the bar denotes ordinary complex conjugation. Using this
involution, one can impose the condition of invariance under
\eqref{involution} on sections \eqref{sections2} of the
bundle \eqref{bundle}. Due to \eqref{sections} and \eqref{three},
moduli $(x^{\alpha\ald})$ of such sections
satisfy the equations
\begin{equation}\label{realsections}
x^{2\dot{2}}=\bar{x}^{1\dot{1}}=:-(\eps x^4+\di x^3)~~~\mbox{and}~~~
x^{2\dot{1}}=\eps\bar{x}^{1\dot{2}}=:-\eps(x^2-\di x^1)~,
\end{equation}
where $(x^\mu)$ are the real coordinates with $\mu=1,...,4$. On
the other hand, holomorphic sections of the bundle \eqref{bundle}
which are invariant under the involution $\tau_0$ are parametrized
by real coordinates
\begin{equation}\nonumber
x^{\alpha\ald}=\bar{x}^{\alpha\ald}
\end{equation}
\begin{equation}\label{realcoor}
\Rightarrow~~
x^{1\dot{1}}=x^4+x^1~,~~~
x^{1\dot{2}}=x^2-x^3~,~~~
x^{2\dot{1}}=x^2+x^3~,~~~
x^{2\dot{2}}=x^4-x^1~.
\end{equation}

\smallskip\noindent{\bf Metric on the moduli space of real
curves.} On the space $\FR^4$ of real holomorphic curves
$\CPP^1_x\embd\CP^3$, one can introduce the metric
\begin{equation}\label{realmetric}
\dd s^2=\det(\dd x^{\alpha\ald})=g_{\mu\nu}\dd x^\mu\dd x^\nu
\end{equation}
with $g=\diag(+1,+1,+1,+1)$ for the involution $\tau_{-1}$ on
$\CP^3$ and $g=\diag(-1,-1,+1,+1)$ for $\tau_{1}$ (and $\tau_0$)
and $g=(g_{\mu\nu})$. Thus, the moduli space of real rational
curves of degree one in $\CP^3$ is the Euclidean space\footnote{In
our notation, $\FR^{p,q}=(\FR^{p+q},g)$ is the space $\FR^{p+q}$
with the metric
$g=\diag(\underbrace{-1,...,-1}_{q},\underbrace{+1,...,+1}_{p})$.}
$\FR^{4,0}$ or the Kleinian space $\FR^{2,2}$.

Note that, in the Euclidean case, the twistor space $\CP^3$ is the
space
\begin{equation}\label{twp}
\FR^4\times \CPP^1\cong\CP^3
\end{equation}
with the coordinates $(x^\mu,\lambda_\pm)$ and one can define a
trivial nonholomorphic fibration
\begin{equation}\label{fibration1}
\pi:~\CP^3\rightarrow \FR^4
\end{equation}
over the space $\FR^4$ with real coordinates $x=(x^\mu)$.
Therefore, on the patches $\CU_+$ and $\CU_-$ covering the space
$\CP^3$, one can use the coordinates $(x,\lambda_+)$ and
$(x,\lambda_-)$, respectively. In the Euclidean case, the double
fibration \eqref{dblfibration} simplifies to the fibration
\eqref{fibration1} since $\pi_1^{-1}(\FR^4)\cong\CP^3\subset\CF^5$
and therefore the restriction of the projection $\pi_2$ to
$\pi_1^{-1}(\FR^4)$ is a bijection.

The twistor correspondence for the Kleinian case is more
complicated. In particular, we have
\begin{equation}
\FR^4\times(\CPP^1\backslash S^1)\cong\CP^3\backslash\CP_0
\end{equation}
instead of the diffeomorphism \eqref{twp} and one should consider
the space $\tilde{\CP}^3:=\CP^3\backslash\CP_0$ with
$\CP_0=\left.\CP^3\right|_{|\lambda_\pm|=1}\cong\FR^4\times S^1$
instead of $\CP^3$ in \eqref{fibration1}. For more details on the
Kleinian case $\eps=+1$, see appendix C. {}To smoothen the
discussion in the following, we ignore this subtlety and use
always \eqref{fibration1} implying the restriction to the space
$\tilde{\CP^3}$ in all necessary cases. Furthermore, we will call
matrix-valued functions $\tau_\eps$-regular, if they are
regular\footnote{By `regular', we mean smooth and having
nonvanishing determinant.} for all values of $\lambda\in D$ in the
case $\eps=-1$ and regular for $\lambda\in D$ with $|\lambda|\neq
1$ in the case $\eps=+1$ (and also for the real structure
$\tau_0$), where $D\subseteq\CPP^1$ is the domain under
consideration.

\smallskip\noindent{\bf Vector fields.} On the complex manifold
$\CP^3$, we have the natural basis $(\dpar/\dpar
\bz_\pm^\alpha,\dpar/\dpar \bz^3_\pm)$ in the space of
antiholomorphic vector fields with
\begin{equation}\label{dertraforules}
\der{\bz_+^\alpha}=\frac{1}{\bz_+^3}\der{\bz^\alpha_-}~~~\mbox{and}
~~~\der{\bz_+^3}=-\frac{1}{(\bz_+^3)^2}\der{\bz^3_-}-\frac{1}{\bz_+^3}\bz_-^\alpha\der{\bz_-^\alpha}
\end{equation}
on the intersection $\CU_+\cap\CU_-$. In the coordinates
$(x,\lambda_+)$ on $\CU_+$ we have\footnote{In the case $\eps=+1$,
this transformation is only valid for $|\lambda_\pm|\neq 1$. For
more details, see appendix C.}
\begin{equation}\label{ident}
\begin{split}
\der{\bz_+^1}&=\gamma_+\left(\der{x^{2\dot{2}}}-\lambda_+
\der{x^{2\dot{1}}}\right)=\gamma_+\lambda_+^\ald\der{x^{2\ald}}
=:\gamma_+\bar{V}_2^+~,\\
\der{\bz_+^2}&=\eps\gamma_+\left(\der{x^{1\dot{2}}}-\lambda_+
\der{x^{1\dot{1}}}\right)=\eps\gamma_+\lambda_+^\ald\der{x^{1\ald}}
=:\eps\gamma_+\bar{V}_1^+~,\\
\der{\bz_+^3}&=\der{\bl_+}-\eps\gamma_+
x^{\alpha\dot{1}}\bar{V}_\alpha^+=: \bar{V}_3^+-\eps\gamma_+
x^{\alpha\dot{1}}\bar{V}_\alpha^+,
\end{split}
\end{equation}
where
\begin{equation}\label{gamma+}
\gamma_+=\frac{1}{1-\eps\lambda_+\bl_+}=\frac{1}{\lambda^\ald_+\hat{\lambda}^+_\ald}~~~\mbox{and}~~~(\lambda_+^\ald)=
(\eps^{\ald\bed}\lambda^+_\bed)=\left(\begin{array}{c}-\lambda_+\\1
\end{array}\right)
\end{equation}
with the convention\footnote{See appendix B.} $\eps^{12}:=-\eps^{21}=-1$.
Thus the vector fields
\begin{equation}\label{basis+}
\bar{V}_\alpha^+=\lambda_+^\ald\dpar_{\alpha\ald}~~~\mbox{and}~~
\bar{V}_3^+=\der{\bl_+}
\end{equation}
form a basis of vector fields of type (0,1) over $\CU_+\subset\CP^3$ in
coordinates $(x,\lambda_+)$, where
$\dpar_{\alpha\ald}:=\dpar/\dpar x^{\alpha\ald}$. The explicit
form of the basis of vector fields of type (0,1) on the open set
$\CU_-\subset\CP^3$ follows from the transformation rules
\eqref{dertraforules},
\begin{equation}\label{basis-}
\bar{V}_\alpha^-=\lambda_-^\ald\dpar_{\alpha\ald}~~~
\mbox{and}~~~\bar{V}_3^-=\der{\bl_-}~,
\end{equation}
where one introduces additionally
\begin{equation}\label{gamma-}
\gamma_-=-\eps\frac{1}{1-\eps\lambda_-\bl_-}=\frac{1}{\lambda^\ald_-
\hat{\lambda}^-_\ald}~~~\mbox{and}~~~
(\lambda^\ald_-)=\left(\begin{array}{c} -1 \\ \lambda_-
\end{array}\right)~.
\end{equation}
Note that the bases \eqref{basis+} and \eqref{basis-} of vector
fields on $\CP^3$ are holonomic, i.e.\ they commute pairwise.
Furthermore, in the case $\eps=+1$, the identification
\eqref{ident} only holds on $\tilde{\CP}^3$.

\section{The twistor description of self-dual Yang-Mills
fields}\label{SDYMfields}

{\bf Holomorphic bundles over the twistor space.} Consider a rank
$n$ holomorphic vector bundle $E$ over the twistor space $\CP^3$.
This bundle is defined by a holomorphic transition function
$f_{+-}$ on the intersection $\CU_+\cap\CU_-$ of the two patches
covering $\CP^3=\CU_+\cup\CU_-$, i.e.\ the function $f_{+-}$ takes
values in the group of nonsingular $n\times n$ matrices
annihilated by the vector fields \eqref{basis+} of type (0,1):
\begin{equation}\label{holomorphicity}
\bar{V}_\alpha^+f_{+-}=0=\dpar_{\bl_+}f_{+-}~.
\end{equation}
In the twistor approach to self-duality, it is assumed that $E$ is
topologically trivial and its restriction to any projective line
$\CPP^1_x\embd\CP^3$ is holomorphically trivial. These two
conditions imply that there are regular matrix-valued functions,
$\psi_+$ on $\CU_+$ and $\psi_-$ on $\CU_-$, such that
\begin{equation}\label{trafofunc}
f_{+-}=\psi_+^{-1}\psi_-
\end{equation}
and
\begin{equation}\label{holo2}
\dpar_{\bl_+}\psi_+=0=\dpar_{\bl_-}\psi_-~,
\end{equation}
i.e.\ $\psi_+$ and $\psi_-$ are smooth functions of $x\in \FR^4$
and holomorphic functions of $\lambda_+$ and $\lambda_-$,
respectively.

\smallskip\noindent{\bf Gauge potentials.} {}From the
holomorphicity condition \eqref{holomorphicity} together with
\eqref{trafofunc} it follows that
\begin{equation}\label{gluing}
\psi_+\bar{V}_\alpha^+\psi_+^{-1}=\psi_-\bar{V}_\alpha^+\psi_-^{-1}
\end{equation}
on $\CU_+\cap\,\CU_-$. One can expand $\psi_+$, $\psi_+^{-1}$ and
$\psi_-$, $\psi_-^{-1}$ as power series in $\lambda_+$ and
$\lambda_-=\lambda_+^{-1}$, respectively. Upon substituting the
expansions into equations \eqref{gluing}, one sees that both sides
in \eqref{gluing} must be linear in $\lambda_+$ and one can
introduce Lie-algebra valued fields $(A_{\alpha\ald})$ depending
only on $x$ by the formul\ae
\begin{equation}\label{contractedA}
\lambda^\ald_+ \,
A_{\alpha\ald}=\lambda_+^\ald\,\psi_+\,\dpar_{\alpha\ald}\,\psi_+^{-1}=
\lambda_+^\ald\,\psi_-\,\dpar_{\alpha\ald}\,\psi_-^{-1}~,
\end{equation}
where $(\lambda_+^\ald)$ is given in \eqref{gamma+}. The
matrix-valued functions $(A_{\alpha\ald}(x))$ can be identified
with the components of a gauge potential $A_{\alpha\ald}\dd
x^{\alpha\ald}$ on $\FR^4$ with the metric
$(g_{\mu\nu})=\diag(-\eps,-\eps,+1,+1)$. Note that
$A_{\alpha\ald}\dd x^{\alpha\ald}$ will be an antihermitean
$n\times n$ matrix if $\psi_\pm$  satisfies the following
condition\footnote{Here $\dagger$ means hermitean conjugation.}:
\begin{equation}
\psi_+^{-1}(x,\lambda_+)=\left(\psi_-\left(x,\frac{\eps}{\bl_-}\right)
\right)^\dagger~.
\end{equation}
The antihermitean gauge potential components can be calculated
from \eqref{contractedA} as
\begin{equation}\label{Acompsah}
A_{1\dot{2}}=\psi_+\dpar_{1\dot{2}}\psi_+^{-1}|^{\phantom{\pm}}_{\lambda_+=0}=
-\eps A_{2\dot{1}}^\dagger~,~~~
A_{2\dot{2}}=\psi_+\dpar_{2\dot{2}}\psi_+^{-1}|^{\phantom{\pm}}_{\lambda_+=0}=
-A_{1\dot{1}}^\dagger~.
\end{equation}
Combining \eqref{gluing} and \eqref{contractedA}, we
introduce matrix-valued functions\footnote{Here
$V\lrcorner\,A^{0,1}$ denotes the interior product of a vector
field $V$ and a $(0,1)$-form $A^{0,1}$.}
\begin{equation}\label{cntrctdA}
A_\alpha^+:={\bar{V}^+_\alpha}\lrcorner\,A^{0,1}=
\psi_+\bar{V}_\alpha^+\psi_+^{-1}= \lambda_+^\ald A_{\alpha\ald}~,
\end{equation}
where $\lambda_+^\ald$ are given in \eqref{gamma+}. Analogously,
one can introduce the component $A_{\bl_+}$, but it will vanish
as\footnote{Note also that $A^+_\alpha$ and $A_{\bl_+}$ are
components of $A^{0,1}$.}
\begin{equation}\label{holo3}
A_{\bl_+}=\psi_+\dpar^{\phantom{\pm}}_{\bl_+}\psi_+^{-1}=0~.
\end{equation}

\smallskip\noindent{\bf Linear system and SDYM equations.} Let us
rewrite \eqref{cntrctdA} together with \eqref{holo2} in the
form
\begin{eqnarray}\label{actionVF1}
(\bar{V}_\alpha^++A_\alpha^+)\psi_+&=&0~,\\\label{actionVF2}
\dpar_{\bl_+}\psi_+&=&0
\end{eqnarray}
with similar equations for $\psi_-$. The compatibility conditions
of this linear system are
\begin{equation}
[\bar{V}_\alpha^++A_\alpha^+\,,\bar{V}_\beta^++A_\beta^+]=
\lambda_+^\ald\lambda_+^\bed
[\dpar_{\alpha\ald}+A_{\alpha\ald}\,,\dpar_{\beta\bed}+A_{\beta\bed}]=:
\lambda_+^\ald\lambda_+^\bed F_{\alpha\ald,\beta\bed}=0~.
\end{equation}
{}To be satisfied for all $(\lambda_+^\ald)$, this equation has to
vanish to all orders in $\lambda_+$ separately, from which we
obtain the self-dual Yang-Mills (SDYM) equations
\begin{equation}
F_{1\dot{1},2\dot{1}}=0~,~~~F_{1\dot{2},2\dot{2}}=0~,~~~
F_{1\dot{1},2\dot{2}}+F_{1\dot{2},2\dot{1}}=0
\end{equation}
for a gauge potential $(A_{\alpha\ald})$. It is convenient to
introduce the notation
\begin{equation}
F_{\alpha\ald,\beta\bed}=\eps_{\alpha\beta}f_{\ald\bed}+\
\eps_{\ald\bed}f_{\alpha\beta}~,
\end{equation}
in which the SDYM equations are rewritten as
\begin{equation}
f_{\ald\bed}=0~.
\end{equation}

\smallskip\noindent{\bf Gauge equivalent linear systems.} Note
that in \eqref{trafofunc}, we have chosen special
trivia\-li\-zations\footnote{On the bundle $E$, the \v{C}ech fibre
coordinates $\chi_\pm (z_\pm^\alpha , \lambda_\pm )$ are related
by $\chi_+=f_{+-}\chi_-$ on $\CU_+\cap\CU_-$. At the same time,
the matrices $\psi_+$ and $\psi_-$ are matrix fundamental
solutions of eqs.\ \eqref{actionVF1}, \eqref{actionVF2}, i.e.\ the
columns of $\psi_\pm$ form smooth frame fields for $E$ over
$\CU_\pm$. In other words, regular matrix-valued functions
$\psi_\pm$ define a trivialization of $E$ over $\CU_\pm$ such that
$\psi_\pm\chi_\pm$ are sections of $E$ holomorphic w.r.t. a new
complex structure (defined by $\bar\partial+A^{0,1}$). They are
related by $\psi_+\chi_+=\psi_-\chi_-$ on $\CU_+\cap\CU_-$.}
$\psi_\pm$ of $E$ over $\CU_\pm$ such that \eqref{holo2} and
therefore \eqref{holo3} were satisfied. However, one can consider
more general trivializations $\{\hat{\psi}_\pm\}$ of $E$ such that
\begin{equation}\label{trafo4}
f_{+-}=\psi_+^{-1}\psi_-=\hat{\psi}_+^{-1}\hat{\psi}_-
\end{equation}
and $\dpar_{\bl_\pm}\hat{\psi}_\pm\neq 0$, i.e.\
$\hat{\psi}_\pm=\hat{\psi}_\pm(x,\lambda_\pm,\bl_\pm)$ are regular
matrix-valued functions smooth in all coordinates on $\CU_\pm$.
{}From \eqref{trafo4} it follows that
\begin{equation}
\varphi:=\psi_+\hat{\psi}_+^{-1}=\psi_-\hat{\psi}_-^{-1}
\end{equation}
is a globally defined regular matrix-valued function on $\CP^3$
and therefore the above two trivializations are related by the
gauge transformation
\begin{equation}
\psi_\pm~\mapsto~\hat{\psi}_\pm=\varphi^{-1}\psi_\pm~.
\end{equation}

In general trivializations $\{\hat{\psi}_\pm\}$, we have
\begin{align}\label{triv1}
\hat{A}^+_\alpha&:=\hat{\psi}_+\bar{V}_\alpha^+\hat{\psi}_+^{-1}=
\hat{\psi}_-\bar{V}_\alpha^+\hat{\psi}_-^{-1}=
\varphi^{-1}(\psi_\pm\bar{V}_\alpha^+\psi_\pm^{-1})\varphi+
\varphi^{-1}\bar{V}_\alpha^+\varphi=
\varphi^{-1}A_\alpha^+\varphi+\varphi^{-1}\bar{V}_\alpha^+\varphi~,\\
\label{triv2}
\hat{A}_{\bl_+}&:=\hat{\psi}_+\dpar^{\phantom{\pm}}_{\bl_+}\hat{\psi}_+^{-1}=
\hat{\psi}_-\dpar^{\phantom{\pm}}_{\bl_+}\hat{\psi}_-^{-1}=
\varphi^{-1}\dpar^{\phantom{\pm}}_{\bl_+}\varphi~,
\end{align}
where the last equality in \eqref{triv2} follows from
\eqref{actionVF2} and \eqref{holo3}. Equations \eqref{triv1} and
\eqref{triv2} can be rewritten in the form of a linear system
\begin{eqnarray}\label{linsys1}
(\bar{V}_\alpha^++\hat{A}_\alpha^+)\hat{\psi}_+&=&0~,\\\label{linsys2}
(\dpar_{\bl_+}+\hat{A}_{\bl_+})\hat{\psi}_+&=&0~,
\end{eqnarray}
which is gauge equivalent to the linear system
\eqref{actionVF1}, \eqref{actionVF2}.

The compatibility conditions of this linear system are in fact the
field equations of hCS theory on the space $\CP^3$, and, e.g.\ on
$\CU_+$, they take the form
\begin{eqnarray}
\bar{V}_\alpha^+\hat{A}_\beta^+-
\bar{V}_\beta^+\hat{A}_\alpha^++
[\hat{A}_\alpha^+,\hat{A}_\beta^+]&=&0~,\\
\dpar_{\bl_+}\hat{A}_\alpha^+- \bar{V}_\alpha^+\hat{A}_{\bl_+}+
[\hat{A}_{\bl_+},\hat{A}_\alpha^+]&=&0~.
\end{eqnarray}
Thus we have a relation between hCS on $\CP^3$ and SDYM on
 the moduli space $\FR^4$ of real holomorphic sections of the
fibration $\CP^3\rightarrow \CPP^1$.

More explicitly, from the formul\ae{}
\eqref{cntrctdA} and \eqref{holo3}, we obtain
\begin{equation}\label{pwtrafo1}
A_{\alpha\dot{1}}=-\oint_{S^1}\frac{\dd
\lambda_+}{2\pi\di}\frac{A_\alpha^+}{\lambda_+^2}~~~\mbox{and}~~~
A_{\alpha\dot{2}}=\oint_{S^1}\frac{\dd
\lambda_+}{2\pi\di}\frac{A_\alpha^+}{\lambda_+}~,
\end{equation}
where the contour $S^1=\{\lambda_+\in\CPP^1:|\lambda_+|=r<1\}$
encircles $\lambda_+=0$. Using \eqref{cntrctdA}, one can easily
show the equivalence of \eqref{pwtrafo1} to \eqref{Acompsah}. The
formul\ae{} \eqref{pwtrafo1} define the Penrose-Ward transform
\begin{equation}
\CP\mathcal{W}:(A^+_\alpha,A_{\bl_+}\hspace{-4pt}=0)\mapsto(A_{\alpha\ald})
\end{equation}
which together with a preceding gauge transformation
\begin{equation}
(\hat{A}^+_\alpha,\hat{A}_{\bl_+})\stackrel{\varphi}{\longmapsto}(A^+_\alpha,A_{\bl_+}\hspace{-4pt}=0)
\end{equation}
maps solutions $(\hat{A}_\alpha^+,\hat{A}_{\bl_+})$ of the field
equations of hCS theory on $\CP^3$ to solutions $(A_{\alpha\ald})$
of the SDYM equations on $\FR^4$. Conversely, by formul\ae{}
\eqref{cntrctdA} and \eqref{holo3}, any solution
$(A_{\alpha\ald})$ of the SDYM equations corresponds to a solution
$(\hat{A}_\alpha^+,\hat{A}_{\bl_+})$ of the field equations of hCS
theory on $\CP^3$ which directly defines the inverse Penrose-Ward
transform $\CP\mathcal{W}^{-1}$. Note that gauge
transformations\footnote{There are two gauge transformations for
gauge potentials on two different spaces present in the
discussion.} of $(\hat{A}^+_\alpha,\hat{A}_{\bl_+})$ on $\CP^3$
and $(A_{\alpha\ald})$ on $\FR^4$ do not change the transition
function $f_{+-}$ of the holomorphic bundle $E\rightarrow\CP^3$.
Therefore, we have a one-to-one correspondence between gauge
equivalence classes of solutions (i.e.\ points of the moduli
spaces) to the field equations of hCS theory\footnote{Note that in
this twistor correspondence, it is assumed that for solutions of
hCS theory, there exists a gauge in which $A_{\bl_\pm}{=}0$. This
is equivalent to the holomorphic triviality of the bundle
$E\rightarrow\CP^3$ on $\CPP^1_x\embd\CP^3$. Solutions
$(A_\alpha^\pm , A_{\bl_\pm}{=}0)$
 form a subset in the
set of all solutions of the hCS theory on $\CP^3$ and we imply the
restriction to this subset when speaking of solutions to hCS
theory in the context of a twistor correspondence.} on $\CP^3$ and
the SDYM equations on $\FR^4$.

Recall that the trivializations defined by \eqref{trafofunc} and
\eqref{holo2} correspond to holomorphic triviality of the bundle
$E|_{\CPP^1_x}$ for any $\CPP^1_x\embd\CP^3$. Similarly, we may
consider restrictions of $E$ to fibres $\FC^2_\lambda$ of the
fibration $\CP^3\rightarrow \CPP^1$. All these restrictions are
holomorphically trivial due to the contractibility of
$\FC^2_\lambda$ for any $\lambda\in \CPP^1$. Therefore there exist
regular matrix-valued functions
$\tilde{\psi}_\pm(z_\pm^\alpha,\lambda_\pm,\bl_\pm)$ depending
holomorphically on $z_\pm^\alpha$ (and nonholomorphically on
$\lambda_\pm$) such that
\begin{equation}\label{trafofunc2}
f_{+-}=\psi_+^{-1}\psi_-=\tilde{\psi}_+^{-1}\tilde{\psi}_-~,
\end{equation}
and
$\tilde{\varphi}:=\psi_+\tilde{\psi}_+^{-1}=\psi_-\tilde{\psi}^{-1}_-$
defines a gauge transformation
\begin{equation}\label{gaugetrafo}
(A_\alpha^+\cb
A_{\bl_+}\hspace{-4pt}=0)~\stackrel{\tilde{\varphi}}{\longmapsto}~
(\tilde{A}_\alpha^+=0\cb\tilde{A}_{\bl_+})
\end{equation}
to a special trivialization in which only
$\tilde{A}_{\bl_+}\neq 0$ and $\bar{V}^+_\alpha(
\tilde{A}_{\bl_+})=0$. Certainly, both trivializations in
\eqref{trafofunc2} and \eqref{gaugetrafo} are gauge equivalent to
the general one given by \eqref{trafo4}-\eqref{linsys2}. This kind
of equivalence will be particularly useful in the super case.

\section{Supertwistor geometry}\label{secSTwistors}

{\bf Coordinates.} A super extension of the twistor
space $\CPP^3$ is the supermanifold $\CPP^{3|\CN}$ with
homogeneous coordinates $(\omega^\alpha,\lambda_\ald,\eta_i)$
subject to the identification
$(\omega^\alpha,\lambda_\ald,\eta_i)\sim
(t\,\omega^\alpha,t\,\lambda_\ald,t\,\eta_i)$ for any nonzero complex
scalar $t$. Here $(\omega^\alpha,\lambda_\ald)$ are homogeneous
coordinates on $\CPP^3$ and $\eta_i$ with $i=1,...,\CN$ are Gra\ss
mann variables. Interestingly, this supertwistor space is a
Calabi-Yau supermanifold in the case $\CN{=}4$ and one may consider
B-type open topological strings living in this space
\cite{Witten:2003nn}.

As in section 2, we consider the space
$\CP^3=\CPP^3\backslash\CPP^1=\CO(1)\oplus\CO(1)$
and its super extension $\CP^{3|\CN}$ covered by two
patches, $\CP^{3|\CN}=\CPP^{3|\CN}\backslash\CPP^{1|\CN}=
\hat{\CU}_+\cup\hat{\CU}_-$, with even
coordinates \eqref{coords} and odd coordinates
\begin{equation}\label{ferm1}
\eta_i^+=\frac{\eta_i}{\lambda_{\dot{1}}}~~\mbox{on}~~\hat{\CU}_+
~~~\mbox{and}~~~\eta_i^-=\frac{\eta_i}{\lambda_{\dot{2}}}~~
\mbox{on}~~\hat{\CU}_-
\end{equation}
related by
\begin{equation}\label{ferm2}
\eta^+_i=z_+^3\eta^-_i
\end{equation}
on $\hat{\CU}_+\cap\hat{\CU}_-$. We see from \eqref{ferm1} and
\eqref{ferm2} that the fermionic coordinates are sections
of\/\footnote{The operator $\Pi$ inverts the parity of fibre
coordinates, see appendix B.} $\Pi\CO(1)$. The supermanifold
$\CP^{3|\CN}$ is fibred over $\CPP^1$,
\begin{equation}\label{superbundle}
\CP^{3|\CN}\rightarrow\CPP^{1|0}~,
\end{equation}
with superspaces $\FC^{2|\CN}_\lambda$ as fibres over
$\lambda\in\CPP^{1|0}=\CPP^1$. We also have a fibration
\begin{equation}\label{superbundle2}
\CP^{3|\CN}\rightarrow\CPP^{1|\CN}
\end{equation}
with $\FC_{\lambda,\eta}^{2|0}$ as fibres over $(\lambda,\eta)
\in\CPP^{1|\CN}$.

\smallskip\noindent{\bf Spaces of supermoduli and chirality.}
Holomorphic sections of the bundle \eqref{superbundle} are
rational curves $\CPP^1_{x_R,\eta}\embd\CP^{3|\CN}$ defined by the
equations\footnote{Here $\top$ means the transpose.}
\begin{align}\nonumber
z_+^\alpha&=x^{\alpha\ald}_R\,\lambda_\ald^+\,,~~~
\eta_i^+=\eta_i^\ald\lambda_\ald^+~~~\mbox{for
}(\lambda_\ald^+)^\top=(1,\lambda_+),~~\lambda_+\in
U_+~,\\\label{supercoords}
z_-^\alpha&=x^{\alpha\ald}_R\,\lambda_\ald^-\,,~~~
\eta_i^-=\eta_i^\ald\lambda_\ald^-~~~\mbox{for
}(\lambda_\ald^-)^\top=(\lambda_-,1),~~\lambda_-\in U_-
\end{align}
and parametrized by supermoduli
$(x_R,\eta)=(x_R^{\alpha\ald},\eta_i^\ald)\in\FC^{4|2\CN}=:\CM_R^{4|2\CN}$.
The space $\CM_R^{4|2\CN}$ is called anti-chiral superspace.
Equations \eqref{supercoords} define a supertwistor correspondence
via a double fibration
\begin{equation}\label{superdblfibration}
\begin{picture}(50,40)
\put(0.0,0.0){\makebox(0,0)[c]{$\CP^{3|\CN}$}}
%\put(37.0,0.1){\makebox(0,0)[c]{$\Leftrightarrow$}}
\put(74.0,0.0){\makebox(0,0)[c]{$\CM_R^{4|2\CN}$}}
\put(42.0,37.0){\makebox(0,0)[c]{$\CF_R^{5|2\CN}$}}
\put(7.0,20.0){\makebox(0,0)[c]{$\pi_2$}}
\put(65.0,20.0){\makebox(0,0)[c]{$\pi_1$}}
\put(25.0,27.0){\vector(-1,-1){18}}
\put(47.0,27.0){\vector(1,-1){18}}
\end{picture}
\end{equation}
where $\CF_R^{5|2\CN}:=\CM_R^{4|2\CN}\times \CPP^1$, a point
$(x_R,\eta)=(x_R^{\alpha\ald},\eta_i^\ald)\in\CM_R^{4|2\CN}$
corresponds to the sphere
$\CPP^1_{x_R,\eta}=\pi_2(\pi_1^{-1}(x_R,\eta))\embd \CP^{3|\CN}$,
and a point
$\pZ=(z_\pm^\alpha,\lambda_\pm,\eta_i^\pm)\in\CP^{3|\CN}$
corresponds to a null $\beta_R$-superplane
$\pi_1(\pi_2^{-1}(\pZ))\embd \CM_R^{4|2\CN}$ of dimension
$(2|\CN)$.

Holomorphic sections of the bundle \eqref{superbundle2} are spaces
$\CPP^{1|\CN}_{x_L,\theta}\embd\CP^{3|\CN}$ defined by the
equations
\begin{equation}\label{supercoordsleft}
z^\alpha_\pm=x_L^{\alpha\ald}\lambda_\ald^\pm-2\theta^{\alpha
i}\eta^\pm_i~~~\mbox{with}~~(\lambda_\ald^\pm,\eta^\pm_i)\in
\hat{\CU}_\pm\cap\CPP^{1|\CN}
\end{equation}
and parametrized by supermoduli
$(x_L,\theta)=(x^{\alpha\ald}_L,\theta^{\alpha
i})\in\FC^{4|2\CN}=:\CM_L^{4|2\CN}$. The space $\CM_L^{4|2\CN}$ is
called chiral superspace. Equations \eqref{supercoordsleft}
define another supertwistor correspondence,
\begin{equation}\label{superdblfibration1.5}
\begin{picture}(50,40)
\put(0.0,0.0){\makebox(0,0)[c]{$\CP^{3|\CN}$}}
%\put(37.0,0.1){\makebox(0,0)[c]{$\Leftrightarrow$}}
\put(74.0,0.0){\makebox(0,0)[c]{$\CM_L^{4|2\CN}$}}
\put(42.0,37.0){\makebox(0,0)[c]{$\CF_L^{5|3\CN}$}}
\put(7.0,20.0){\makebox(0,0)[c]{$\pi_2$}}
\put(65.0,20.0){\makebox(0,0)[c]{$\pi_1$}}
\put(25.0,27.0){\vector(-1,-1){18}}
\put(47.0,27.0){\vector(1,-1){18}}
\end{picture}
\end{equation}
where $\CF_L^{5|3\CN}:=\CM_L^{4|2\CN}\times\CPP^{1|\CN}$, and a
point $(x_L,\theta)=(x_L^{\alpha\ald},\theta^{\alpha
i})\in\CM_L^{4|2\CN}$ corresponds to the supersphere
$\CPP^{1|\CN}_{x_L,\theta}=\pi_2(\pi_1^{-1}(x_L,\theta))\embd
\CP^{3|\CN}$, and a point $\pZ\in\CP^{3|\CN}$ corresponds to a
null $\beta_L$-superplane
$\pi_1(\pi_2^{-1}(\pZ))\embd\CM_L^{4|2\CN}$ of dimensions
$(2|2\CN)$.

{}From \eqref{supercoords} and \eqref{supercoordsleft} it follows
that
\begin{equation}\label{coords4}
x_R^{\alpha\ald}=x^{\alpha\ald}-\theta^{\alpha
i}\eta^\ald_i\vspace{-0.3cm}
\end{equation}
and
\begin{equation}\label{coords4L}
x_L^{\alpha\ald}=x^{\alpha\ald}+\theta^{\alpha i}\eta^\ald_i~,
\end{equation}
where $(x^{\alpha\ald})\in\FC^{4|0}$ are ``symmetric'' (nonchiral)
bosonic coordinates. In the following, we will continue labelling
chiral objects by a subscript $L$ for left-handed or chiral and a
subscript $R$ for right-handed or anti-chiral ones. Substituting
\eqref{coords4} into \eqref{supercoords}, we obtain the equations
\begin{equation}
z_\pm^\alpha=x^{\alpha\ald}\lambda^\pm_\ald-\theta^{\alpha
i}\eta_i^\ald\lambda_\ald^\pm~~~\mbox{and}~~~\eta_i^\pm=\eta_i^\ald\lambda_\ald^\pm
\end{equation}
defining degree one curves $\CPP^1_{x,\theta,\eta}\embd
\CP^{3|\CN}$ parametrized by supermoduli
$(x,\theta,\eta)=(x^{\alpha\ald},\theta^{\alpha
i},\eta^\ald_i)\in\FC^{4|4\CN}$. Therefore we obtain a double
fibration
\begin{equation}\label{superdblfibration2}
\begin{picture}(50,40)
\put(0.0,0.0){\makebox(0,0)[c]{$\CP^{3|\CN}$}}
%\put(37.0,0.1){\makebox(0,0)[c]{$\Leftrightarrow$}}
\put(74.0,0.0){\makebox(0,0)[c]{$\FC^{4|4\CN}$}}
\put(42.0,37.0){\makebox(0,0)[c]{$\CF^{5|4\CN}$}}
\put(7.0,20.0){\makebox(0,0)[c]{$\pi_2$}}
\put(65.0,20.0){\makebox(0,0)[c]{$\pi_1$}}
\put(25.0,27.0){\vector(-1,-1){18}}
\put(47.0,27.0){\vector(1,-1){18}}
\end{picture}
\end{equation}
with coordinates
\begin{eqnarray}
&(x^{\alpha\ald}\cb\lambda_\ald^\pm\cb\theta^{\alpha
i}\cb\eta_i^\ald)~~~\mbox{on}~~\CF^{5|4\CN}:=\FC^{4|4\CN}
\times\CPP^1~,\\\label{coordsc1}
&(x^{\alpha\ald}\cb\theta^{\alpha
i}\cb\eta^\ald_i)~~~\mbox{on}~~\FC^{4|4\CN}~,
\end{eqnarray}
and
\begin{equation}\label{coords6}
z_\pm^\alpha=(x^{\alpha\ald}-\theta^{\alpha i}\eta^\ald_i)
\lambda_\ald^\pm~,~~~\lambda_\ald^\pm~,~~~
\eta_i^\pm=\eta_i^\ald\lambda_\ald^\pm~~~\mbox{on}~~\CP^{3|\CN}~.
\end{equation}
The definition of the projection $\pi_1$ is obvious and $\pi_2$ is
defined by \eqref{coords6} and \eqref{coords4}.

The double fibration \eqref{superdblfibration2} generalizes both
\eqref{superdblfibration} and \eqref{superdblfibration1.5} and
defines the following twistor correspondence: a point
$(x,\theta,\eta)=(x^{\alpha\ald},\theta^{\alpha
i},\eta_i^\ald)\in\FC^{4|4\CN}$ corresponds to the projective line
$\CPP^1_{x,\theta,\eta}=\pi_2(\pi_1^{-1}(x,\theta,\eta))\embd\CP^{3|\CN}$,
and a point $\pZ\in\CP^{3|\CN}$ corresponds to a totally
null $\beta$-superplane
$\pi_1(\pi_2^{-1}(\pZ))\embd\FC^{4|4\CN}$ of dimensions
$(2|3\CN)$.

\smallskip\noindent{\bf Vector fields.} Note that one can project
from $\CF^{5|4\CN}$ onto $\CP^{3|\CN}$ in two steps: first from
$\CF^{5|4\CN}$ onto $\CF^{5|2\CN}_R$, which is given in
coordinates by
\begin{equation}\label{projection1}
(x^{\alpha\ald}\cb\lambda_\ald^\pm\cb\theta^{\alpha i}\cb\eta^\ald_i)
\rightarrow (x_R^{\alpha\ald}\cb\lambda_\ald^\pm\cb\eta_i^\ald)
\end{equation}
with the $x_R^{\alpha\ald}$ from \eqref{coords4}, and then from
$\CF_R^{5|2\CN}$ onto $\CP^{3|\CN}$, which is given in coordinates
by
\begin{equation}\label{projection2}
(x_R^{\alpha\ald}\cb\lambda_\ald^\pm\cb\eta_i^\ald)\rightarrow
(x_R^{\alpha\ald}\lambda_\ald^\pm\cb\lambda_\ald^\pm\cb\eta_i^\ald
\lambda_\ald^\pm)~.
\end{equation}
The tangent spaces to the $(0|2\CN)$-dimensional leaves of the
fibration \eqref{projection1} are spanned by the vector fields
\begin{equation}\label{vectorfields0}
D_{\alpha i}=\der{\theta^{\alpha
i}}+\eta_i^\ald\der{x^{\alpha\ald}}=:
\dpar_{\alpha i}+\eta_i^\ald\dpar_{\alpha\ald}
\end{equation}
on $\FC^{4|4\CN}\subset\CF^{5|4\CN}$. The coordinates
$x_R^{\alpha\ald}$, $\lambda_\ald^\pm$ and $\eta^\ald_i$ belong to
the kernel of these vector fields which are also tangent to the
fibres of the projection $\FC^{4|4\CN}\rightarrow \FC^{4|2\CN}$
onto the anti-chiral superspace. The tangent spaces to the
$(2|\CN)$-dimensional leaves of the projection \eqref{projection2}
are spanned by the vector fields\footnote{For the definition of
$\lambda_\pm^\ald$, see section \ref{secTwistorspace}}
\begin{eqnarray}\label{vectorfields1}
\bar{V}_\alpha^\pm&=&\lambda_\pm^\ald\dpar^R_{\alpha\ald}~,\\
\label{vectorfields2}
\bar{\partial}_\pm^i&=&\lambda_\pm^\ald\, \dpar_\ald^i~~~\mbox{with}~~
\dpar_\ald^i=\der{\eta^\ald_i}~,
\end{eqnarray}
where $\dpar^R_{\alpha\ald}=\dpar/\dpar x_R^{\alpha\ald}$.

\smallskip\noindent{\bf Twistor correspondence for a real
superspace.}
Let us now discuss the action of the anti-linear involutions
\eqref{trafo1}-\eqref{trafo3} on fermionic coordinates. In the
Kleinian case, we can simply define
\begin{equation}\label{tau1}
\tau_1\left(\begin{array}{c}\theta^{1i} \\ \theta^{2i}
\end{array}\right)=
\left(\begin{array}{c}\bar{\theta}^{2i} \\ \bar{\theta}^{1i}
\end{array}\right)~,~~~
\tau_1\left(\begin{array}{c}\eta^{\dot{1}}_{i} \\[0.8mm] \eta^{\dot{2}}_{i}
\end{array}\right)=
\left(\begin{array}{c}\bar{\eta}^{\dot{2}}_{i} \\[0.8mm]
\bar{\eta}^{\dot{1}}_{i}
\end{array}\right)
\end{equation}
and
\begin{equation}\label{tau2}
\tau_0(\theta^{\alpha i})=\bar{\theta}^{\alpha i}~~~\mbox{and}~~~
\tau_0(\eta^{\ald}_i)=\bar{\eta}^{\ald}_i~,
\end{equation}
which matches the definition for commuting spinors.

In the Euclidean case $\eps=-1$, we can only fix a real structure
on the fermionic coordinates if the number of supersymmetries
$\CN$ is even (see e.g.\ \cite{Kotrla:1984ky, Lukierski:1986jw}).
In these cases, one groups together the fermionic coordinates in
pairs of two and defines matrices
\begin{equation}
(\epsilon^r_s):=\left(\begin{array}{cc} 0 & -1 \\ 1 & 0
\end{array}\right)~,~~r,s=1,2~~~\mbox{and}~~~
(T^i_j):=\left(\begin{array}{ccc} \epsilon & 0 \\
0 & \epsilon
\end{array}\right)~,~~i,j=1,...,4~.
\end{equation}
 The action of $\tau_{-1}$ is then given by
\begin{equation}\label{tau3}
\tau_{-1}\left(\begin{array}{cc}\theta^{11} & \theta^{1 2} \\
\theta^{2 1} & \theta^{2 2}
\end{array}\right)=\left(\begin{array}{cc}
0 & -1 \\ 1 & 0
\end{array}\right)
\left(\begin{array}{cc}\bar{\theta}^{1 1} & \bar{\theta}^{1 2} \\
\bar{\theta}^{2 1} & \bar{\theta}^{2 2}
\end{array}\right)\left(\begin{array}{cc}
0 & -1 \\ 1 & 0
\end{array}\right)
\end{equation}
for $\CN{=}2$ and
\begin{equation}\label{tau4}
\tau_{-1}\left(\begin{array}{ccc}\theta^{11} & \cdots &\theta^{1 4} \\
\theta^{2 1} & \cdots & \theta^{2 4}
\end{array}\right)=\left(\begin{array}{cc}
0 & -1 \\ 1 & 0
\end{array}\right)
\left(\begin{array}{ccc}\bar{\theta}^{1 1} & \cdots &\bar{\theta}^{1 4} \\
\bar{\theta}^{2 1} & \cdots & \bar{\theta}^{2 4}
\end{array}\right)\left(\begin{array}{cccc}
0 & -1 & 0 & 0 \\ 1 & 0 & 0 & 0 \\
0 & 0 & 0 & -1\\
0 & 0 & 1 & 0
\end{array}\right)
\end{equation}
for $\CN{=}4$. The last equation can also be written in components as
\begin{equation}
\tau_{-1}(\theta^{\alpha i})=-\eps^{\alpha\beta}
T^i_j\bar{\theta}^{\beta j}~,
\end{equation}
where there is a summation over $\beta$ and $j$. The same
definition applies to $\eta^\ald_i$:
\begin{equation}\label{tau5}
\tau_{-1}(\eta_i^\ald)=-\eps^{\ald\bed} T_i^j\bar{\eta}_{j}^\bed~.
\end{equation}
In general, we define furthermore
$\tau(\xi^1\xi^2)=\tau(\xi^1)\tau(\xi^2)$ for the product of any two
Gra\ss mann variables\footnote{See appendix B.}. These conventions
imply
\begin{equation}
\overline{\der{\xi}}=\der{\bar{\xi}}
\end{equation}
for any Gra{\ss}mann variable $\xi$.

Using the involutions \eqref{tau1} and \eqref{tau2}, we can impose
the following Majorana conditions on the odd variables
$\theta^{\alpha i}$ and $\eta^\ald_i$:
\begin{align}\label{majorana1}
\tau_1(\theta^{\alpha i})=\theta^{\alpha i}~~~\mbox{and}~~~
\tau_1(\eta^\ald_i)=\eta^\ald_i~~&\Leftrightarrow~~
\theta^{2i}=\bar{\theta}^{1i}~~~\mbox{and}~~~
\eta_i^{\dot{2}}=\bar{\eta}_i^{\dot{1}}~,\\
\label{majorana1.5}
\tau_0(\theta^{\alpha i})=\theta^{\alpha i}~~~\mbox{and}~~~
\tau_0(\eta^\ald_i)=\eta^\ald_i~~&\Leftrightarrow~~
\theta^{\alpha i}=\bar{\theta}^{\alpha i}~~~\mbox{and}~~~
\eta_i^{\dot{\alpha}}=\bar{\eta}_i^{\dot{\alpha}}~.
\end{align}
For the Euclidean case \eqref{tau3}-\eqref{tau5}, the reality
conditions
\begin{equation}\label{realcond}
\tau_{-1}(\theta^{\alpha i})=\theta^{\alpha i}~~~\mbox{and}~~~
\tau_{-1}(\eta_i^\ald)=\eta_i^\ald
\end{equation}
can be read off from \eqref{tau3} and \eqref{tau4} for $\CN{=}2$
and $\CN{=}4$, respectively. For instance, for $\CN{=}2$ we have
\begin{equation}\label{realcnd}
\theta^{22}=-\bar{\theta}^{11}\quad\mbox{and}\quad
\theta^{21}=\bar{\theta}^{12}~.
\end{equation}
Imposing the reality conditions \eqref{realsections} or
\eqref{realcoor} on $x^{\alpha\ald}$ and corresponding conditions
from \eqref{majorana1}-\eqref{realcnd} on $\theta^{\alpha i}$ and
$\eta_i^\ald$, we obtain real chiral coordinates
$x_L^{\alpha\ald}$ and $x_R^{\alpha\ald}$ for all three real
structures. Note that contrary to the Minkowski signature $(3,1)$,
the variables $\theta^{\alpha i}$ and $\eta^{\ald}_i$ are
independent for both signatures (4,0) and (2,2).

We concentrate now on the real cases defined by the involutions
$\tau_1$ and $\tau_{-1}$. Note in advance that complexified
self-dual $\CN$-extended SYM theory can be described by the
diagram \eqref{superdblfibration} with coordinates
$(x_R^{\alpha\ald},\eta^\ald_i)$ on $\CM_R^{4|2\CN}$ and
$(z_\pm^\alpha,\lambda_\ald^\pm,\eta^\pm_i)=
(x_R^{\alpha\ald}\lambda^\pm_\ald,\lambda_\ald^\pm,
\eta_i^\ald\lambda_\ald^\pm)$ on $\CP^{3|\CN}$. After imposing the
reality condition \eqref{realsections} and \eqref{majorana1} or
\eqref{realcond}, the coordinates $(x_R^{\alpha\ald},\eta_i^\ald)$
belong to the {\em real} anti-chiral superspace
$\FR^{4|2\CN}=:\CR_R^{4|2\CN}$ of real dimension\footnote{Note
that in Minkowski signature, chiral and anti-chiral superspaces
are always complex.} $(4|2\CN)$. We keep the coordinates
$\lambda_\ald^\pm$ complex\footnote{In the Euclidean case, there
are no fixed points of $\tau_{-1}$ on the Riemann sphere
$\CPP^1\ni(\lambda^\pm_\ald)$ and therefore $\lambda^\pm_\ald$
must be complex. The case of signature (2,2) is discussed in the
appendices C and D.} and therefore the supertwistor space
$\CP^{3|\CN}$ has complex dimension $(3|\CN)$.

{}For coordinates $x_R^{\alpha\ald}$ satisfying
\eqref{realsections}, the vector fields \eqref{vectorfields1} and
\begin{equation}\label{pole}
\bar{V}_3^\pm =\dpar_{\bl_\pm}
\end{equation}
can be identified with bosonic vector fields of type $(0,1)$ on
$\CP^{3|\CN}$ similar to the vector fields \eqref{basis+} and
\eqref{basis-} on $\CP^{3|0}$. In the Euclidean case, this is due
to the fact that as a real supermanifold, $\CP^{3|\CN}$ is
diffeomorphic\footnote{In the case $\eps=+1$ (and also for the
real structure $\tau_0$), there is a diffeomorphism of an open
subset $\tilde{\CP}^{3|\CN}$ of $\CP^{3|\CN}$ onto the space
$\CR_R^{4|2\CN}\times(\CPP^1\backslash S^1)$. See appendix C for
more details.} to the space $\CR_R^{4|2\CN}\times \CPP^1$,
\begin{equation}
\CP^{3|\CN}\cong\CR_R^{4|2\CN}\times
\CPP^1=\pi^{-1}_1(\CR_R^{4|2\CN})\subset
\CF^{5|2\CN}_R~,
\end{equation}
where $\pi_1$ is one of the projections in the diagram
\eqref{superdblfibration}. In other words, the map $\pi_2$,
restricted to $\pi_1^{-1}(\CR_R^{4|2\CN})$, is one-to-one (cf.
\eqref{twp} for the purely bosonic case). Moreover,
\eqref{vectorfields2} become odd vector fields of type $(0,1)$ on
$\CP^{3|\CN}$ annihilating all complex coordinates on this space.
For example, for $\tau_1$-real vector fields with
$|\lambda_\pm|\neq 1$, we have
\begin{align}
&\bar{\dpar}_+^i=\lambda_+^\ald\der{\eta^\ald_i}=
\der{\bar{\eta}_i^{\dot{1}}}-\lambda_+
\der{\eta_i^{\dot{1}}}~,\\
&\bar{\dpar}_+^i\eta^+_j=0\quad\mbox{and}\quad
\bar{\dpar}^i_+\bar{\eta}^+_j=\gamma_+^{-1}\delta^i_j~,
\end{align}
where ${\eta}^+_i={\eta}_i^{\dot{1}}+\lambda_+\bar{\eta}_i^{\dot{1}}$
and $\gamma_+$ is given in \eqref{gamma+}. Similar formul\ae{}
can be written down for $\tau_{-1}$-real and $\tau_0$-real cases.
Thus, in the real setup, the double fibration
\eqref{superdblfibration} simplifies to the nonholomorphic
fibration
\begin{equation}\label{supfib1}
\pi:~\CP^{3|\CN}~\rightarrow~\CR^{4|2\CN}_R~,
\end{equation}
where $(3|\CN)$ stands for complex and $(4|2\CN)$ for real
dimensions. Fibres over a point $(x_R,\eta)\in\CR^{4|2\CN}_R$
in the fibration \eqref{supfib1} are real holomorphic sections
$\CPP^1_{x_R,\eta}\embd\CP^{3|\CN}$ described by
\eqref{supercoords} with real $(x_R,\eta)\in\CR_R^{4|2\CN}$.

\section{The supertwistor description of self-dual
$\CN$-extended super Yang-Mills theory}\label{secSSDYM}

{\bf Super self-duality for extended supersymmetry.} Self-dual
Yang-Mills (SDYM) fields on $\FR^{4,0}$ and $\FR^{2,2}$ are
solutions to the self-duality equations
\begin{equation}\label{SDcond}
F_{\mu\nu}=\frac{1}{2}\eps_{\mu\nu\rho\sigma}F^{\rho\sigma}~~~
\mbox{or}~~~F=*F~,
\end{equation}
which is equivalently written in spinor notation as
\begin{equation}
f_{\ald\bed}:=-\frac{1}{2}\eps^{\alpha\beta}(\dpar_{\alpha\ald}A_{\beta\bed}-
\dpar_{\beta\bed}A_{\alpha\ald}
+[A_{\alpha\ald},A_{\beta\bed}])=0~.
\end{equation}
Solutions to these equations form a subset of the solution space
of Yang-Mills theory. Thus, a possible supersymmetric extension of
the self-duality equations can be obtained by taking the full set
of SYM field equations and imposing certain constraints on them.
These constraints have to include \eqref{SDcond} and keep the
resulting set of equations invariant under supersymmetry
transformations. This works for SYM theories with $\CN\leq 3$, and
the field content of the full $\CN$-extended SYM theory splits
into a self-dual supermultiplet and an anti-self-dual
supermultiplet. For $\CN{=}4$, the situation is more complicated,
as the SYM multiplet $(f_{\alpha\beta},\chi^{\alpha
i},\phi^{ij},\tilde{\chi}_{\ald i},f_{\ald\bed})$, where the
fields have the helicities $(+1,+\frac{1}{2},0,-\frac{1}{2},-1)$,
is irreducible. By introducing an additional field $G_{\ald\bed}$
with helicity -1, which takes in some sense the place of
$f_{\ald\bed}$, one can circumvent this problem (see e.g.\
\cite{Siegel:1992za,Devchand:1996gv}). Using this field, one can
even obtain the $\CN{=}4$ super SDYM equations of motion from an
action principle \cite{Siegel:1992za}. Note that although the
field content appearing in the corresponding Lagrangian is
$(f_{\alpha\beta},\chi^{\alpha i},\phi^{ij},\tilde{\chi}_{\ald
i},f_{\ald\bed},G_{\ald\bed})$, $f_{\ald\bed}$ vanishes due to the
SDYM equations of motion and the supermultiplet of non-trivial
fields is $(f_{\alpha\beta},\chi^{\alpha
i},\phi^{ij},\tilde{\chi}_{\ald i},G_{\ald\bed})$. These degrees
of freedom match exactly those of the full $\CN{=}4$ SYM and often
-- by a slight abuse of language -- it is stated that they are the
same. Following this line, one can even consider the full
$\CN{=}4$ SYM theory and $\CN{=}4$ SDYM as the same theories on
linearized level, which are only distinguished by different
interactions.

{}For all values $0\leq\CN\leq4$, supersymmetrically extended self-dual SYM
theories can be nicely described in terms of twistor theory, which
will be shown in this section.

\smallskip\noindent{\bf Holomorphic bundles and gauge potentials.} The twistor
description of complexified self-dual $\CN$-extended SYM theory is
known and based on the diagram \eqref{superdblfibration2} (see
e.g.\ \cite{Semikhatov:wj,Volovich:ii,Tafel:1985qk}) or
(implicitly) on the diagram \eqref{superdblfibration} (see e.g.\
\cite{Devchand:1992st,Devchand:1996gv}). Here, we consider
self-dual $\CN$-extended SYM theory in the real setting based on
the fibration \eqref{supfib1}. Namely, let us consider a
holomorphic bundle $\CE$ over the supertwistor space
$\CP^{3|\CN}=\hat{\CU}_+\cup\hat{\CU}_-$ without $\RZ_2$-grading
on the fibres and with the coordinates \eqref{coords},
\eqref{ferm1}. As usual, the bundle $\CE\rightarrow \CP^{3|\CN}$
is defined by a holomorphic transition function annihilated by the
vector fields \eqref{vectorfields1}, \eqref{vectorfields2},
\eqref{pole} of type $(0,1)$ on $\CP^{3|\CN}$,
\begin{eqnarray}\label{5.3}
\bar{V}_\alpha^+f_{+-}&=&0~,\\
\dpar_{\bl_+} f_{+-}&=&0
\end{eqnarray}
and
\begin{equation}\label{5.5}
\bar{\dpar}^i_+f_{+-}=0~.
\end{equation}

Further, it is assumed that the restriction of $\CE$ to any
projective line $\CPP^1_{x_R,\eta}\embd\CP^{3|\CN}$ is
holomorphically trivial and therefore there exist regular
matrix-valued functions
$\psi_+(x_R^{\alpha\ald},\lambda_+,\eta^\ald_i)$ on $\hat{\CU}_+$
and $\psi_-(x_R^{\alpha\ald},\lambda_-,\eta^\ald_i)$ on
$\hat{\CU}_-$ such that
\begin{equation}\label{strafo0}
f_{+-}=\psi_+^{-1}\psi_-
\end{equation}
and
\begin{equation}\label{Donpsi}
\dpar_{\bl_\pm}\psi_\pm=0~.
\end{equation}
This trivialization is similar to \eqref{trafofunc}, \eqref{holo2}
in the purely bosonic case, and using arguments identical to those
from section 3, one can introduce matrix-valued components of a
gauge potential,
\begin{eqnarray}
\CA_\alpha^+:=\bar{V}_\alpha^+\lrcorner\,\CA&=&
\psi_+\bar{V}_\alpha^+\psi_+^{-1}=\psi_-\bar{V}^+_\alpha\psi_-^{-1}=
\lambda_+^\ald\CA_{\alpha\ald} (x_R,\eta)~,\\
\label{A3cpt} \CA_{\bl_+}:=\dpar_{\bl_+}\lrcorner\,\CA&=&
\psi_+\dpar_{\bl_+}\psi_+^{-1}=\psi_-\dpar_{\bl_+}\psi_-^{-1}=0~,\\
\CA^i_+:=\bar{\dpar}^i_+\lrcorner\,\CA&=&
\psi_+\dparb_+^i\psi_+^{-1}=\psi_-
\dparb_+^i\psi_-^{-1}=\lambda_+^\ald\CA_\ald^i(x_R,\eta)~,
\end{eqnarray}
where $(x_R,\eta)=(x_R^{\alpha\ald},\eta^\ald_i)$, and a linear
system
\begin{eqnarray}\label{slinsys1}
(\bar{V}_\alpha^++\CA_\alpha^+)\psi_+&=&0~,\\
\dpar_{\bl_+}\psi_+&=&0~,\\\label{slinsys4}
(\bar{\dpar}_+^i+\CA_+^i)\psi_+&=&0
\end{eqnarray}
of differential equations equivalent to the existence of
holomorphic sections of the bundle $\CE$.

\smallskip\noindent{\bf Super self-duality.} The compatibility
conditions of the linear system \eqref{slinsys1}-\eqref{slinsys4}
read
\begin{equation}
[\nabla_{\alpha\ald},\nabla_{\beta\bed}]+[\nabla_{\alpha\bed},
\nabla_{\beta\ald}]=0~,~~~
[\nabla_{\ald}^i,\nabla_{\beta\bed}]+[\nabla_{\bed}^i,\nabla_{\beta\ald}]=0~,
~~~\{\nabla_\ald^i,\nabla_\bed^j\}+\{\nabla_\bed^i,\nabla_\ald^j\}=0~,
\label{compcon}
\end{equation}
where we have introduced covariant derivatives
\begin{equation}
\nabla_{\alpha\ald}:=\dpar_{\alpha\ald}^R +\CA_{\alpha\ald}~~~\mbox{and}~~~
\nabla_\ald^i:=\dpar_\ald^i+\CA_\ald^i~.
\end{equation}

The compatibility conditions \eqref{compcon} suggest the
introduction of the following self-dual super gauge field strengths:
\begin{equation}
[\nabla_{\alpha\ald},\nabla_{\beta\bed}]=\eps_{\ald\bed}
f_{\alpha\beta}(x_R,\eta)~,~~~
[\nabla^i_{\ald},\nabla_{\beta\bed}]=\eps_{\ald\bed}
f^i_{\beta}(x_R,\eta)~,~~~
\{\nabla^i_{\ald},\nabla^j_{\bed}\}=\eps_{\ald\bed}
f^{ij}(x_R,\eta)~,
\end{equation}
where $f^{ij}$ is antisymmetric and $f_{\alpha\beta}$ is symmetric.

Let us focus on the case $\CN{=}4$ and discuss the cases $\CN{<}4$
later on. The set of physical fields for $\CN{=}4$ SYM theory
consists of the self-dual and the anti-self-dual field strengths
of a gauge potential $\CA_{\alpha\ald}$, four spinors
$\chi^i_\alpha$ together with four spinors $\tilde{\chi}_{\ald
i}\sim\eps_{ijkl}\tilde{\chi}_\ald^{jkl}$ of opposite chirality
and six real (or three complex) scalars
$\phi^{ij}=\phi^{[ij]}$. For $\CN{=}4$ super SDYM, the multiplet
is joined by an additional spin-one field
$G_{\ald\bed}\sim\eps_{ijkl}G^{ijkl}_{\ald\bed}$, as discussed
before. Now the above super gauge field strengths contain in their
expansion exactly these fields. The lowest component of
$f_{\alpha\beta}$, $f_\alpha^i$ and $f^{ij}$ will be the SDYM
field strength, the spinor field $\chi_\alpha^i$ and the scalars
$\phi^{ij}$, respectively. By using Bianchi identities for the
self-dual super gauge field strengths, one
obtains \cite{Devchand:1996gv} successively the superfield
expansions and the field equations for the physical field
content\footnote{The fields are scaled to match the discussion
following \eqref{expAa}, \eqref{expAl}.},
\begin{eqnarray}\nonumber
&&f_{\ald\bed}=-\frac{1}{2}\eps^{\alpha\beta}(\dpar_{\alpha\ald}A_{\beta\bed}-
\dpar_{\beta\bed}A_{\alpha\ald}
+[A_{\alpha\ald},A_{\beta\bed}])=0~,\\
\nonumber
&&\nabla_{\alpha\ald}\chi^{\alpha i}=0~,\\
\label{SDYMeom}
&&\Box\phi^{ij}+2\eps\{\chi^{\alpha i},\chi^j_{\alpha}\}=0~,\\
\nonumber
&&\nabla_{\alpha\ald}\tilde{\chi}^{\ald ijk}-
\eps\,[\chi^{[i}_{\alpha},\phi^{jk]}]=0~,\\
\nonumber
&&\eps^{\ald\dot{\gamma}}\nabla_{\alpha\ald}G^{[ijkl]}_{\dot{\gamma}
\dot{\delta}}+
12\eps\{\chi^{[i}_{\alpha},\tilde{\chi}^{jkl]}_{\dot{\delta}}\}
-18\eps\,[\phi^{[ij},\nabla_{\alpha\dot{\delta}}\phi^{kl]}]=0~,
\end{eqnarray}
where $\Box:=\nabla_{\alpha\ald}\nabla^{\alpha\ald}$ and the
antisymmetrizations $[i...j]$ are defined to have
weight\footnote{So $[ijk]:=ijk+jki+kij-kji-jik-ikj$ and $[ijkl]$
is obtained by contracting with $\eps_{ijkl}$.} one. Note that by
construction, all fields depend on the coordinates
$x_R^{\alpha\ald}$. As we will see soon, \eqref{SDYMeom} is in
some sense an $\CN$-independent formulation~\cite{Devchand:1996gv}
of the field equations of super SDYM theory in which the cases
$\CN{<}4$ are governed by the first $\CN{+}1$ equations of
\eqref{SDYMeom}, where $f_{\ald\bed}=0$ is counted as one equation
and so on. In the case $\CN{=}4$, one can introduce ``dualized''
fields
\begin{equation}
\phi_{ij}:=\frac{1}{2!}\eps_{ijkl}\phi^{kl}~,~~~
\tilde{\chi}^\ald_i:=\frac{1}{3!}\eps_{ijkl}\tilde{\chi}^{\ald
jkl}~~~ \mbox{and}~~~
G_{\ald\bed}:=\frac{1}{4!}\eps_{ijkl}G_{\ald\bed}^{ijkl}~,
\end{equation}
for which the equations of motion take the form:
\begin{eqnarray}\nonumber
&&f_{\ald\bed}=0~,\\\nonumber
&&\nabla_{\alpha\ald}\chi^{\alpha i}=0~,\\\label{SDYMeom2}
&&\Box\phi^{ij}+2\eps\{\chi^{\alpha i},\chi^j_{\alpha}\}=0~,\\\nonumber
&&\nabla_{\alpha\ald}\tilde{\chi}_i^{\ald}-
2\eps\,[\chi^j_{\alpha},\phi_{ij}]=0~,\\\nonumber
&&\eps^{\ald\dot{\gamma}}\nabla_{\alpha\ald}G_{\dot{\gamma}
\dot{\delta}}+ 3\eps\{\chi^{i}_{\alpha},\tilde{\chi}_{\dot{\delta}i}\}
-\frac{3}{2}\eps\,[\phi_{ij},\nabla_{\alpha\dot{\delta}}\phi^{ij}]=0~.
\end{eqnarray}
After rescaling some of the fields as
\begin{equation}
\chi^i_\alpha\rightarrow\frac{1}{2}\chi^i_\alpha~,~~~
\phi^{ij}\rightarrow\frac{1}{2}\phi^{ij}~,~~~
\tilde{\chi}_{\ald i}\rightarrow-\frac{1}{2}\tilde{\chi}_{\ald
i}~,~~~
G_{\ald\bed}\rightarrow\frac{3}{2}G_{\ald\bed}~,
\end{equation}
the equations \eqref{SDYMeom2} are the field equations of the
Lagrangian for $\CN{=}4$ self-dual SYM given in
\cite{Witten:2003nn}.

\smallskip\noindent{\bf Gauge equivalent trivializations.} We have
described the twistor correspondence between holomorphic bundles
$\CE$ over the supertwistor space $\CP^{3|\CN}$ trivial on
projective lines in $\CP^{3|\CN}$ and solutions to the field
equations of self-dual $\CN$-extended SYM theory on the space
$\FR^4$ with metric $(g_{\mu\nu})=\diag(-\eps,-\eps,+1,+1)$. The
derivation of the $\CN{=}4$ super self-duality equations
\eqref{SDYMeom} is based on trivializations of $\CE$ over
$\hat{\CU}_\pm$ such that eqs.\ \eqref{Donpsi} and therefore
\eqref{A3cpt} is satisfied. However, there are other convenient
trivializations of $\CE$ over $\hat{\CU}_\pm$ such that the
compatibility conditions of the corresponding linear system are
described by holomorphic Chern-Simons theory
\cite{Witten:1992fb,Witten:2003nn} on the supertwistor space.
Namely, since restrictions of the bundle $\CE$ to
$(2|\CN)$-dimensional leaves of the fibration \eqref{superbundle}
are trivial\footnote{Note that fibres $\FC_\lambda^{2|\CN}$ over
$\lambda\in\CPP^1$ in the fibration $\CP^{3|\CN}\rightarrow\CPP^1$
are exactly the $\beta_R$-superplanes introduces in section 4.
Super self-duality is equivalent in the discussed superfield
formulation to flatness of super Yang-Mills fields on these
$\beta_R$-superplanes.}, there exist
$\tau_\eps$-regular\footnote{See the definition on p.\ 6.}
matrix-valued functions $\hat{\psi}_\pm$ on $\tilde{\CU}_\pm$ such
that
\begin{equation}\label{strafo}
f_{+-}=\psi_+^{-1}\psi_-=\hat{\psi}_+^{-1}\hat{\psi}_-
\end{equation}
and
\begin{equation}\label{sDonpsi}
\bar{\dpar}^i_\pm\hat{\psi}_\pm=0~.
\end{equation}
This implies $\hat{\psi}_\pm=\hat{\psi}_\pm(x_R^{\alpha\ald},
\lambda_\pm, \bl_\pm, \eta^\pm_i)$. These trivializations are
analogous to the trivializations \eqref{trafo4}-\eqref{linsys2} in
the $\CN{=}0$ case. Note that similarly to \eqref{trafofunc2} in
the purely bosonic case, one can also choose the trivializations
$\tilde{\psi}_\pm(z^\alpha_\pm,\lambda_\pm,\bl_\pm,\eta_i^\pm)$
but we will not consider them here.

\smallskip\noindent{\bf Linear system.} It follows from \eqref{strafo}
that
\begin{equation}
\varphi:=\psi_+\hat{\psi}_+^{-1}=\psi_-\hat{\psi}_-^{-1}
\end{equation}
is a matrix-valued function (superfield) generating a gauge
transformation\footnote{The function $\varphi$ is
$\tau_\eps$-regular, and in particular, it can be singular on
$\CP_{0,\CN}:=\CP^{3|\CN}|_{|\lambda_\pm|=1}$ in the Kleinian case
$\eps=+1$, see appendix D.}
\begin{eqnarray}\label{5.24}
\psi_\pm&\mapsto&\hat{\psi}_\pm=\varphi^{-1}\psi_\pm~,\\\label{5.25}
\CA_\alpha^\pm&\mapsto&\hat{\CA}_\alpha^\pm=
\varphi^{-1}\CA_\alpha^\pm\varphi+
\varphi^{-1}\bar{V}^\pm_\alpha\varphi=
\hat{\psi}_\pm\bar{V}^\pm_\alpha\hat{\psi}^{-1}_\pm~,\\\label{5.26}
0=\CA_{\bl_\pm}&\mapsto&\hat{\CA}_{\bl_\pm}=
\varphi^{-1}\dpar^{\phantom{\pm}}_{\bl_\pm}\varphi=
\hat{\psi}_\pm\dpar^{\phantom{\pm}}_{\bl_\pm}
\hat{\psi}_\pm^{-1}~,\\\label{5.27}
\CA^i_\pm&\mapsto&\hat{\CA}^i_\pm= \varphi^{-1}\CA^i_\pm\varphi+
\varphi^{-1}\bar{\dpar}_\pm^i\varphi=
\hat{\psi}_\pm\bar{\dpar}_\pm^i\hat{\psi}^{-1}_\pm=0~,
\end{eqnarray}
where we used \eqref{Donpsi}. Thus, we obtain the linear system
\begin{eqnarray}\label{slinsysb1}
(\bar{V}_\alpha^++\hat{\CA}_\alpha^+)\hat{\psi}_+&=&0~,\\
(\dpar^{\phantom{\pm}}_{\bl_+}+\hat{\CA}_{\bl_+})\hat{\psi}_+&=&0~,\\
\label{slinsysb4}
\bar{\dpar}^i_+\hat{\psi}_+&=&0~,
\end{eqnarray}
which is gauge equivalent to the linear system
\eqref{slinsys1}-\eqref{slinsys4}. For $\hat{\psi}_-$ we have
similar equations, which are obtained by changing ``+'' to ``--''
in \eqref{slinsysb1}-\eqref{slinsysb4}. In the case $\CN{=}0$, this
linear system coincides with \eqref{linsys1}, \eqref{linsys2}.

\smallskip\noindent{\bf Super hCS theory.} The
compatibility conditions of the linear differential equations
\eqref{slinsysb1}-\eqref{slinsysb4} are the field equations of hCS
theory on the supertwistor space\footnote{More accurately, in the
case $\eps=+1$ (and also for the real structure $\tau_0$)
$\hat{\CA}^{0,1}$ (and thus hCS theory) is defined only on the
subset $\tilde{\CP}^{3|\CN}$ of $\CP^{3|\CN}$ for which
$|\lambda_\pm|\neq 1$. See appendix D for more details.}
$\CP^{3|\CN}$. On $\hat{\CU}_+$ they read
\begin{eqnarray}\label{shCS1}
\bar{V}_\alpha^+\hat{\CA}_\beta^+-
\bar{V}_\beta^+\hat{\CA}_\alpha^++
[\hat{\CA}_\alpha^+,\hat{\CA}_\beta^+]&=&0~,\\
\label{shCS2} \dpar_{\bl_+}\hat{\CA}_\alpha^+-
\bar{V}_\alpha^+\hat{\CA}_{\bl_+}+
[\hat{\CA}_{\bl_+},\hat{\CA}_\alpha^+]&=&0
\end{eqnarray}
and similarly on $\hat{\CU}_-$. Here $\hat{\CA}_\alpha^+$ and
$\hat{\CA}_{\bl_+}$ are functions of
$(x_R^{\alpha\ald},\lambda_+,\bl_+,\eta^+_i)$. These equations are
equivalent to the equations of self-dual $\CN$-extended SYM theory
on $\FR^4$ which form a subset of equations \eqref{SDYMeom}. As
already mentioned, the most interesting case is $\CN{=}4$ since
the supertwistor space $\CP^{3|4}$ is a CY supermanifold and one
can derive equations \eqref{shCS1}, \eqref{shCS2} from a
manifestly Lorentz invariant action \cite{Witten:2003nn,
Sokatchev:1995nj}. For this reason, we concentrate on the
equivalence with self-dual SYM for the case $\CN{=}4$; for other
values of $\CN$, the derivation goes along the same lines.

Recall that $\hat{\CA}_\alpha$ and $\hat{\CA}_\bl$ are sections of
the bundles\footnote{The bundle $\bar{\CO}(n)$ is the complex
conjugate to $\CO(n)$.} $\CO(1)\otimes\FC^2$ and $\bar{\CO}(-2)$
over $\CPP^1$ since the vector fields in \eqref{5.25} take values
in $\CO(1)$ and the holomorphic cotangent bundle of $\CPP^1$ is
$\CO(-2)$. {}Together with the fact that $\eta^\pm_i$'s take
values in the bundle $\Pi\CO(1)$, this fixes the dependence of
$\hat{\CA}^\pm_\alpha$ and $\hat{\CA}_{\bl_\pm}$ on
$\lambda^\ald_\pm$ and $\hat{\lambda}^\ald_\pm$. In the case
$\CN{=}4$, this dependence takes the form
\begin{eqnarray}\label{expAa}
\hat{\CA}_\alpha^+&=&\lambda_+^\ald\,
A_{\alpha\ald}(x_R)+\eta_i^+\chi^i_\alpha(x_R)+
\gamma_+\,\frac{1}{2!}\,\eta^+_i\eta^+_j\,\hat{\lambda}^\ald_+\,
\phi_{\alpha \ald}^{ij}(x_R)+\\
\nonumber
&&+\gamma_+^2\,\,\frac{1}{3!}\,\eta^+_i\eta^+_j\eta^+_k\,\hat{\lambda}_+^\ald\,
\hat{\lambda}_+^\bed\,
\tilde{\chi}^{ijk}_{\alpha\ald\bed}(x_R)+\gamma_+^3\,\frac{1}{4!}\,
\eta^+_i\eta^+_j\eta^+_k\eta^+_l\,
\hat{\lambda}_+^\ald\,\hat{\lambda}_+^\bed\,\hat{\lambda}_+^{\dot{\gamma}}\,
G^{ijkl}_{\alpha\ald\bed\dot{\gamma}}(x_R)~,\\
\label{expAl}
\hat{\CA}_{\bl_+}&=&\gamma_+^2\,\frac{1}{2!}\,\eta^+_i\eta^+_j\,\phi^{ij}(x_R)+
\gamma_+^3\,\frac{1}{3!}\,\eta^+_i\eta^+_j\eta^+_k\,\hat{\lambda}_+^\ald\,
\tilde{\chi}^{ijk}_{\ald}
(x_R)+\\
\nonumber
&&+\gamma_+^4\,\frac{1}{4!}\,\eta^+_i\eta^+_j\eta^+_k\eta^+_l\,
\hat{\lambda}_+^\ald\,\hat{\lambda}_+^\bed
G^{ijkl}_{\ald\bed}(x_R)
\end{eqnarray}
and similar for $\hat{\CA}_\alpha^-, \hat{\CA}_{\bl_-}$. Here,
again,
$A_{\alpha\ald}\cb\chi_\alpha^i\cb\phi^{ij}\cb\tilde{\chi}_{\ald
i}$ is the ordinary field content of $\CN{=}4$ super Yang-Mills
theory and the field $G_{\ald\bed}$ is the auxiliary field arising
in the $\CN{=}4$ self-dual case, as discussed above.  It follows
from \eqref{shCS2}-\eqref{expAl} that\footnote{We use the
symmetrization $(\cdot)$ with weight one, e.g.\
$(\ald\bed)=\ald\bed+\bed\ald$}
\begin{equation}\label{expA2}
\phi^{ij}_{\alpha\ald}=-\nabla_{\alpha\ald}\phi^{ij}\ ,\quad
\tilde{\chi}^{ijk}_{\alpha\ald\bed}=-\frac{1}{4}\nabla_{\alpha(\ald}
\tilde{\chi}^{ijk}_{\bed)}\quad\mbox{and}\quad
G^{ijkl}_{\alpha\ald\bed\dot{\gamma}}=-\frac{1}{18}\nabla_{\alpha(\ald}
G^{ijkl}_{\bed\dot{\gamma})}\ ,
\end{equation}
i.e.\ these fields do not contain additional degrees of freedom.
The expansion \eqref{expAa}, \eqref{expAl} together with the field
equations \eqref{shCS1}, \eqref{shCS2} reproduces exactly
equations \eqref{SDYMeom}.

Consider now the cases $\CN{<}4$. Since the $\eta^+_i$'s are Gra\ss
mann variables and thus nilpotent, the expansion \eqref{expAa} and
\eqref{expAl} will only have terms up to order $\CN$ in the
$\eta^+_i$'s. This exactly reduces the expansion to the
appropriate field content for $\CN$-extended super SDYM theory:
\begin{eqnarray}
\CN=0 & & A_{\alpha\ald}\\
\CN=1 & & A_{\alpha\ald},~~\chi^i_\alpha~~~\mbox{with}~~i=1\\
\CN=2 & &
A_{\alpha\ald},~~\chi^i_\alpha,~~\phi^{[ij]}~~~\mbox{with}~~i,j=1,2\\
\CN=3 & &
A_{\alpha\ald},~~\chi^i_\alpha,~~\phi^{[ij]},~~\chi^{[ijk]}_\ald~~~
\mbox{with}~~i,j,\ldots =1,2,3\\
\CN=4 & &
A_{\alpha\ald},~~\chi^i_\alpha,~~\phi^{[ij]},~~\chi^{[ijk]}_\ald,~~
G_{\ald\bed}^{[ijkl]}~~~\mbox{with}~~i,j,\ldots =1,\ldots ,4~.
\end{eqnarray}
One should note that the antisymmetrization $[\cdot]$ leads to a
different number of fields depending on the range of $i$. For
example, in the case $\CN{=}2$, there is only one real scalar
$\phi^{12}$, while for $\CN{=}4$ there exist six real
scalars. Inserting such a truncated expansion for $\CN{<}4$ into the
field equations \eqref{shCS1} and \eqref{shCS2}, we obtain the
first $\CN{+}1$ equations of \eqref{SDYMeom}, which is the
appropriate set of equations for $\CN{<}4$ super SDYM theory.

{}To sum up, we have described a one-to-one correspondence between
gauge equivalence classes of solutions to the $\CN$-extended SDYM
equations on $({\FR^4},g)$ with $g=\diag(-\eps,-\eps,+1,+1)$ and
equivalence classes\footnote{Two holomorphic bundles
$(\CE,f_{+-})$ and $(\CE',f'_{+-})$ over $\CP^{3|\CN}$ are called
equivalent, if there exist regular matrix-valued functions $h_+$
and $h_-$, which are holomorphic on $\hat{\CU}_+$ and
$\hat{\CU}_-$, respectively, such that the transition functions
$f_{+-}$ and $f_{+-}'$ are related by
$f'_{+-}=h_+^{-1}f_{+-}h_-$.} of holomorphic vector bundles $\CE$
over the supertwistor space $\CP^{3|\CN}$ such that the bundles
$\CE$ are holomorphically trivial on each projective line
$\CPP^1_{x_R,\eta}$ in $\CP^{3|\CN}$. In other words, there is a
bijection between the moduli spaces of hCS theory on $\CP^{3|\CN}$
and the one of self-dual $\CN$-extended SYM theory on $(\FR^4,g)$.
It is assumed that appropriate reality conditions are imposed. The
Penrose-Ward transform and its inverse are defined by the
formul\ae{} \eqref{expAa}-\eqref{expA2}. In fact, these
formul\ae{} relate solutions of the equations of motion of hCS
theory on $\CP^{3|\CN}$ to those of self-dual $\CN$-extended SYM
theory on $(\FR^4,g)$. One can also write integral formul\ae{} of
type \eqref{pwtrafo1} but we refrain from doing this.

\section{Dual supertwistors and $\CN$-extended anti-self-duality}

\smallskip\noindent{\bf Coordinates.} In section
4, we described the supertwistor space $\CPP^{3|\CN}$ as the space
of $(1|0)$-dimensional subspaces in the space $\FC^{4|\CN}$. Its dual
supermanifold can be defined as a space of ${(3|\CN)}$-dimensional
planes in $\FC^{4|\CN}$ parametrized by homogeneous coordinates
$(\mu_\alpha,\sigma^\ald,\theta^i)$ subject to the identification
$(\mu_\alpha,\sigma^\ald,\theta^i)\sim(t\mu_\alpha,t\sigma^\ald,t\theta^i)$
for any nonzero complex number $t$. We again have the
supermanifold\footnote{We use the subscript $*$ to denote the dual
supertwistor space, its subspaces and its preimages under
projections $\pi_2$.} $\CPP^{3|\CN}_*$ and the space
$\CP_*^{3|\CN}=\CPP^{3|\CN}_*\backslash\CPP^{1|\CN}_*=
(\CO(1)\oplus\CO(1))_*^{3|\CN}= \hat{\CV}_+\cup\hat{\CV}_-$ with
coordinates
\begin{eqnarray}\label{ip1}
&&w_+^\ald=\frac{\sigma^\ald}{\mu_1}~,~~
w_+^{\dot3}=\zeta_+=\frac{\mu_2}{\mu_1}~~ \mbox{and} ~~
\theta_+^i=\frac{\theta^i}{\mu_1}~~ \mbox{on} ~~ \hat{\CV}_+~,\\
\label{ip2}
&&w_-^\ald=\frac{\sigma^\ald}{\mu_2}~,~~
w_-^{\dot3}=\zeta_-=\frac{\mu_1}{\mu_2}~~ \mbox{and} ~~
\theta_-^i=\frac{\theta^i}{\mu_2}~~ \mbox{on} ~~ \hat{\CV}_-~,\\
\label{ip3}
&&w_+^\ald=\zeta_+w_-^\ald~,~~~\zeta_+=\zeta_-^{-1}~, ~~~
\theta_+^i=\zeta_+\theta^i_-~~~\mbox{on}~~~\hat{\CV}_+\cap\hat{\CV}_-~.
\end{eqnarray}

Note that $\zeta_\pm$ are coordinates on the patches
$V_\pm=\hat{\CV}_\pm\cap\CPP^1_*$ covering the base\footnote{Recall
that we use the subscript $*$ in $\CPP^1_*$ for distinguishing this
Riemann sphere with the coordinates $\zeta_\pm$ from the Riemann
sphere $\CPP^1$ with the coordinates $\lambda_\pm$.}
$\CPP^1_*=\CPP^{1|0}_*$ of the holomorphic vector bundle
\begin{equation}\label{dualsection}
\CP^{3|\CN}_*\rightarrow\CPP^{1|0}_*~.
\end{equation}
Sections of this bundle (degree one holomorphic curves
$\CPP^1_{x_L,\theta}\embd\CP_*^{3|\CN}$) are defined by the
equations
\begin{equation}\label{ssections}
w_\pm^\ald=x_L^{\alpha\ald}\mu_\alpha^\pm~,~~~
\theta_\pm^i=\theta^{\alpha i}\mu_\alpha^\pm~~~
\mbox{with}~~~(\mu_\alpha^+)=\left(\begin{array}{c}1\\\zeta_+
\end{array}\right),~~~(\mu_\alpha^-)=\left(\begin{array}{c}\zeta_-\\1
\end{array}\right),
\end{equation}
and parametrized by supermoduli
$(x_L,\theta)=(x^{\alpha\ald}_L,\theta^{\alpha i})\in\FC^{4|2\CN}
=:\CM^{4|2\CN}_L$.
Real sections are obtained by imposing the reality conditions
\eqref{realsections} or \eqref{realcoor}. We denote the space
of such sections by $\CR^{4|2\CN}_{L}$ ($=\FR^{4|2\CN}$).

Similarly to the supertwistor case, we can introduce a fibration
\begin{equation}\label{anotherfibration}
\CP^{3|\CN}_*\rightarrow\CPP^{1|\CN}_*
\end{equation}
whose holomorphic sections
$\CPP^{1|\CN}_{x_R,\eta}\embd\CP^{3|\CN}_*$ are defined by the
equations
\begin{equation}\label{supersectionsL}
w_\pm^\ald=x_R^{\alpha\ald}\mu_\alpha^\pm - 2\eta_i^\ald\theta_\pm^i
\end{equation}
and parametrized by supermoduli
$(x_R,\eta)=(x_R^{\alpha\ald},\eta^\ald_i)\in\FC^{4|2\CN}=:\CM_R^{4|2\CN}$.
The real subspace $\CR^{4|2\CN}_{R}$ was considered in section 4.
So for dual supertwistors, we can again introduce double
fibrations similar to \eqref{superdblfibration} and
\eqref{superdblfibration1.5} simply by substituting the spaces
$\CF_{R}^{5|2\CN}$, $\CF_{L}^{5|3\CN}$ and $\CP^{3|\CN}$ by the
spaces $\CF_{R*}^{5|2\CN}:=\CM_{R}^{4|2\CN}\times\CPP^1_*$,
$\CF_{L*}^{5|3\CN}:=\CM_{L}^{4|2\CN}\times\CPP^{1|\CN}_*$ and
$\CP^{3|\CN}_*\supset\CPP^1_*$, respectively.

{}From \eqref{ssections} and \eqref{supersectionsL}, we obtain again
formul\ae{} \eqref{coords4}, \eqref{coords4L} and the equations
\begin{equation}\label{supersectionsL2}
w_\pm^\ald=(x^{\alpha\ald}+\theta^{\alpha
i}\eta_i^\ald)\mu_\alpha^\pm~~~\mbox{and}~~~ \theta^i_\pm
=\theta^{\alpha i}\mu_\alpha^\pm
\end{equation}
defining projective lines
$\CPP^1_{x,\theta,\eta}\embd\CP^{3|\CN}_*$ parametrized by
supermoduli $(x,\theta,\eta)=(x^{\alpha\ald},\theta^{\alpha
i},\eta^\ald_i)\in\FC^{4|4\CN}$. Via the dual supertwistor
correspondence, a point
$\pZ=(w^\ald_\pm,\zeta_\pm,\theta_\pm^i)\in\CP^{3|\CN}_*$
corresponds to a totally null $\alpha_L$-, $\alpha_R$- or
$\alpha$-superplane in $\CM^{4|2\CN}_L$, $\CM^{4|2\CN}_R$ or
$\FC^{4|4\CN}$ with dimension $(2|\CN)$, $(2|2\CN)$ or $(2|3\CN)$,
respectively.

\smallskip\noindent{\bf Vector fields.} Using \eqref{supersectionsL2}
and \eqref{ssections}, one can introduce a double fibration
\begin{equation}\label{superdblfibration4}
\begin{picture}(50,50)
\put(0.0,0.0){\makebox(0,0)[c]{$\CP_*^{3|\CN}$}}
%\put(37.0,0.1){\makebox(0,0)[c]{$\Leftrightarrow$}}
\put(74.0,0.0){\makebox(0,0)[c]{$\FC^{4|4\CN}$}}
\put(42.0,37.0){\makebox(0,0)[c]{$\CF_{*}^{5|4\CN}$}}
\put(7.0,20.0){\makebox(0,0)[c]{$\pi_2$}}
\put(65.0,20.0){\makebox(0,0)[c]{$\pi_1$}}
\put(25.0,27.0){\vector(-1,-1){18}}
\put(47.0,27.0){\vector(1,-1){18}}
\end{picture}
\end{equation}
for the complex nonchiral case and the fibration
\begin{equation}\label{superdblfibration5}
\pi:~\CP_*^{3|\CN}\ \to\ \CR^{4|2\CN}_{L}
\end{equation}
in the case of $(x^{\alpha\ald})$ and $(\theta^{\alpha i},\eta^\ald_i)$
satisfying the reality conditions induced by $\tau_\eps$ as
discussed in section 4. The superspace $\FC^{4|4\CN}$ and its real
subspace $\CR^{4|2\CN}_{L}$ are the same as in section 4 and  parametrized
by the same coordinates. We have coordinates
\begin{eqnarray}
&&(x^{\alpha\ald},\, \mu_\alpha^\pm\cb \theta^{\alpha i}\cb \eta^\ald_i)~~~
\mbox{on}~~~\CF_*^{5|4\CN}~,\\
&&(x_L^{\alpha\ald}\mu_\alpha^\pm\cb \mu_\alpha^\pm\cb \theta^{\alpha i}\mu_\alpha^\pm)~~~
\mbox{on}~~~\CP_*^{3|\CN}
\end{eqnarray}
with obvious projections in the fibrations
\eqref{superdblfibration4} and \eqref{superdblfibration5}.

It is not difficult to see that the tangent spaces of the
$(2|3\CN)$-dimensional leaves of the projection
$\pi_2:\CF_{*}^{5|4\CN}\rightarrow\CP_*^{3|\CN}$ from
\eqref{superdblfibration4} are spanned by the vector fields
\begin{equation}\label{vectorfield613}
D_\ald^i=\dpar_\ald^i+\theta^{\alpha i}\dpar_{\alpha\ald}~,
\end{equation}
\begin{equation}
D^\pm_i=\mu_\pm^\alpha D_{\alpha i}~~~\mbox{with}~~D_{\alpha
i}=\dpar_{\alpha i}+\eta^\ald_i\dpar_{\alpha\ald}~,~~
(\mu_+^\alpha)=\left(\begin{array}{c}-\zeta_+\\ 1
\end{array}\right),~~(\mu_-^\alpha)=\left(\begin{array}{c}
-1\\\zeta_-
\end{array}\right)~,
\end{equation}
and
\begin{equation}\label{pip1}
\bar{V}^\pm_\ald=\mu_\pm^\alpha\,\dpar_{\alpha\ald}^L~~~\mbox{with}~~
\dpar_{\alpha\ald}^L=\frac{\dpar}{\dpar x^{\alpha\ald}_L}~.
\end{equation}

In the real case, when the coordinates
$(x_L^{\alpha\ald}\cb\theta^{\alpha i})$ belong to the {\it real}
chiral superspace $\CR^{4|2\CN}_{L}$, we have the fibration
\eqref{superdblfibration5}, and the vector fields \eqref{pip1} and
\begin{equation}\label{pip2}
\bar{V}^\pm_{\dot{3}}=\dpar_{\bar{\zeta}_\pm}
\end{equation}
can be identified with bosonic vector fields\footnote{In the case
$\eps=+1$, this identification only holds for
$\tilde{\CP}^{3|\CN}_*=\left.\CP^{3|\CN}_*\right|_{|\zeta_\pm|\neq1}$.}
of type (0,1) on $\CP_*^{3|\CN}$ similar to the vector fields
\eqref{vectorfields1} and \eqref{pole} as discussed in section 4
for the self-dual case. As an odd vector field of type (0,1) on
$\CP^{3|\CN}_*$ we have
\begin{equation}\label{pip3}
\bar\partial^+_i = \mu^\alpha_+\frac{\partial}{\partial\theta^{\alpha i}}
\quad \mbox{on} \quad \hat\CV_+\qquad \mbox{and}\qquad
\bar\partial^-_i = \mu^\alpha_-\frac{\partial}{\partial\theta^{\alpha i}}
\quad \mbox{on} \quad \hat\CV_-\ .
\end{equation}
These bosonic and fermionic vector fields of type (0,1) on $\CP_*^{3|\CN}$
annihilate all complex coordinates \eqref{ip1}, \eqref{ip2} (or, equivalently,
\eqref{ssections}) on $\CP_*^{3|\CN}$.

\smallskip\noindent{\bf Anti-self-dual gauge fields.} Consider a
holomorphic vector bundle $\CE$ over the space
$\CP^{3|\CN}_*=\hat{\CV}_+\cup\hat{\CV}_-$ defined by a transition
function $f_{+-}$ on $\hat{\CV}_+\cap\hat{\CV}_-$. On
$\CP^{3|\CN}_{*}$, we have
\begin{equation}
\bar{V}_\ald^+ f_{+-}=0~,~~~\dpar_{\bs_+}
f_{+-}=0~~~\mbox{and}~~~
\bar{\dpar}_i^+ f_{+-}=0~
\end{equation}
since $f_{+-}$ is holomorphic. Let us consider trivializations
$\psi_\pm$ over $\hat{\CV}_\pm$ similar to \eqref{strafo0} and
\eqref{Donpsi}, i.e.\ such that
\begin{equation}
f_{+-}=\psi^{-1}_+\psi_-~~~\mbox{and}~~
\dpar_{\bs_\pm}\psi_\pm=0~.
\end{equation}
{}From these equations, we obtain matrix-valued components of a
super gauge potential one-form,
\begin{eqnarray}\label{asdAacp}
\CA_\ald^+:=\bar{V}_\ald^+\lrcorner\,\CA&=&
\psi_+\bar{V}_\ald^+\psi_+^{-1}=\psi_-\bar{V}^+_\ald\psi_-^{-1}=
\mu_+^\alpha\CA_{\alpha\ald}
(x_L,\theta)~,\\
\label{asdA3cpt}
\CA_{\bs_+}:=\dpar_{\bs_+}\lrcorner\,\CA&=&
\psi_+\dpar_{\bar{\zeta}_+}\psi_+^{-1}=
\psi_-\dpar_{\bar{\zeta}_+}\psi_-^{-1}=0~,\\
\label{asdAacpt}
\CA_i^+:={\bar{\dpar}_i^+}\lrcorner\,\CA&=&\psi_+\bar{\dpar}^+_i
\psi_+^{-1}=\psi_- \bar{\dpar}_i^+\psi_-^{-1}=
\mu_+^\alpha\CA_{\alpha i}(x_L,\theta)~,
\end{eqnarray}
and a linear system
\begin{eqnarray}\label{asdslinsys1}
(\bar{V}_\ald^++\CA_\ald^+)\psi_+&=&0~,\\
\dpar_{\bs_+}\psi_+&=&0~,\\\label{asdslinsys4}
(\bar{\dpar}^+_i+\CA^+_i)\psi_+&=&0~,
\end{eqnarray}
where the last equalities in \eqref{asdAacp} and \eqref{asdAacpt}
follow from the generalized Liouville's theorem on
$\CPP^1_*$.
The compatibility conditions of the linear system
\eqref{asdslinsys1}-\eqref{asdslinsys4} read
\begin{equation}\label{asdcompcon}
[\nabla_{\alpha\ald},\nabla_{\beta\bed}]+[\nabla_{\beta\ald},
\nabla_{\alpha\bed}]=0~,~~~
[\nabla_{\alpha i},\nabla_{\beta\bed}]+[\nabla_{\beta i},
\nabla_{\alpha\bed}]=0~,~~~
\{\nabla_{\alpha i},\nabla_{\beta j}\}+\{\nabla_{\beta i},
\nabla_{\alpha j}\}=0~,
\end{equation}
where we have -- as before -- introduced covariant derivatives
\begin{equation}
\nabla_{\alpha\ald}:=\dpar_{\alpha\ald}^L+\CA_{\alpha\ald}~~~\mbox{and}~~~
\nabla_{\alpha i}:=\dpar_{\alpha i}+\CA_{\alpha i}~.
\end{equation}
Equations \eqref{asdcompcon} are the anti-self-dual $\CN$-extended
SYM equations in superspace formulation\footnote{Note that fibres
$\FC^{2|\CN}_\zeta$ over $\zeta\in\CPP^1_*$ in the fibration
\eqref{dualsection} are the $\alpha_L$-superplanes introduced in
section 5. {}From \eqref{asdslinsys1}-\eqref{asdcompcon} it
follows that $\CN$-extended anti-self-duality in superfield
formulation is equivalent to flatness of super Yang-Mills fields
on these complex $(2|\CN)$-dimensional $\alpha_L$-superplanes.}.

As in the self-dual case, one can rewrite \eqref{asdcompcon}
in component fields. The full set of equations of motion for
$\CN{=}4$ is
\begin{eqnarray}\nonumber
&&f_{\alpha\beta}:=-\frac{1}{2}\eps^{\ald\bed}(\dpar_{\alpha\ald}A_{\beta\bed}-
\dpar_{\beta\bed}A_{\alpha\ald}
+[A_{\alpha\ald},A_{\beta\bed}])=0~,\\\nonumber
&&\nabla_{\alpha\ald}\tilde{\chi}^{\ald}_i=0~,\\\label{asdSDYMeom}
&&\Box\phi_{ij}+2\eps\{\tilde{\chi}^{\ald}_i,\tilde{\chi}_{\ald j}\}=0~,\\
\nonumber
&&\nabla_{\alpha\ald}\chi^{\alpha}_{ijk}-\eps\,[\tilde{\chi}_{\ald[i},
\phi_{jk]}]=0~,\\\nonumber
&&\eps^{\alpha\gamma}\nabla_{\alpha\ald}G_{\gamma\delta[ijkl]}+12\eps
\{\tilde{\chi}_{\ald[i},\chi_{\delta jkl]}\}
-18\eps[\phi_{[ij},\nabla_{\delta\ald}\phi_{kl]}]=0~,
\end{eqnarray}
and the cases $\CN{<}4$ are governed by the first $\CN{+}1$
equations.

\smallskip\noindent{\bf Gauge equivalent trivializations.} One can again
consider trivializations similar to \eqref{strafo},
\eqref{sDonpsi} and transform the linear system
\eqref{asdslinsys1}-\eqref{asdslinsys4} to a gauge equivalent
linear system. Namely, there exist $\tau_\eps$-regular
matrix-valued functions $\hat{\psi}_\pm$ on
$\hat{\CV}_\pm\subset\CP_{*}^{3|\CN}$ such that
\begin{equation}
f_{+-}=\psi_+^{-1}\psi_-=\hat{\psi}_+^{-1}\hat{\psi}_-~~~\mbox{and}~~~
\bar{\dpar}_i^\pm\hat{\psi}_\pm=0~~~\mbox{with}
~~~\bar{\dpar}_i^\pm:=\mu_\pm^\alpha\dpar_{\alpha i}~,
\end{equation}
and therefore the matrix-valued functions
\begin{equation}
\varphi:=\psi_+\hat{\psi}_+^{-1}=\psi_-\hat{\psi}_-^{-1}
\end{equation}
generate gauge transformations
\begin{eqnarray}
\CA_\ald^\pm&\mapsto&\hat{\CA}_\ald^\pm=
\varphi^{-1}\CA_\ald^\pm\varphi+
\varphi^{-1}\bar{V}^\pm_\ald\varphi=
\hat{\psi}_\pm\bar{V}^\pm_\ald\hat{\psi}^{-1}_\pm~,\\
0=\CA_{\bs_\pm}&\mapsto&\hat{\CA}_{\bs_\pm}=
\varphi^{-1}\dpar_{\bs_\pm}\varphi=\hat{\psi}_\pm\dpar_{\bs_\pm}
\hat{\psi}_\pm^{-1}~,\\
\CA_i^\pm&\mapsto&\hat{\CA}_i^\pm=
\varphi^{-1}\CA_i^\pm\varphi+
\varphi^{-1}\bar{\dpar}^\pm_i\varphi=
\hat{\psi}_\pm\bar{\dpar}^\pm_i\hat{\psi}^{-1}_\pm=0~.
\end{eqnarray}
For the new linear system\footnote{On $\hat{\CV}_-$, the
equations are similar.}
\begin{eqnarray}\label{asdslinsysb1}
(\bar{V}_\ald^++\hat{\CA}_\ald^+)\hat{\psi}_+&=&0~,\\
(\dpar_{\bs_+}+\hat{\CA}_{\bs_+})\hat{\psi}_+&=&0~,\\\label{asdslinsysb4}
\bar{\dpar}^+_i\hat{\psi}_+&=&0~,
\end{eqnarray}
the compatibility conditions read
\begin{eqnarray}\label{asdshCS1}
\bar{V}_\ald^+\hat{\CA}_\bed^+-
\bar{V}_\bed^+\hat{\CA}_\ald^++
[\hat{\CA}_\ald^+,\hat{\CA}_\bed^+]=0~,\\
\label{asdshCS2} \dpar_{\bs_+}\hat{\CA}_\ald^+-
\bar{V}_\ald^+\hat{\CA}_{\bs_+}+
[\hat{\CA}_{\bs_+},\hat{\CA}_\ald^+]=0~.
\end{eqnarray}

The dependence of $\hat{\CA}_\ald^+$ and
$\hat{\CA}_{\bs_+}$ on $\zeta_+$ and $\bs_+$ is fixed by
their transformation properties similarly to the self-dual case
\eqref{expAa} and \eqref{expAl}. Namely, for $\CN{=}4$ we have
\begin{eqnarray}\label{asdexpAa}
\hat{\CA}_\ald^+&=&\mu_+^\alpha\,
A_{\alpha\ald}(x_L)+\theta^i_+\,\tilde{\chi}_{\ald i}(x_L)+
\nu_+\,\frac{1}{2!}\,\theta^i_+\theta^j_+\,\hat{\mu}^\alpha_+\,
\phi_{\alpha \ald ij}(x_L)+\\
\nonumber
&&+\nu_+^2\,\,\frac{1}{3!}\,\theta_+^i\theta_+^j\theta_+^k\,\hat{\mu}_+^\alpha\,
\hat{\mu}_+^\beta\, \chi_{\alpha\ald\beta
ijk}(x_L)+\nu_+^3\,\frac{1}{4!}
\theta_+^i\theta_+^j\theta_+^k\theta_+^l\,
\hat{\mu}^\alpha_+\,\hat{\mu}^\beta_+\,\hat{\mu}^{\gamma}_+\,
G_{\alpha\ald\beta\gamma ijkl}(x_L)~,\\\label{asdexpAl}
\hat{\CA}_{\bs_+}&=&\nu_+^2\,\frac{1}{2!}\theta_+^i\theta_+^j\,\phi_{ij}(x_L)+
\nu_+^3\,\frac{1}{3!}\theta_+^i\theta_+^j\theta_+^k\,\hat{\mu}_+^\alpha\,
\chi_{\alpha ijk}(x_L)+\\ \nonumber &&
+\nu_+^4\,\frac{1}{4!}\theta_+^i\theta_+^j\theta_+^k\theta_+^l\,
\hat{\mu}^\beta_+\,\hat{\mu}^\gamma_+\, G_{\beta\gamma
ijkl}(x_L)~,
\end{eqnarray}
where
\begin{equation}
(\mu_+^\alpha)=\left(\begin{array}{c}-\zeta_+\\ 1
\end{array}\right)~,~~~(\hat{\mu}_+^\alpha)=\left(\begin{array}{c}\eps\\
-\bs_+\end{array}\right)~~~\mbox{and}~~
\nu_+=\frac{1}{1-\eps\zeta_+\bs_+},~~~\zeta_+\in
V_+\subset\CPP^1_*~.
\end{equation}
Substituting \eqref{asdexpAa} and \eqref{asdexpAl} into the field
equations \eqref{asdshCS1} and \eqref{asdshCS2} of hCS theory on
the supertwistor space $\CP_*^{3|4}$, one obtains the field
equations \eqref{asdSDYMeom} for $\CN{=}4$. The appropriate
truncation for $\CN{<}4$ is done exactly as in the self-dual case:
from the nilpotency of the $\theta^i_+$'s it follows that there
are less fields in the expansions \eqref{asdexpAa} and
\eqref{asdexpAl}, which, in turn, yields the first $\CN{+}1$
equations of \eqref{asdSDYMeom}.

Again, we have described a one-to-one correspondence between gauge
equivalence classes of solutions to the $\CN$-extended
anti-self-dual Yang-Mills equations on $(\FR^4,g)$ and equivalence
classes of holomorphic vector bundles $\CE$ over the dual
supertwistor space $\CP_*^{3|\CN}$, analogously to the self-dual
case. The Penrose-Ward transform here is given by the formul\ae{}
\eqref{asdexpAa}-\eqref{asdexpAl}.

\section{A quadric in a product of supertwistor spaces}

\smallskip\noindent{\bf Coordinates on
$\CP^{3|3}\times\CP^{3|3}_*$.} Let us fix $\CN{=}3$ and consider the
direct product $\CP^{3|3}\times\CP^{3|3}_*$ of supertwistor spaces
which is an open subset in the supermanifold
$\CPP^{3|3}\times\CPP_*^{3|3}$. This subset can be covered by four
patches $\,\hat{\CU}_a\,$, $a=1,...,4$, defined as
\begin{equation}
\hat{\CU}_1:=\hat{\CU}_+\times \hat{\CV}_+~,~~~
\hat{\CU}_2:=\hat{\CU}_-\times \hat{\CV}_+~,~~~
\hat{\CU}_3:=\hat{\CU}_+\times \hat{\CV}_-~,~~~
\hat{\CU}_4:=\hat{\CU}_-\times \hat{\CV}_-~,
\end{equation}
where $\hat{\CU}_\pm$ and $\hat{\CV}_\pm$ are coordinate patches
on $\CP^{3|3}$ and $\CP^{3|3}_*$, respectively, described in
sections 4 and 6. Thus, we have
\begin{equation}
\CP^{3|3}\times\CP^{3|3}_*={\textstyle \bigcup_{a=1}^4}\,\hat{\CU}_a
\end{equation}
with coordinates
\begin{align}\nonumber
(z^\alpha_{(1)}\cb\lambda_\ald^{(1)}\cb\eta_i^{(1)}\cb
w_{(1)}^\ald\cb \mu_\alpha^{(1)}\cb\theta_{(1)}^i)=
(z^\alpha_{+}\cb\lambda_\ald^{+}\cb\eta_i^{+}\cb
w_{+}^\ald\cb
\mu_\alpha^{+}\cb\theta_{+}^i)~~~&\mbox{on}~~~\hat{\CU}_1~,\\\nonumber
(z^\alpha_{(2)}\cb\lambda_\ald^{(2)}\cb\eta_i^{(2)}\cb
w_{(2)}^\ald\cb \mu_\alpha^{(2)}\cb\theta_{(2)}^i)=
(z^\alpha_{-}\cb\lambda_\ald^{-}\cb\eta_i^{-}\cb
w_{+}^\ald\cb
\mu_\alpha^{+}\cb\theta_{+}^i)~~~&\mbox{on}~~~\hat{\CU}_2~,\\\nonumber
(z^\alpha_{(3)}\cb\lambda_\ald^{(3)}\cb\eta_i^{(3)}\cb
w_{(3)}^\ald\cb \mu_\alpha^{(3)}\cb\theta_{(3)}^i)=
(z^\alpha_{+}\cb\lambda_\ald^{+}\cb\eta_i^{+}\cb
w_{-}^\ald\cb
\mu_\alpha^{-}\cb\theta_{-}^i)~~~&\mbox{on}~~~\hat{\CU}_3~,\\\label{ambicoords}
(z^\alpha_{(4)}\cb\lambda_\ald^{(4)}\cb\eta_i^{(4)}\cb
w_{(4)}^\ald\cb \mu_\alpha^{(4)}\cb\theta_{(4)}^i)=
(z^\alpha_{-}\cb\lambda_\ald^{-}\cb\eta_i^{-}\cb
w_{-}^\ald\cb
\mu_\alpha^{-}\cb\theta_{-}^i)~~~&\mbox{on}~~~\hat{\CU}_4~.
\end{align}
On nonempty intersections
\begin{align}\nonumber
&\hat{\CU}_1\cap\hat{\CU}_2=(\hat{\CU}_+\cap\hat{\CU}_-)\times\hat{\CV}_+~,~~~
\hat{\CU}_1\cap\hat{\CU}_3=\hat{\CU}_+\times(\hat{\CV}_+\cap\hat{\CV}_-)~,\\
&\hat{\CU}_2\cap\hat{\CU}_4=\hat{\CU}_-\times(\hat{\CV}_+\cap\hat{\CV}_-)~,~~~
\hat{\CU}_3\cap\hat{\CU}_4=(\hat{\CU}_+\cap\hat{\CU}_-)\times\hat{\CV}_-~,~~~\\
\nonumber
&\hat{\CU}_1\cap\hat{\CU}_4=(\hat{\CU}_+\cap\hat{\CU}_-)\times(\hat{\CV}_+
\cap\hat{\CV}_-)=\hat{\CU}_2\cap\hat{\CU}_3~,
\end{align}
these coordinates are transformed in an obvious way which follows
from the transformations of the coordinates on $\CP^{3|3}$ and
$\CP_*^{3|3}$. Recall that the coordinates \eqref{ambicoords} are
obtained from homogeneous coordinates
\begin{equation}
[\,\omega^\alpha\cb
\lambda_\ald\cb\eta_i\,;\,\sigma^\ald\cb\mu_\alpha\cb\theta^i\,]
\end{equation}
on $\CPP^{3|3}\times\CPP^{3|3}_*$.

\smallskip\noindent{\bf A quadric.} Let us consider a quadric
$\mathbbm{L}^{5|6}$ in $\CPP^{3|3}\times\CPP^{3|3}_*$ defined by
the equation\footnote{The space $\CPP_*^{3|3}$ is the space of
$(3|3)$-planes in $\FC^{4|3}$. Each such plane is naturally
described by a ray, i.e.\ a $(1|0)$-dimensional subspaces of
$\FC^{4|3}$, orthogonal to the plane. Thus the space
$\CPP^{3|3}_*$ is biholomorphic to $\CPP^{3|3}$, which is the
space of rays in $\FC^{4|3}$. The quadric is exactly the
appropriate orthogonality condition between elements of both
projective spaces.}
\begin{equation}
\omega^\alpha\mu_\alpha-\sigma^\ald\lambda_\ald+2\theta^i\eta_i=0~.
\end{equation}
Consider also an open subset $\CL^{5|6}$ in $\mathbbm{L}^{5|6}$
defined as the quadric
$\mathbbm{L}^{5|6}\cap(\CP^{3|3}\times\CP^{3|3}_*)$ in
$\CP^{3|3}\times\CP^{3|3}_*$. The supermanifold $\CL^{5|6}$ is
covered by four patches
\begin{equation}
\CU_a:=\hat{\CU}_a\cap
\mathbbm{L}^{5|6}~,~~~\CL^{5|6}={\textstyle \bigcup_{a=1}^4}\,\CU_a~,
\end{equation}
and defined on $\hat{\CU}_a$ by the equation
\begin{equation}\label{quadric2}
z^\alpha_{(a)}\mu_\alpha^{(a)}-w_{(a)}^\ald\lambda_\ald^{(a)}+
2\theta_{(a)}^i\eta_i^{(a)}=0~,
\end{equation}
where no summation over the index $a$ is implied.

Note that one can introduce a holomorphic projection
\begin{equation}\label{ambifibration}
p:~\CL^{5|6}\rightarrow\CPP^1\times\CPP^1_*
\end{equation}
with $(3|6)$-dimensional fibres. Then $\CL^{5|6}$ can be
identified with a quotient bundle whose body $\CL^5$ is
\begin{equation}
\CL^5=\left(\CO(1,0)\oplus\CO(1,0)\oplus\CO(0,1)\oplus\CO(0,1)\right)/\CO(1,1)~.
\end{equation}
Here $\CO(n,0)$ is the bundle $\CO(n)$ over the projective space
$\CPP^1$ with homogeneous coordinates $[\lambda_\ald]$, $\CO(0,n)$
is the bundle $\CO(n)$ over the projective space $\CPP^1_*$ with
homogeneous coordinates $[\mu_\alpha]$ and $\CO(m,n)$ is the line
bundle $\CO(m,0)\otimes\CO(0,n)$ over $\CPP^1\times\CPP^1_*$. The
quotient by $\CO(1,1)$ arises from the quadric condition
\eqref{quadric2} (see e.g.\ \cite{Khenkin}). The base
$\CPP^1\times \CPP^1_*$ of the fibration \eqref{ambifibration} is
covered by four patches
\begin{equation}\label{4patches}
V_a\,:=\,\CU_a\cap(\CPP^1\times\CPP^1_*)
\end{equation}
with coordinates $(\lambda_{(a)},\zeta_{(a)})$ on an open set
$V_a\subset\CPP^1\times\CPP^1_*$. Recall that
$\CL^{5|6}=\bigcup_{a=1}^4\,\CU_a$ and
\begin{align}\nonumber
(\lambda_\ald^{(1)})^\top=(1\cb\lambda_{(1)})=(1\cb\lambda_+)~,~~~
(\lambda_\ald^{(2)})^\top=(\lambda_{(2)}\cb 1)=(\lambda_-\cb1)~,\\\nonumber
(\lambda_\ald^{(3)})^\top=(1\cb\lambda_{(3)})=(1\cb\lambda_+)~,~~~
(\lambda_\ald^{(4)})^\top=(\lambda_{(4)}\cb 1)=(\lambda_-\cb1)~,\\\nonumber
(\mu_\alpha^{(1)})^\top=(1\cb\zeta_{(1)})=(1\cb\zeta_+)~,~~~
(\mu_\alpha^{(2)})^\top=(1\cb\zeta_{(2)})=(1\cb\zeta_+)~,\\
(\mu_\alpha^{(3)})^\top=(\zeta_{(3)}\cb 1)=(\zeta_-\cb1)~,~~~
(\mu_\alpha^{(4)})^\top=(\zeta_{(4)}\cb 1)=(\zeta_-\cb1)~.
\end{align}
We denote by $z^A_{(a)}$ with $A=1,2,3$ bosonic coordinates
on the fibres over $V_a$ in the bundle \eqref{ambifibration}.
Additionally, we use odd variables $\theta^i_{(a)}$ and
$\eta^{(a)}_i$ as the fermionic coordinates on these
fibres.

\smallskip\noindent{\bf Moduli of complex submanifolds.}
Holomorphic sections over $V_a$ of the bundle
\eqref{ambifibration} are spaces $\CL^{2|0}_{x,\theta,\eta}\cong
\CPP^1\times\CPP^1_*$ defined by the equations
\begin{equation}\label{ambisections}
z^\alpha_{(a)}=x_R^{\alpha\ald}\lambda_\ald^{(a)}~,~~~
w^\ald_{(a)}=x_L^{\alpha\ald}\mu_\alpha^{(a)}~,~~~
\theta^i_{(a)}=\theta^{\alpha i}\mu_\alpha^{(a)}~,~~~
\eta^{(a)}_i=\eta_i^\ald\lambda_\ald^{(a)}~~~\mbox{with}
~~~(\lambda_\ald^{(a)},\mu_\alpha^{(a)})\in V_a~.
\end{equation}
These sections are not independent due to equation
\eqref{quadric2}, which is solved by the choice
\begin{equation}\label{ambcoord3}
x_R^{\alpha\ald}=x^{\alpha\ald}-\theta^{\alpha i}\eta_i^\ald~~~
\mbox{and}~~~x_L^{\alpha\ald}=x^{\alpha\ald}+\theta^{\alpha
i}\eta^\ald_i~.
\end{equation}
We may choose three independent functions from
$z_{(a)}^\alpha=x_R^{\alpha\ald}\lambda_\ald^{(a)}$ and
$w_{(a)}^\ald=x_L^{\alpha\ald}\mu_\alpha^{(a)}$ and raise them to
the coordinates $z^A_{(a)}$ on the fibres.
The local sections \eqref{ambisections} are properly glued on $V_a\cap
V_b\neq \varnothing$ and therefore define a global holomorphic
section of the bundle \eqref{ambifibration} parametrized by
supermoduli $(x,\theta,\eta)=(x^{\alpha\ald}\cb\theta^{\alpha i}\cb
\eta_i^\ald)\in\FC^{4|12}$ due to \eqref{ambcoord3}.

Equations \eqref{ambisections} with $a=1,...,4$ define a
supertwistor correspondence between $\CL^{5|6}$ and $\FC^{4|12}$
via a double fibration
\begin{equation}\label{ambidblfibration2}
\begin{picture}(80,40)
\put(0.0,0.0){\makebox(0,0)[c]{$\CL^{5|6}$}}
%\put(32.0,0.1){\makebox(0,0)[c]{$\Leftrightarrow$}}
\put(64.0,0.0){\makebox(0,0)[c]{$\FC^{4|12}$}}
\put(32.0,33.0){\makebox(0,0)[c]{$\CF^{6|12}$}}
\put(7.0,18.0){\makebox(0,0)[c]{$\pi_2$}}
\put(55.0,18.0){\makebox(0,0)[c]{$\pi_1$}}
\put(25.0,25.0){\vector(-1,-1){18}}
\put(37.0,25.0){\vector(1,-1){18}}
\end{picture}
\end{equation}
where $\CF^{6|12}:=\FC^{4|12}\times\CPP^1\times\CPP^1_*$. Namely, a
point $(x,\theta,\eta)\in\FC^{4|12}$ corresponds to
\begin{equation}
\CL^{2|0}_{x,\theta,\eta}=\pi_2(\pi_1^{-1}(x,\theta,\eta))\cong\CPP^1
\times\CPP^1_*\,\embd\,
\CL^{5|6}
\end{equation}
and a point $\ell\in\CL^{5|6}$ corresponds to a
$(1|6)$-dimensional super null line (an intersection of the
$\alpha$- and $\beta$-superplanes described in the previous
sections) $\pi_1(\pi_2^{-1}(\ell))\embd\FC^{4|12}$. Note that for
the coordinates \eqref{ambicoords} satisfying the quadric equation
\eqref{quadric2}, we have \cite{Harnad:1984vk}
\begin{align}\nonumber
&\pi_2^*z^\alpha_{(a)}=x_R^{\alpha\ald}\lambda_\ald^{(a)}~,~~~
\pi_2^*\lambda_\ald^{(a)}=\lambda_\ald^{(a)}~,~~~
\pi_2^*w^\ald_{(a)}=x_L^{\alpha\ald}\mu_\alpha^{(a)}~,~~~
\pi_2^*\mu_\alpha^{(a)}=\mu_\alpha^{(a)}~,\\
&\pi_2^*\theta^i_{(a)}=\theta^{\alpha
i}\mu_\alpha^{(a)}~~~\mbox{and}~~~
\pi_2^*\eta^{(a)}_i=\eta_i^\ald\lambda_\ald^{(a)}~,
\end{align}
where $\pi_2^*$ means the pull-back to $\CF^{6|12}$. The
supermanifold $\CF^{6|12}$ is covered by four patches
\begin{equation}\label{ambfgcoord1}
\tilde{\CU}_a\,=\,\FC^{4|12}\times V_a
\end{equation}
with coordinates
\begin{equation}\label{ambfgcoord2}
(x^{\alpha\ald}\cb\theta^{\alpha i}\cb\eta^\ald_i\cb
\lambda_\ald^{(a)}\cb\mu_\alpha^{(a)})~~~\mbox{equivalent to}~~~
(x^{\alpha\ald}\cb\theta^{\alpha i}\cb\eta^\ald_i\cb
\lambda_{(a)}\cb\zeta_{(a)})~.
\end{equation}
This supermanifold is obviously projected onto $\FC^{4|12}$ with
coordinates
\begin{equation}
(x^{\alpha\ald}\cb\theta^{\alpha i}\cb\eta^\ald_i)
\end{equation}
and onto $\CL^{5|6}$ with coordinates
\begin{equation}\label{ambcoord2}
z^\alpha_{(a)}=x_R^{\alpha\ald}\lambda_\ald^{(a)}~,~~\lambda_\ald^{(a)}~,~~
\eta_i^{(a)}=\eta_i^\ald\lambda_\ald^{(a)}~~~\mbox{and}~~~
w_{(a)}^\ald=x_L^{\alpha\ald}\mu_\alpha^{(a)}~,~~\mu_\alpha^{(a)}~,~~
\theta^i_{(a)}=\theta^{\alpha i} \mu_\alpha^{(a)}~,
\end{equation}
which are not all independent but equivalent to
$(z_{(a)}^A,\lambda_{(a)},\zeta_{(a)},\theta^i_{(a)},\eta_i^{(a)})$.
Note that we are considering the {\em complex} superspace
$\FC^{4|12}$. Appropriate reality conditions will be discussed
later on.

\smallskip\noindent{\bf Vector fields.} The tangent spaces to the
$(1|6)$-dimensional leaves of the fibration
\begin{equation}
\pi_2:~\CF^{6|12}\rightarrow\CL^{5|6}
\end{equation}
from \eqref{ambidblfibration2} are spanned by the holomorphic
vector fields
\begin{align}\label{ambivec1}
W^{(a)}&:=\mu_{(a)}^\alpha\lambda_{(a)}^\ald\dpar_{\alpha\ald}~,\\
D^i_{(a)}&=\lambda_{(a)}^\ald D_\ald^i~~~\mbox{and}~~~
D^{(a)}_i=\mu_{(a)}^\alpha D_{\alpha i}~,
\end{align}
which are properly glued on
$\tilde{\CU}_a\cap\tilde{\CU}_b\neq\varnothing$ into global vector
fields on $\CF^{6|12}$. Here $D_{\alpha i}$ and $D_\ald^i$ are
vector fields given by \eqref{vectorfields0} and
\eqref{vectorfield613}. We shall also consider the antiholomorphic
part
\begin{equation}\label{dparF}
\dparb_{\CF}=\dd \bar{x}^{\alpha\ald}
\der{\bar{x}^{\alpha\ald}}+\dd \bl_{(a)} \der{\bl_{(a)}}+\dd
\bs_{(a)}\der{\bs_{(a)}}+\dd\bar{\theta}^{i\alpha}\der{\bar{\theta}^{i\alpha}}+
\dd\bar{\eta}_i^\ald\der{\bar{\eta}^\ald_i}
\end{equation}
of the exterior derivative $\dd$ on $\tilde{\CU}_a\,$.

\section{Holomorphic Chern-Simons theory on the quadric}

\smallskip\noindent{\bf Holomorphic vector bundles over $\CL^{5|6}$.} For
defining a holomorphic rank $n$ vector bundle $\CE$ over
$\CL^{5|6}$, one should consider a covering $\{\CU_a\}$ of
$\CL^{5|6}$ and a collection $\{f_{ab}\}$ of holomorphic $n\times
n$ matrices (\v{C}ech 1-cocycle) on nonempty intersections
$\CU_a\cap\CU_b$ such that
\begin{equation}\label{cocyclecond}
f_{ab}f_{bc}=f_{ac}
\end{equation}
on $\CU_a\cap\CU_b\cap\CU_c\neq\varnothing$. We restrict ourselves
to topologically trivial bundles $\CE\rightarrow\CL^{5|6}$, i.e.\
those for which there exists a collection $\{\hat{\psi}_a\}$ of
regular matrix-valued functions (\v{C}ech 0-cochain) such that
\begin{equation}\label{ambsplit}
f_{ab}=\hat{\psi}_a^{-1}\hat{\psi}_b
\end{equation}
on any nonempty intersection $\CU_a\cap\CU_b$. Since $f_{ab}$ is
holomorphic on a trivialization of $\CE$ over $\CU_a$, we have
\begin{equation}\label{ambtriv1}
\dpar_{\bar{z}^A_{(a)}} f_{ab}=0~,~~~
\dpar_{\bl_{(a)}}f_{ab}=0=\dpar_{\bs_{(a)}}f_{ab}
\end{equation}
plus the trivial equations
\begin{equation}\label{ambtriv2}
\der{\bar{\theta}^i_{(a)}}f_{ab}=0=\der{\bar{\eta}_i^{(a)}}f_{ab}~.
\end{equation}
Equations \eqref{ambtriv1} and \eqref{ambtriv2} imply that
$f_{ab}$ is a function of the coordinates
$(z^A_{(a)},\theta^i_{(a)},\eta_i^{(a)})$ on $\CL^{5|6}$.
Equivalently, the $f_{ab}$'s are arbitrary functions of the
coordinates \eqref{ambcoord2} restricted only by the algebraic
equations \eqref{cocyclecond}. However, if one finds a splitting
\eqref{ambsplit} then \eqref{cocyclecond} is automatically
satisfied.

{}From \eqref{ambsplit} and \eqref{ambtriv1} it follows that
\begin{align}\label{ambpot1}
\hat{\psi}_a\dpar_{\bar{z}_{(a)}^A}\hat{\psi}_a^{-1}=
\hat{\psi}_b\dpar_{\bar{z}_{(a)}^A}\hat{\psi}_b^{-1}&=:\hat{\CA}_{\bar{z}_{(a)}^A}~,\\
\hat{\psi}_a\dpar_{\bl_{(a)}}\hat{\psi}_a^{-1}=
\hat{\psi}_b\dpar_{\bl_{(a)}}\hat{\psi}_b^{-1}&=:\hat{\CA}_{\bl_{(a)}}~,\\
\label{ambpot4} \hat{\psi}_a\dpar_{\bs_{(a)}}\hat{\psi}_a^{-1}=
\hat{\psi}_b\dpar_{\bs_{(a)}}\hat{\psi}_b^{-1}&=:\hat{\CA}_{\bs_{(a)}}~,
\end{align}
where $\hat{\CA}_{\bar{z}_{(a)}^A}$, $\hat{\CA}_{\bl_{(a)}}$ and
$\hat{\CA}_{\bs_{(a)}}$ are components of a $(0,1)$-form
$\hat{\CA}^{0,1}$ on $\CL^{5|6}$ restricted to an open set $\CU_a$
and having zero components along $\dpar/\dpar
\bar{\theta}_{(a)}^i$ and $\dpar/\dpar\bar{\eta}_i^{(a)}$ since we
assume $\dpar \hat{\psi}_a/\dpar \bar{\theta}^i_{(a)}=\dpar
\hat{\psi}_a/\dpar \bar{\eta}_i^{(a)}=0$, i.e.
\begin{equation}
\left.\hat{\CA}^{0,1}\right|_{\CU_a}=\hat{\CA}^{(a)}=
\hat{\CA}_{\bar{z}^A_{(a)}}\dd\bar{z}^A_{(a)}+
\hat{\CA}_{\bl_{(a)}}\dd\bl_{(a)}+
\hat{\CA}_{\bs_{(a)}}\dd\bs_{(a)}~.
\end{equation}
On nonempty intersections $\CU_a\cap\CU_b$, we have
\begin{equation}\label{ambAint}
\hat{\CA}^{0,1}|^{\phantom{\pm}}_{\CU_a}=
\hat{\CA}^{0,1}|^{\phantom{\pm}}_{\CU_b}
\end{equation}
and therefore \eqref{ambpot1}-\eqref{ambpot4} define the $(0,1)$ part of a global
gauge potential on the bundle $\CE$ over the supermanifold
$\CL^{5|6}$.

\smallskip\noindent{\bf Super hCS theory on $\CL^{5|6}$.} Let us introduce the notation
\begin{equation}
\bar{\dpar}_\CL=\dd\bar{z}^A_{(a)}\dpar_{\bar{z}^A_{(a)}}
+\dd\bl_{(a)}\dpar_{\bl_{(a)}}+\dd\bs_{(a)}\dpar_{\bs_{(a)}}~.
\end{equation}
Then the definitions \eqref{ambpot1}-\eqref{ambAint} of the super
gauge potential $\hat{\CA}^{0,1}=\{\hat{\CA}^{(a)}\}$ can be
rewritten as a linear equation
\begin{equation}\label{l56lineq}
\bar{\dpar}_\CE\hat{\psi}_a:=(\bar{\dpar}_\CL+\hat{\CA}^{(a)})\hat{\psi}_a=0
\end{equation}
on unknown regular matrix-valued functions\footnote{Recall that
$\der{\bar{\theta}^i_{(a)}}\hat{\psi}_a=\der{\bar{\eta}^{(a)}_i}\hat{\psi}_a=0$.}
$\hat{\psi}_a$ for given
$\hat{\CA}^{(a)}=\hat{\CA}^{0,1}|^{\phantom{\pm}}_{\CU_a}$,
$a=1,...,4$. This linear equation reads in components as
\begin{align}\label{amblinsys1}
&(\dpar_{{\bar{z}_{(a)}^A}}+
\hat{\CA}_{\bar{z}_{(a)}^A})\hat{\psi}_a=0~,\\
&(\dpar_{\bl_{(a)}}+
\hat{\CA}_{\bl_{(a)}})\hat{\psi}_a=0~,\\\label{amblinsys4}
&(\dpar_{\bs_{(a)}}+
\hat{\CA}_{\bs_{(a)}})\hat{\psi}_a=0~.
\end{align}
The compatibility condition of this linear system
is the equation
\begin{equation}\label{ambieom}
\bar{\dpar}_\CL\hat{\CA}^{(a)}+\hat{\CA}^{(a)}\wedge\hat{\CA}^{(a)}=0
\end{equation}
which is simply the field equation\footnote{Note that solutions to
these equations can be constructed via a generalization of the
Riemann-Hilbert problem based on the equivalence of the \v{C}ech
and the Dolbeault description of holomorphic bundles
\cite{Popov:1998fb, Ivanova:2000af}.}
 $\left.\CF^{0,2}\right|_{\CU_a}=0$ of hCS theory on the
supermanifold $\CL^{5|6}$.

\smallskip\noindent{\bf A special gauge.} Note that restrictions
of a vector bundle $\CE\rightarrow\CL^{5|6}$ to the fibres
$\FC^{3|6}_{\lambda,\zeta}$ of the bundle \eqref{ambifibration}
are holomorphically trivial since all these fibres are
contractible. Therefore, there exist trivializations
$\tilde{\psi}_a$ of $\CE$ over $\CU_a$ such that
\begin{equation}\label{ambitrans}
f_{ab}=\hat{\psi}^{-1}_a\hat{\psi}_b=\tilde{\psi}^{-1}_a\tilde{\psi}_b~~~
\mbox{on}~~~\CU_a\cap\CU_b\neq\varnothing
\end{equation}
and
\begin{equation}\label{ambiVonpsi}
\dpar_{\bar{z}_{(a)}^A}\tilde{\psi}_a=0~~~\mbox{on}~~~\CU_a~.
\end{equation}
{}From \eqref{ambitrans} and \eqref{ambiVonpsi}, it follows that
the globally defined regular matrix-valued function
\begin{equation}\label{ambigaugegelt}
\tilde{\varphi}=\hat{\psi}_a\tilde{\psi}_a^{-1}=\hat{\psi}_b\tilde{\psi}_b^{-1}
\end{equation}
on $\CL^{5|6}$ is an element of the group of gauge
transformations, whose action yields
\begin{align}\label{ambispgauge1}
\hat{\CA}_{\bar{z}^A_{(a)}}&\mapsto\tilde{\CA}_{\bar{z}^A_{(a)}}
=\tilde{\varphi}\hat{\CA}_{\bar{z}^A_{(a)}}\tilde{\varphi}^{-1}+
\tilde{\varphi}\dpar_{\bar{z}^A_{(a)}}\tilde{\varphi}^{-1}
=\tilde{\psi}_a\dpar_{\bar{z}^A_{(a)}}\tilde{\psi}_a^{-1}=0~,\\
\hat{\CA}_{\bl_{(a)}}&\mapsto\tilde{\CA}_{\bl_{(a)}}=
\tilde{\varphi}\hat{\CA}_{\bl_{(a)}}\tilde{\varphi}^{-1}+
\tilde{\varphi}\dpar_{\bl_{(a)}}\tilde{\varphi}^{-1}=
\tilde{\psi}_a\dpar_{\bl_{(a)}}\tilde{\psi}^{-1}_a~,\\\label{ambispgauge4}
\hat{\CA}_{\bs_{(a)}}&\mapsto\tilde{\CA}_{\bs_{(a)}}=
\tilde{\varphi}\hat{\CA}_{\bs_{(a)}}\tilde{\varphi}^{-1}+
\tilde{\varphi}\dpar_{\bs_{(a)}}\tilde{\varphi}^{-1}=
\tilde{\psi}_a\dpar_{\bs_{(a)}}\tilde{\psi}^{-1}_a~.
\end{align}
The linear system \eqref{amblinsys1}-\eqref{amblinsys4} is
correspondingly transformed to
\begin{align}\label{ambsplinsys1}
\dpar_{\bar{z}_{(a)}^A}\tilde{\psi}_a&=0~,\\\label{ambsplinsys2}
(\dpar_{\bl_{(a)}}+\tilde{\CA}_{\bl_{(a)}})\tilde{\psi}_a&=0~,\\
\label{ambsplinsys3}
(\dpar_{\bs_{(a)}}+\tilde{\CA}_{\bs_{(a)}})\tilde{\psi}_a&=0~,
\end{align}
which in coordinate independent form reads
\begin{equation}
(\bar{\dpar}_\CL+\tilde{\CA}^{(a)})\tilde{\psi}_{a}=0~.
\end{equation}

In this new gauge described by equations
\eqref{ambitrans}-\eqref{ambigaugegelt}, the field equations
\begin{equation}
\bar{\dpar}_\CL\tilde{\CA}^{(a)}+\tilde{\CA}^{(a)}\wedge\tilde{\CA}^{(a)}=0
\end{equation}
of super hCS theory on $\CL^{5|6}$ are simplified to the equations
\begin{equation}
\dpar_{\bar{z}_{(a)}^A}\tilde{\CA}_{\bl_{(a)}}=0=
\dpar_{\bar{z}_{(a)}^A}\tilde{\CA}_{\bs_{(a)}}
\end{equation}
and the equation
\begin{equation}
\dpar_{\bl_{(a)}}\tilde{\CA}_{\bs_{(a)}}-
\dpar_{\bs_{(a)}}\tilde{\CA}_{\bl_{(a)}}+
[\tilde{\CA}_{\bl_{(a)}},\tilde{\CA}_{\bs_{(a)}}]=0
\end{equation}
on an open set $V_a=\CU_a\cap\left(\CPP^1\times\CPP^1_*\right)\subset\CL^{5|6}$
with gluing conditions
\begin{equation}
\tilde{\CA}_{\bl_{(a)}}\dd\bl_{(a)}+
\tilde{\CA}_{\bs_{(a)}}\dd\bs_{(a)}=
\tilde{\CA}_{\bl_{(b)}}\dd\bl_{(b)}+
\tilde{\CA}_{\bs_{(b)}}\dd\bs_{(b)}
\end{equation}
on nonempty intersections $V_a\cap V_b$ with $a,b=1,...,4$. Thus,
three of five components of the gauge potential
$\tilde{\CA}^{0,1}$ on $\CL^{5|6}$ satisfying equation
\eqref{ambieom} are gauge equivalent to zero. However, even after
transforming to the gauge
\eqref{ambispgauge1}-\eqref{ambispgauge4} the new gauge potential
$\tilde{\CA}^{0,1}$ contains information on the extra dimensions
by depending holomorphically on the coordinates $z^A_{(a)}$,
$\theta^i_{(a)}$ and $\eta_i^{(a)}$. Note that one can also choose
$\{\tilde{\psi}_a\}$ such that a linear combination of components
$\tilde{\CA}_{\bar{z}^A_{(a)}}$, say a component
\begin{equation}
\tilde{\CA}_{\bar{\rho}_{(a)}}=\bar{X}^{(a)}\lrcorner\,\tilde{\CA}^{0,1}=
\bar{X}_{(a)}^A \tilde{\CA}_{\bar{z}^A_{(a)}}
\end{equation}
along a vector field
$\bar{X}^{(a)}=\bar{X}^A_{(a)}\dpar_{\bar{z}^A_{(a)}}=\dpar_{\bar{\rho}_{(a)}}$,
will be nonzero. In such a gauge, one will have three nonzero
components $\tilde{\CA}_{\bl_{(a)}}$, $\tilde{\CA}_{\bs_{(a)}}$
and $\tilde{\CA}_{\bar{\rho}_{(a)}}$, which may in principle be
used for constructing an action of type \eqref{ShCSaction} on the
supermanifold $\CL^{5|6}$.

\section{Supertwistors and the full $\CN{=}4$ super Yang-Mills theory}

In this section we shall consider $\CN{=}3$ SYM theory which is
known to be equivalent to the $\CN{=}4$ SYM
theory when formulated on $\FR^4$. More explicitly, we
shall consider the integrability of super Yang-Mills fields on
super null lines, which turns out to be equivalent to the equations
of motion of $\CN{=}3$ SYM theory
\cite{Witten:1978xx,Volovich:ii,Harnad:1984vk,Tafel:1985qk,Harnad:1988rs},
and its relation with super hCS theory on a $(5|6)$-dimensional supermanifold.

\smallskip\noindent{\bf Pulled-back bundle.} Let us consider a
holomorphic vector bundle $\CE\rightarrow\CL^{5|6}$ and the
pulled-back bundle $\pi_2^*\CE$ over the supermanifold
$\CF^{6|12}$ with a covering $\{\tilde{\CU}_a\}$ given by
\eqref{ambfgcoord1} and \eqref{ambfgcoord2}.
Pull-backs\footnote{For simplicity, we denote the pulled-back
transition functions also by $f_{ab}$, slightly abusing notation.
The same holds true for functions $\hat{\psi}_a$ and
$\tilde{\psi}_a$.} of the transition functions $\{f_{ab}\}$ of
$\CE$ to $\pi_2^*\CE$ are constant along the fibres of $\pi_2$,
i.e.
\begin{equation}\label{ambWcond1} W^{(a)}f_{ab}=D^i_{(a)}f_{ab}=
D_i^{(a)}f_{ab}=0~,
\end{equation}
and due to $\pi_2^*\dparb_{\CL}=\dparb_{\CF}\circ\pi_2^*$, they
also satisfy $\pi_2^*\dparb_\CL f_{ab}:=\bar{\dpar}_\CF f_{ab}=0$,
where $\dparb_\CF$ is the antiholomorphic part \eqref{dparF} of
the exterior derivative on $\CF^{6|12}$. On the pulled-back
trivializations $\{\tilde{\psi}_a\}$ from \eqref{ambitrans} and
\eqref{ambiVonpsi}, one can also impose the conditions
\begin{equation}\label{ambWcond2}
W^{(a)}\tilde{\psi}_a=0~,~~~D^i_{(a)}\tilde{\psi}_a=0~~~\mbox{and}~~~
D_i^{(a)}\tilde{\psi}_a=0
\end{equation}
since the fibres of $\pi_2$ are contractible\footnote{The same
conditions \eqref{ambWcond2} can be imposed on the pulled-back
trivializations $\{\hat{\psi}_a\}$ from \eqref{ambsplit}.}. So,
for $\pi_2^*\CE$ we have transition functions $\{f_{ab}\}$ split
as $f_{ab}=\tilde{\psi}_a^{-1}\tilde{\psi}_b$ on
$\tilde{\CU}_a\cap\tilde{\CU}_b\subset\CF^{6|12}$ and satisfying
equations \eqref{ambWcond1}. Moreover, regular matrix-valued
functions $\tilde{\psi}_a$'s defining a trivialization of
$\pi_2^*\CE$ over $\tilde{\CU}_a$ satisfy \eqref{ambWcond2} with
$a=1,...,4$.

\smallskip\noindent{\bf Holomorphic triviality on subspaces.} Let
us now consider holomorphic vector bundles $\CE$ over $\CL^{5|6}$
such that their restriction to any submanifold
$\CL^{2|0}_{x,\theta,\eta}\cong\CPP^1\times\CPP^1_*$ in
$\CL^{5|6}$ is holomorphically trivial, or, equivalently, such
that $\pi_2^*\CE$ is trivial along the fibres of $\pi_1$. For such
bundles, there exist trivializations $\{\psi_a\}$ of $\pi_2^*\CE$
over $\tilde{\CU}_a$ such that
\begin{equation}\label{ambpbtrans}
f_{ab}=\tilde{\psi}_a^{-1}\tilde{\psi}_b=\psi_a^{-1}\psi_b~~~
\mbox{on}~~~\CU_a\cap\CU_b\neq\varnothing
\end{equation}
and
\begin{equation}\label{extholcond}
\dparb_{\CF}\,\psi_a=0~,
\end{equation}
i.e.\ the regular matrix-valued functions $\psi_a$ are holomorphic
in the coordinates on $\CF^{6|12}$. It follows from
\eqref{ambpbtrans} that
\begin{equation}\label{ambiphi}
\phi:=\psi_a\tilde{\psi}_a^{-1}=\psi_b\tilde{\psi}_b^{-1}
\end{equation}
is a globally defined regular matrix-valued (super)function on
$\CF^{6|12}$ which generates gauge transformations
\begin{equation}
\tilde{\psi}_a\mapsto\psi_a=\phi\tilde{\psi}_a~~~\mbox{for}~~a=1,...,4~,
\end{equation}
%\vspace{-0.9cm}
such that
\begin{align}\label{ambpbgauge1}
\tilde{\CA}^{(a)}&\mapsto
\CA^{(a)}=\phi\tilde{\CA}^{(a)}\phi^{-1}+\phi\bar{\dpar}_\CF\phi^{-1}=0
\end{align}
and
\begin{align}\label{ambpbgauge5}
0=&\tilde{\psi}_aD^i_{(a)}\tilde{\psi}_a^{-1}=:\tilde{\CA}^i_{(a)}\mapsto \CA^i_{(a)}=\phi D^i_{(a)}\phi^{-1}=
\psi_a D^i_{(a)}\psi^{-1}_a=\psi_b D^i_{(a)}\psi^{-1}_b=
\lambda^\ald_{(a)}A_\ald^i(x,\theta,\eta)~,\\\label{ambpbgauge6}
0=&\tilde{\psi}_aD_i^{(a)}\tilde{\psi}_a^{-1}=:\tilde{\CA}_i^{(a)}\mapsto \CA_i^{(a)}=\phi D_i^{(a)}\phi^{-1}=
\psi_a D_i^{(a)}\psi^{-1}_a=\psi_b D_i^{(a)}\psi^{-1}_b=
\mu^\alpha_{(a)}A_{\alpha i}(x,\theta,\eta)~,\\\label{ambpbgauge7}
0=&\tilde{\psi}_a W\tilde{\psi}_a^{-1}=:\tilde{\CA}_{w_{(a)}}\mapsto \CA_{w_{(a)}}=\phi
W^{(a)}\phi^{-1}= \psi_a W^{(a)}\psi^{-1}_a =\psi_b
W^{(a)}\psi^{-1}_b=\mu^\alpha_{(a)}
\lambda^\ald_{(a)}A_{\alpha\ald}(x,\theta,\eta)~,
\end{align}
where $(x,\theta,\eta)=(x^{\alpha\ald}\cb\theta^{\alpha
i}\cb\eta_i^\ald)$. Note that the last equalities in
\eqref{ambpbgauge5}-\eqref{ambpbgauge7} follow from a
ge\-nera\-lized Liouville theorem on $\CPP^1\times \CPP^1_*$ which says
that $\CA_{(a)}^i$ is a local section of the bundle $\CO(1,0)$,
$\CA_i^{(a)}$ is a local section of the bundle $\CO(0,1)$ and
$\CA_{w_{(a)}}$ is a local section of the bundle $\CO(1,1)$ over
$\CPP^1\times \CPP^1_*$.

\smallskip\noindent{\bf Linear system.} Equations
\eqref{ambpbgauge5}-\eqref{ambpbgauge7} can be rewritten as the
linear system
\begin{align}\label{amblinsys21}
&\mu_{(a)}^\alpha\lambda_{(a)}^\ald(\dpar_{\alpha\ald}+
A_{\alpha\ald})\psi_a=0~,\\
&\lambda_{(a)}^\ald(D^i_{\ald}+A^i_{\ald})\psi_a=0~,\\\label{amblinsys24}
&\mu_{(a)}^\alpha(D_{\alpha i}+A_{\alpha i})\psi_a=0~.
\end{align}
By construction, the new linear system
\eqref{amblinsys21}-\eqref{amblinsys24} together with
\eqref{extholcond} is gauge equivalent to the linear system
\eqref{ambsplinsys1}-\eqref{ambsplinsys3} together with
\eqref{ambWcond2}.

\smallskip\noindent{\bf Full $\CN{=}3$ SYM equations.} The linear
system \eqref{amblinsys21}-\eqref{amblinsys24} has been known for
a long time \cite{Witten:1978xx, Volovich:ii, Tafel:1985qk, Harnad:1988rs}.
Its compatibility conditions read
\begin{equation}\label{ambcompcond1}
\{\nabla_\ald^i,\nabla_\bed^j\}+\{\nabla_\bed^i,\nabla_\ald^j\}=0~,~~~
\{\nabla_{\alpha i},\nabla_{\beta j}\}+\{\nabla_{\beta i},
\nabla_{\alpha j}\}=0~,~~~
\{\nabla_{\alpha i},\nabla^j_{\ald}\}-2\delta_i^j\nabla_{\alpha\ald}=0~,
\end{equation}
where
\begin{equation}
\nabla_{\alpha\ald}=\dpar_{\alpha\ald}+A_{\alpha\ald}~,~~
\nabla_{\alpha i}=D_{\alpha i}+A_{\alpha i}~~~\mbox{and}~~~
\nabla_\ald^i=D_\ald^i+A_\ald^i
\end{equation}
are gauge covariant derivatives in the superspace $\FC^{4|12}$.
Equations \eqref{ambcompcond1} for the components\linebreak
$(A_{\alpha\ald}(x,\theta,\eta)$, $A_{\alpha i}(x,\theta,\eta)$,
$A_\ald^i(x,\theta,\eta))$ of a superconnection are equivalent to
the equations of motion of the full $\CN{=}3$ SYM theory
\cite{Witten:1978xx, Harnad:1984vk, Harnad:1988rs}. Using the
expansions of the superfields (i.e.\ of the components of the
superconnection) in the odd variables, one can rewrite
\eqref{ambcompcond1} as equations on a supermultiplet of ordinary
fields. Moreover, these equations turn out to be equivalent to the
equations of motion of $\CN{=}4$ SYM theory in ordinary space.

Summarizing, we have (implicitly) described the Penrose-Ward
transform
\begin{equation}
\CP\mathcal{W}:(\hat{\CA}^{0,1})~
\mapsto~(f_{\alpha\beta},\chi_{\alpha
i},\phi_{ij},\tilde{\chi}_\ald^i,f_{\ald\bed})
\end{equation}
which maps solutions of hCS theory on $\CL^{5|6}$ to solutions of
$\CN{=}4$ SYM theory on $\FC^4$. Note that the existence of a
gauge in which $\CA_{\bl_\pm}=0=\CA_{\bs_\pm}$ is equivalent to
holomorphic triviality of the bundle $\CE\to\CL^{5|6}$ on
$\CL^{2|0}_{x,\theta,\eta}\cong\CPP^1\times\CPP^1_*\embd\CL^{5|6}$.
Note furthermore that the moduli space of such bundles is a subset
of the moduli space of all topologically trivial holomorphic
bundles $\CE$ over $\CL^{5|6}$ \cite{Manin:ds, Khenkin}. This
means that the solution space of hCS theory on $\CL^{5|6}$ is
larger than that of $\CN{=}4$ SYM theory.

\smallskip\noindent{\bf Looking for an action.} We saw that the full set
of equations of motion for $\CN{=}3$ SYM theory is encoded in the
equation $\CF^{0,2}=0$ on the supermanifold $\CL^{5|6}$ which is
the quadric in an open subset of $\CPP^{3|3}\times\CPP^{3|3}_*$.
One might wonder whether there is some action principle for super
hCS theory on this space. For complex three-dimensional
supermanifolds, this is the hCS action \eqref{ShCSaction}. In the
case of the $(5|6)$-dimensional supermanifold $\CL^{5|6}$, the
situation is less clear. Recall that this space is a Calabi-Yau
supermanifold and thus it comes with a holomorphic volume form
$\Omega^{5|6}$. Therefore, a possible ansatz is
\begin{equation}\label{ansatz}
S=\int\Omega^{5|6}\wedge\tr\left(\hat{\CA}^{0,1}\wedge\bar{\dpar}\hat{\CA}^{0,1}+
\frac{2}{3}\hat{\CA}^{0,1}\wedge \hat{\CA}^{0,1}\wedge
\hat{\CA}^{0,1}\right)\wedge \omega^{0,2}~,
\end{equation}
where we abbreviated $\hat{\CA}^{0,1}=\hat{\CA}^{0,1|0,0}$ and
$\omega^{0,2}=\omega^{0,2|0,0}$. For this ansatz to be correct,
$\omega^{0,2}$ must be nowhere vanishing (otherwise the total
measure would be degenerate). Furthermore, the partial integration
used for deriving the equations of motion demands that
$\omega^{0,2}$ is partially closed, i.e.\ it has to satisfy the
equation $\hat{\CA}^{0,1}\wedge\bar{\dpar}\omega^{0,2}=0$. It is
not clear whether such a $(0,2)$-form exists on $\CL^{5|6}$. Even
less clear is the relation of the action \eqref{ansatz} with
string field theory for the target space $\CL^{5|6}$. Therefore we
leave this discussion to forthcoming work\footnote{Another
possibility to obtain the equation \eqref{ambieom} is to use an
action of holomorphic BF type theories \cite{Popov:1998fb}.
However, the relation of this kind of action with string field
theory is also unclear.}.

\section{Reality conditions on the quadric}

In the purely bosonic case, one can introduce real (antihermitean)
gauge fields on $\FR^4$ with a metric $g$ of Euclidean signature
$(4,0)$, Kleinian signature $(2,2)$ or Minkowski signature $(3,1)$
by choosing an appropriate real structure on $\FC^4$. However, as
already mentioned in section 4, on the superspace $\FC^{4|4\CN}$
there exists a real structure defining a Euclidean superspace only
for an even number of supersymmetries. For simplicity, we restrict
ourselves here to the Kleinian and Minkowskian cases.

\smallskip\noindent{\bf Real structure $\tau_1$.} The Kleinian
signature $(2,2)$ is related to anti-linear
transformations\footnote{We will not consider the map $\tau_0$
here.} $\tau_1$ of spinors defined in sections 2 and 4. Recall
that
\begin{equation}
\tau_1\left(\begin{matrix}\omega^1\\\omega^2
\end{matrix}\right){=}\left(\begin{matrix}\bar{\omega}^2\\\bar{\omega}^1
\end{matrix}\right)\ ,\quad
\tau_1\left(\begin{matrix}\lambda_{\dot{1}}\\\lambda_{\dot{2}}
\end{matrix}\right){=}\left(\begin{matrix}\bl_{\dot{2}}\\
\bar{\lambda}_{\dot{1}}\end{matrix}\right)\ ,\quad
\tau_1\left(\begin{matrix}\sigma^{\dot{1}}\\\sigma^{\dot{2}}
\end{matrix}\right){=}\left(\begin{matrix}\bar{\sigma}^{\dot{2}}\\
\bar{\sigma}^{\dot{1}}
\end{matrix}\right)\ ,\quad
\tau_1\left(\begin{matrix}\mu_1\\\mu_2
\end{matrix}\right){=}\left(\begin{matrix}\bar{\mu}_2\\\bar{\mu}_1
\end{matrix}\right)\ ,
\end{equation}
and obviously $\tau_1^2=1$. Correspondingly for $(\lambda_\pm$,
$\zeta_\pm)\in\CPP^1\times\CPP^1_*$, we have
\begin{equation}
\tau_1(\lambda_+)=\frac{1}{\bar{\lambda}_+}=\bl_-~,~~~
\tau_1(\zeta_+)=\frac{1}{\bar{\zeta}_+}=\bs_-~
\end{equation}
with stable points
\begin{equation}
\{(\lambda,\,\zeta)\,\in\CPP^1\times\CPP^1_*:\lambda\bl=1,\zeta\bs=1\}=S^1\times
S^1_*\subset\CPP^1\times\CPP^1_*
\end{equation}
parametrizing a torus $S^1\times S^1_*$. For the coordinates
$(x^{\alpha\ald})$, we have
\begin{equation}
\tau_1\left(\begin{array}{cc}x^{1\dot{1}}&x^{1\dot{2}}\\
x^{2\dot{1}}&x^{2\dot{2}}
\end{array}\right)=\left(\begin{array}{cc} 0 & 1\\ 1& 0
\end{array}\right)\left(\begin{array}{cc}
\bar{x}^{1\dot{1}}&\bar{x}^{1\dot{2}}\\
\bar{x}^{2\dot{1}}&\bar{x}^{2\dot{2}}
\end{array}\right)\left(\begin{array}{cc} 0 & 1\\ 1& 0
\end{array}\right)=\left(\begin{array}{cc}
\bar{x}^{2\dot{2}}&\bar{x}^{2\dot{1}}\\
\bar{x}^{1\dot{2}}&\bar{x}^{1\dot{1}}
\end{array}\right)~,
\end{equation}
and the real subspace $\FR^4$ of $\FC^4$ invariant under the
involution $\tau_1$ is defined by the equations
\begin{equation}\label{realcoord}
x^{2\dot{2}}=\bar{x}^{1\dot{1}}=:-(x^4+\di x^3)~~~\mbox{and}~~~
x^{2\dot{1}}=\bar{x}^{1\dot{2}}=:-(x^2-\di x^1)
\end{equation}
with a metric $\dd s^2=\det(\dd x^{\alpha\ald})$ of signature
$(2,2)$. Recall also that
\begin{equation}
\tau_1\left(\begin{array}{c}\theta^{1i} \\ \theta^{2i}
\end{array}\right)=
\left(\begin{array}{c}\bar{\theta}^{2i} \\ \bar{\theta}^{1i}
\end{array}\right)~,~~~
\tau_1\left(\begin{array}{c}\eta^{\dot{1}}_{i} \\[0.8mm] \eta^{\dot{2}}_{i}
\end{array}\right)=
\left(\begin{array}{c}\bar{\eta}^{\dot{2}}_{i} \\[0.8mm] \bar{\eta}^{\dot{1}}_{i}
\end{array}\right)~,
\end{equation}
and therefore real (Majorana) fermions satisfy (cf. \eqref{majorana1})
\begin{equation}\label{majorana2}
\tau_1(\theta)=\theta\,\Leftrightarrow\,\theta^{2i}=\bar{\theta}^{1i}~~~
\mbox{and}~~~
\tau_1(\eta)=\eta\,\Leftrightarrow\,\eta_i^{\dot{2}}=\bar{\eta}^{\dot{1}}_i~.
\end{equation}

\smallskip\noindent{\bf A $\tau_1$-real twistor diagram.}
Imposing conditions \eqref{realcoord} and \eqref{majorana2} for $\CN{=}3$, we
obtain the real superspace $\FR^{4|12}$ as a fixed point set
of the involution $\tau_1:\,\FC^{4|12}\rightarrow\FC^{4|12}.$
Analogously, for the supertwistor space $\CPP^{3|3}$ and its open
subset $\CP^{3|3}$, we obtain real subspaces $\RPS^{3|3}$ and
$\CT^{3|3}$ (cf. \eqref{realtwistor}). Accordingly, a real form
of the superspace $\CF^{6|12}$ is
\begin{equation}
\CF^{6|12}_{\tau_1}:=\FR^{4|12}\times S^1\times S^1_*~,
\end{equation}
and we have a real quadric
\begin{equation}
\CL^{5|6}_{\tau_1}\subset\CT^{3|3}\times\CT^{3|3}_*
\end{equation}
as the subset of fixed points of the involution\footnote{Our
notation is slightly sloppy: We use the same symbol $\tau_1$ for
maps defined on different spaces.} $\tau_1:\,\CL^{5|6}\rightarrow
\CL^{5|6}$. This quadric is defined by equations
\eqref{quadric2}-\eqref{ambcoord3} with $x^{\alpha\ald}$,
$\theta^{\alpha i}$ and $\eta_i^\ald$ satisfying \eqref{realcoord}
and \eqref{majorana2} with
$\lambda_+=\de^{\di\chi_1}=\lambda_-^{-1}$,
$\zeta_+=\de^{\di\chi_2}= \zeta_-^{-1}$, $0\leq\chi_1,\,\chi_2<
2\pi$. Thus, we obtain a real form
\begin{equation}\label{ambdblfibration3}
\begin{picture}(50,40)
\put(0.0,0.0){\makebox(0,0)[c]{$\CL^{5|6}_{\tau_1}$}}
%\put(32.0,0.1){\makebox(0,0)[c]{$\Leftrightarrow$}}
\put(64.0,0.0){\makebox(0,0)[c]{$\FR^{4|12}$}}
\put(37.0,37.0){\makebox(0,0)[c]{$\CF^{6|12}_{\tau_1}$}}
%\put(7.0,20.0){\makebox(0,0)[c]{$\pi_1$}}
%\put(55.0,20.0){\makebox(0,0)[c]{$\pi_2$}}
\put(25.0,27.0){\vector(-1,-1){18}}
\put(37.0,27.0){\vector(1,-1){18}}
\end{picture}\vspace{2mm}
\end{equation}
of the double fibration \eqref{ambidblfibration2}. One can
restrict all (super)functions defined on spaces in
\eqref{ambidblfibration2} to the real subspaces in
\eqref{ambdblfibration3}.

\smallskip\noindent{\bf Reality of fields in the Kleinian case.}
For imposing reality conditions on the functions $\psi_a$ (and
$f_{ab}$) inducing antihermiticity of the fields of $\CN{=}3$ (and
$\CN{=}4$) SYM theory via the twistor correspondence, it is
convenient to consider an open neighborhood (and an analytic continuation
of the functions to a complex domain) of all these real spaces. In fact, for our
purpose it is enough to consider the supermanifold
\begin{equation}
\tilde{\CF}^{8|12}=\FR^{4|12}\times(U_+\cap U_-)\times(V_+\cap V_-)
=\FR^{4|12}\times (V_1\cap V_4)=\FR^{4|12}\times (V_2\cap V_3)    ~,
\end{equation}
where $U_\pm$ and $V_\pm$ cover projective spaces
$\CPP^1=U_+\cup U_-$ and $\CPP^1_*=V_+\cup V_-$ parametrized by
homogeneous coordinates $[\lambda_\ald]$ and $[\mu_\alpha]$,
respectively. Recall that the manifold $\CPP^1\times\CPP^1_*$ is
covered by four patches $V_a$ defined in \eqref{4patches} with
coordinates $(\lambda_{(a)}\cb\zeta_{(a)})$ on $V_a$. The
involution $\tau_1$ interchanges these patches as
$V_1\leftrightarrow V_4$, $V_2\leftrightarrow V_3$ and therefore
\begin{equation}
V_1\cap V_2\leftrightarrow V_4\cap V_3~,~~
V_1\cap V_3\leftrightarrow V_4\cap V_2~,~~
V_1\cap V_4\leftrightarrow V_4\cap V_1~,~~
V_2\cap V_3\leftrightarrow V_3\cap V_2~.
\end{equation}
Considering $(\lambda_{(a)},\zeta_{(a)})\neq 0$, we impose a
reality condition on the complex regular matrix-valued functions
$\psi_a=\psi_a(x^{\alpha\ald},\theta^{\alpha i},
\eta^\ald_i,\lambda_{(a)},\zeta_{(a)})$ by taking them depending
on $\tau_1$-real coordinates $x^{\alpha\ald}$, $\theta^{\alpha
i}$, $\eta^\ald_i$ and satisfying the equations
\begin{align}\nonumber
&\psi_1^\dagger\left(x\cb\theta\cb\eta\cb\frac{1}{\bl_{(1)}}\cb
\frac{1}{\bs_{(1)}}\right)=
\psi_4^{-1}(x\cb\theta\cb\eta\cb\lambda_{(4)}\cb\zeta_{(4)})~,\\
\label{psirealcond}
&\psi_2^\dagger\left(x\cb\theta\cb\eta\cb\frac{1}{\bl_{(2)}}\cb
\frac{1}{\bs_{(2)}}\right)=
\psi_3^{-1}(x\cb\theta\cb\eta\cb\lambda_{(3)}\cb\zeta_{(3)})~,
\end{align}
which lead to the relations
\begin{align}\nonumber
&f^\dagger_{14}\left(...,\frac{1}{\bl_{(1)}},\frac{1}{\bs_{(1)}}\right)=
f_{14}\left(...,{\lambda_{(1)}},{\zeta_{(1)}}\right)~,~~~
f^\dagger_{23}\left(...,\frac{1}{\bl_{(2)}},\frac{1}{\bs_{(2)}}\right)=
f_{23}\left(...,{\lambda_{(3)}},{\zeta_{(3)}}\right)~,\\
&f^\dagger_{12}\left(...,\frac{1}{\bl_{(1)}},\frac{1}{\bs_{(1)}}\right)=
f_{43}\left(...,{\lambda_{(3)}},{\zeta_{(3)}}\right)~,~~~
f^\dagger_{13}\left(...,\frac{1}{\bl_{(1)}},\frac{1}{\bs_{(1)}}\right)=
f_{42}\left(...,{\lambda_{(4)}},{\zeta_{(4)}}\right)~.
\end{align}
Now, using the definitions \eqref{ambpbgauge5}-\eqref{ambpbgauge7}
one can show by direct calculation that the conditions
\eqref{psirealcond} yield antihermitean superconnections and the
real $\CN{=}3,4$ supermultiplet of ordinary fields.

\smallskip\noindent{\bf The Minkowskian involution $\tau_M$.} Let
us consider the supermanifold $\CP^{3|3}\times\CP^{3|3}_*$ with
homogeneous coordinates
$[\omega^\alpha,\lambda_\ald,\eta_i;\mu_\alpha,\sigma^\ald,\theta^i]$.
The antiholomorphic involution
\begin{equation}\label{Minvolution}
\tau_M:\,\CP^{3|3}\times\CP^{3|3}_*\rightarrow\CP^{3|3}\times\CP^{3|3}_*
\end{equation}
appropriate to Minkowski space is defined as the map (see e.g.\
\cite{Manin:ds})
\begin{equation}\label{Mtau}
\tau_M(\omega^\alpha,\lambda_\ald,\eta_i;\mu_\alpha,\sigma^\ald,\theta^i)=
(-\overline{\sigma^\ald},\overline{\mu_\alpha},\overline{\theta^i};
\overline{\lambda_\ald},-\overline{\omega^\alpha},\overline{\eta_i})
\end{equation}
interchanging $\alpha$-superplanes and $\beta$-superplanes. One
sees from \eqref{Mtau} that the real slice in the space
$\CP^{3|3}\times\CP^{3|3}_*$ is defined by the
equations\footnote{Here $\alpha$ and $\ald$ denote {\em the same}
number.}
\begin{equation}\label{Mreal1}
\sigma^\ald=-\overline{\omega^\alpha}~,~~~\mu_\alpha=\overline{\lambda_\ald}
\end{equation}
and
\begin{equation}
\eta_i=\eta_i^\ald\lambda_\ald=\overline{\theta^i}=\overline{\theta^{\alpha
i}\mu_\alpha}=\overline{\theta^{\alpha i}}\overline{{\mu}_\alpha}\quad
 \Rightarrow\quad
\eta_i^\ald=\overline{\theta^{\alpha i}}\ .
\end{equation}
Finally, for coordinates $(x^{\alpha\ald})\in \FC^4$, we have
\begin{equation}
\tau_M(x^{\alpha\bed})=-\bar{x}^{\beta\ald}~,
\end{equation}
and the Minkowskian real slice $\FR^{3,1}\subset\FC^4$ is
parametrized by coordinates
\begin{align}\nonumber
&\hspace{5cm}\left(\begin{matrix}x^{1\dot{1}}&x^{1\dot{2}}\\
x^{2\dot{1}}&x^{2\dot{2}}\end{matrix}\right)^\dagger=-
\left(\begin{matrix}x^{1\dot{1}}&x^{1\dot{2}}\\
x^{2\dot{1}}&x^{2\dot{2}}\end{matrix}\right)\\\label{Mreal4}
&~\Rightarrow~
x^{1\dot{1}}=-\di(x^0+x^3),~
x^{1\dot{2}}=-\di(x^1-\di x^2),~
x^{2\dot{1}}=-\di(x^1+\di x^2),~
x^{2\dot{2}}=-\di(x^0-x^3)
\end{align}
with $(x^0,x^1,x^2,x^3)\in \FR^4$ and (cf. \eqref{realmetric})
\begin{equation}
\dd s^2=\det(\dd
x^{\alpha\ald})~\Rightarrow~g=\diag(-1,+1,+1,+1)~.
\end{equation}
One can also introduce coordinates
\begin{equation}
\tilde{x}^{\alpha\ald}=\di x^{\alpha\ald}~,
\end{equation}
obtaining a metric with signature $(1,3)$. In terms of
$\tilde{x}^{\alpha\ald}$, imaginary units will appear in many
formul\ae{}, which is the common convention on Minkowski space
with signature (1,3), e.g.
\begin{equation} D_{\alpha
i}=\der{\theta^{\alpha i}}+\di\bar{\theta}^\ald_i
\der{\tilde{x}^{\alpha\ald}}~,~~~
\tilde{x}^{\alpha\ald}_R=\tilde{x}^{\alpha\ald}-\di\theta^{\alpha
i}\bar{\theta}^\ald_i~,~~~\mbox{etc.}
\end{equation}
Recall that the involution $\tau_M$ interchanges $\alpha$-superplanes
and $\beta$-superplanes and therefore exchanges opposite helicity
states. It might be identified with a $\RZ_2$-symmetry discussed
recently in the context of mirror symmetry~\cite{Aganagic:2004yh} and
parity invariance~\cite{Witten:2004cp}.

\smallskip\noindent{\bf A $\tau_M$-real twistor diagram.} Recall
that $[\lambda_\ald]$ and $[\mu_\alpha]$ are homogeneous
coordinates on two Riemann spheres and the involution $\tau_M$
maps these spheres one into another. Moreover, fixed points of
the map $\tau_M:\CPP^1\times\CPP^1_*\rightarrow \CPP^1\times\CPP^1_*$
form the Riemann sphere
\begin{equation}
\CPP^1=\diag(\CPP^1\times\overline{\CPP}^1)\ ,
\end{equation}
where $\overline{\CPP}^1 (=\CPP^1_*)$ denotes the Riemann sphere
$\CPP^1$ with the opposite complex structure. Therefore, a real slice in
the space $\CF^{6|12}=\FC^{4|12}\times\CPP^1\times\CPP^1_*$
introduced in \eqref{ambidblfibration2} and characterized as the
fixed point set of the involution $\tau_M$ is the space
\begin{equation}
\CF^{6|12}_{\tau_M}:=\FR^{4|12}\times\CPP^1
\end{equation}
of real dimension $(6|12)$.

The fixed point set of the involution \eqref{Minvolution} is the
diagonal in the space $\CP^{3|3}\times\bar{\CP}^{3|3}$, which can
be identified with the complex supertwistor space $\CP^{3|3}$ of
real dimension $(6|6)$. This involution also picks out a real
quadric $\CL^{5|6}_{\tau_M}$ defined by equations
\eqref{quadric2} and the reality conditions
\eqref{Mreal1}-\eqref{Mreal4}. Thus, we obtain a real version of
the double fibration \eqref{ambidblfibration2},
\begin{equation}\label{ambidblfibration3}
\begin{picture}(80,40)
\put(0.0,0.0){\makebox(0,0)[c]{$\CL_{\tau_M}^{5|6}$}}
%\put(32.0,0.1){\makebox(0,0)[c]{$\Leftrightarrow$}}
\put(64.0,0.0){\makebox(0,0)[c]{$\FR^{4|12}$}}
\put(35.0,37.0){\makebox(0,0)[c]{$\CF_{\tau_M}^{6|12}$}}
\put(7.0,20.0){\makebox(0,0)[c]{$\pi_2$}}
\put(55.0,20.0){\makebox(0,0)[c]{$\pi_1$}}
\put(25.0,27.0){\vector(-1,-1){18}}
\put(37.0,27.0){\vector(1,-1){18}}
\end{picture}
\end{equation}
The dimensions of all spaces in this diagram are
real. For imposing the reality conditions on the superconnection
components, one should proceed analogously to the case of Kleinian
signature. We will not discuss this here.

\section{Conclusions}

In this paper, we considered two examples of the fibration
\begin{equation}\label{conc1}
\pi~:~Z~\rightarrow~X_\tau~,
\end{equation}
which describe self-dual and anti-self-dual $\CN$-extended SYM
theory in four real dimensions. As the supermanifold $Z$, we used
the supertwistor space
$\CP^{3|\CN}=\CPP^{3|\CN}\backslash\CPP^{1|\CN}$ (self-dual case)
and the dual supertwistor space
$\CP^{3|\CN}_*=\CPP^{3|\CN}_*\backslash\CPP^{1|\CN}_*$
(anti-self-dual case) with $0{\le}\CN{\le}4$.
 As the supermanifold $X_\tau$, we chose the
real anti-chiral superspace $\CR_{R}^{4|2\CN}$ (self-dual case)
and the real chiral superspace $\CR_{L}^{4|2\CN}$ (anti-self-dual
case). In both cases, we considered holomorphic Chern-Simons
theory on the supermanifold $Z$ and showed that, by using a gauge
transformation on $Z$, one can bring Witten's form of the hCS field
equations to the previously known constrained equations on the
supercurvature field strength corresponding to $\CN$-extended
self-dual or anti-self-dual SYM theory on $\CR^{4|2\CN}_{R}$ or
$\CR^{4|2\CN}_{L}$ with a metric on the body\footnote{Here the
body is $(\FR^4,g)$. See also appendix B.} of signature $(4,0)$ or
$(2,2)$.

Considering hCS theory on the supertwistor space $\CP^{3|\CN}$, we
gave an explicit expansion of the super gauge potential in
coordinates on $\CPP^{1}\subset\CP^{3|\CN}$ in which the
equivalence of the equations of motion
$\bar{\dpar}\hat{\CA}^{0,1}+\hat{\CA}^{0,1}\wedge\hat{\CA}^{0,1}=0$
to the equations of motion of self-dual $\CN$-extended SYM theory
in four dimensions becomes manifest. All this was translated to
the anti-self-dual case by using the dual supertwistor space
$\CP^{3|\CN}_*$.

We also considered an example of the double fibration
\begin{equation}
\begin{picture}(80,35)
\put(0.0,0.0){\makebox(0,0)[c]{$Z$}}
%\put(32.0,0.1){\makebox(0,0)[c]{$\Leftrightarrow$}}
\put(64.0,0.0){\makebox(0,0)[c]{$X$}}
\put(34.0,33.0){\makebox(0,0)[c]{$Y$}}
\put(7.0,18.0){\makebox(0,0)[c]{$\pi_2$}}
\put(55.0,18.0){\makebox(0,0)[c]{$\pi_1$}}
\put(25.0,25.0){\vector(-1,-1){18}}
\put(37.0,25.0){\vector(1,-1){18}}
\end{picture}
\end{equation}
where $X$ was chosen to be the superspace $\FC^{4|12}$ or its real
version $\FR^{4|12}$ with a metric on the body of signature
$(4,0)$, $(2,2)$ or $(3,1)$. As supermanifold $Z$, we used the
quadric $\CL^{5|6}$ in $\CP^{3|3}\times\CP^{3|3}_*$ or a real
subspace of it with the real structure depending on the signature
of the metric on $\FR^4$. The correspondence space
$Y=\CF^{6|12}=\FC^{4|12}\times\CPP^1\times\CPP^1_*$ was embedded
as a submanifold\footnote{Recall that $Y$ is fibred over $X$ with
fibres $\pi_1^{-1}(x)$ diffeomorphic to submanifolds
$\pi_2(\pi_1^{-1}(x))$ of $Z$ and $Y$ is also fibred over $Z$ with
fibres $\pi_2^{-1}(z)$ which are diffeomorphic to submanifolds
$\pi_1(\pi_2^{-1}(z))$ of $X$, i.e.\ $Y\embd Z\times X$.} in
$Z\times X$ by using the projections $(\pi_1,\pi_2)$. We showed
that, by using a gauge transformation on the correspondence space,
one can bring Witten's form of the hCS field equations to the
well-known constraint equations on the supercurvature field
strength corresponding to full $\CN{=}3$ SYM theory on the
superspace $\FC^{4|12}$ or one of its real subspaces. This theory
is known to be equivalent to $\CN=4$ SYM theory, when formulated
on $\FR^4$.

There are a lot of open problems which deserve further study. On
the field theory side, it is not clear yet how to construct an
action for hCS theory on $\CL^{5|6}$ which will correspond to the
action of $\CN{=}4$ SYM theory. Generalizations of the twistor
correspondence and the Penrose-Ward transform to the string field
theory (SFT) level may also be of interest. This could either be
done in the setting proposed by \cite{Berkovits:2004tx}, although
it seems that due to the off-shell character of SFT one should
employ the more general setting~\cite{Siegel:2004dj}; or one
could concentrate on (an appropriate
extension of) SFT for N=2 string theory.
This theory is known to describe SDYM at tree level~\cite{Ooguri:1991fp}; its
SFT~\cite{berk2} is based on a description of N=2 string theory
as a topological N=4 theory~\cite{berk1}.
This description contains twistors from the
outset: The coordinate $\lambda\in\CPP^1$, the linear system,
integrability and the solution of the equations of motion by
twistor methods were incorporated into the N=2 open SFT in
\cite{Lechtenfeld:2000qj,Lechten3}. However, this theory
reproduces only classical bosonic SDYM theory, its symmetries and
integrability properties~\cite{Lechtenfeld:2000qj,Ivanova:2000zt,Ihl:2002kz}.
Following various proposals, e.g.~\cite{Siegel:1992za,Lu:1995pn,
Bellucci:2001uu,Berkovits:2004ib} (see also references therein),
one can extend it to be
spacetime supersymmetric. This is believed to lead to an explicit relation
between the supersymmetric extension of N=4 topological string
theory and the N=2 topological string (B-type) \cite{Neitzke:2004pf}, but
the picture is far from being complete.

\addcontentsline{toc}{section}{Acknowledgements}

\section*{Acknowledgements}
We are grateful to O. Lechtenfeld, S. Uhlmann and M. Wolf for
many useful comments. This work was partially supported by the
Deutsche Forschungsgemeinschaft (DFG).

\addtocontents{toc}{Appendices}

\appendix
\begin{center}
\Large
Appendices
\end{center}

\renewcommand{\thesection}{\Alph{section}.}
\renewcommand{\theequation}{\thesection\arabic{equation}}

\section{Dictionary: homogeneous $\leftrightarrow$ inhomogeneous coordinates}

The sphere $S^2$ is diffeomorphic to the complex
projective space $\CPP^1$. This space can be para\-metri\-zed
globally by complex homogeneous coordinates $\lambda_{\dot{1}}$ and
$\lambda_{\dot{2}}$ which are not simultaneously zero (in
projective spaces, the origin is excluded). So, the Riemann sphere
$\CPP^1$ can be covered by two coordinate patches
\begin{equation}
U_+=\{\,[\lambda_{\dot{1}},\lambda_{\dot{2}}]\,|\,\lambda_{\dot{1}}\neq 0\,\}
~~~\mbox{and}~~~
U_-=\{\,[\lambda_{\dot{1}},\lambda_{\dot{2}}]\,|\,\lambda_{\dot{2}}\neq
0\,\}~,
\end{equation}
with coordinates
\begin{equation}
\lambda_+:=\frac{\lambda_{\dot{2}}}{\lambda_{\dot{1}}}~~~\mbox{on}~~
U_+~~~\mbox{and}~~~
\lambda_-:=\frac{\lambda_{\dot{1}}}{\lambda_{\dot{2}}}~~~\mbox{on}~~
U_-~.
\end{equation}
On the intersection $U_+\cap U_-$, we get $\lambda_+=1/\lambda_-$.

A global section of the holomorphic line bundle\footnote{See
appendix B.} $\CO(n)$ over $\CPP^1$ exists only for $n\geq 0$.
Over $U_\pm$, it is represented by a polynomial $p^{(n)}_\pm$ of
degree $n$ in the coordinates $\lambda_\pm$ with
$p^{(n)}_+=\lambda_+^n p^{(n)}_-$ on $U_+\cap U_-$. The explicit
expansion will look like
\begin{equation}
p_+^{(n)}=
a_0+a_1\lambda_++a_2\lambda_+^2+...+a_n\lambda_+^n~~~\mbox{and}~~~
p_-^{(n)}=a_0\lambda_-^n+...+a_{n-2}\lambda_-^2+a_{n-1}\lambda_-+a_n~,
\end{equation}
and, multiplying the expansion in $\lambda_+$ by
$\lambda_{\dot{1}}^n$ (or the expansion in $\lambda_-$ by
$\lambda_{\dot{2}}^n$), one obtains a homogeneous polynomial of
degree $n$:
\begin{equation}
a_0\lambda_{\dot{1}}^n+a_1\lambda_{\dot{1}}^{n-1}\lambda_{\dot{2}}+...
+a_{n-1}\lambda_{\dot{1}}\lambda_{\dot{2}}^{n-1}+a_n\lambda_{\dot{2}}^n~=:~
Q^{\ald_1...\ald_n}\lambda_{\ald_1}...\lambda_{\ald_n}~.
\end{equation}

Now let us consider the expansion \eqref{expAa} and \eqref{expAl} of the
super gauge potentials of hCS theory on the supertwistor space.
We get the following list of objects:
\begin{eqnarray}
\eta_i^+& \CO(1) & \eta_i=\lambda_{\dot{1}}\eta_i^+\\
\gamma_+& \CO(-1)\otimes\bar{\CO}(-1) & \gamma=
\frac{1}{\lambda_{\dot{1}}\bl_{\dot{1}}}\;
\gamma_+\left(=\frac{1}{\lambda^\ald\hat{\lambda}_\ald}\right)\\
\hat{\CA}^+_\alpha & \CO(1) &
\hat{\CA}_\alpha=\lambda_{\dot{1}}\;\hat{\CA}_\alpha^+\\
\hat{\CA}_{\bl_+} & \bar{\CO}(-2) &
\hat{\CA}_3=\frac{1}{\bl_{\dot{1}}\bl_{\dot{1}}}\;\hat{\CA}_{\bl_+}~.
\end{eqnarray}
This implies the following expansions in homogeneous coordinates
(cf. \eqref{expAa}, \eqref{expAl}):
\begin{eqnarray}\label{hexpAa}
\hat{\CA}_\alpha&=&\lambda^\ald\,
A_{\alpha\ald}(x_R)+\eta_i\chi^i_\alpha(x_R)+
\gamma\,\frac{1}{2!}\,\eta_i\eta_j\,\hat{\lambda}^\ald\,\phi_{\alpha
\ald}^{ij}(x_R)+\\\nonumber
&&+\gamma^2\,\,\frac{1}{3!}\,\eta_i\eta_j\eta_k\,\hat{\lambda}^\ald\,
\hat{\lambda}^\bed\,
\tilde{\chi}^{ijk}_{\alpha\ald\bed}(x_R)+\gamma^3\,\frac{1}{4!}\,
\eta_i\eta_j\eta_k\eta_l\,
\hat{\lambda}^\ald\,\hat{\lambda}^\bed\,\hat{\lambda}^{\dot{\gamma}}\,
G^{ijkl}_{\alpha\ald\bed\dot{\gamma}}(x_R)~,\\\label{hexpAl}
\hat{\CA}_{3}&=&\gamma^2\,\frac{1}{2!}\,\eta_i\eta_j\,\phi^{ij}(x_R)+
\gamma^3\,\frac{1}{3!}\,\eta_i\eta_j\eta_k\,\hat{\lambda}^\ald\,
\tilde{\chi}^{ijk}_{\ald} (x_R)+\\\nonumber
&&+\gamma^4\,\frac{1}{4!}\,\eta_i\eta_j\eta_k\eta_l\,
\hat{\lambda}^\ald\,\hat{\lambda}^\bed G^{ijkl}_{\ald\bed}(x_R)~.
\end{eqnarray}
For rewriting the equations of motion in terms of this gauge
potential, we also need to rewrite the vector fields
\eqref{vectorfields1} and \eqref{pole} in homogeneous coordinates.
The vector fields along the fibres are easily rewritten,
analogously to the corresponding components of the gauge
potential. The vector field on the sphere can be calculated by
considering $\hat{\CA}_{\bl_+}\dd \bl_+= \hat{\CA}_3
\bar{\Theta}^3$. This implies
$\bar{\Theta}^3=\bl_{\dot{1}}\dd\bl_{\dot{2}}-\bl_{\dot{2}}\dd\bl_{\dot{1}}$,
which has a dual vector field $\bar{V}_3$ defined by
$\bar{V}_3\lrcorner\,\bar{\Theta}^3=1$. Altogether, we obtain the
basis
\begin{equation}
\bar{V}_\alpha=\lambda^\ald\der{x_R^{\alpha\ald}}~~~
\mbox{and}~~~\bar{V}_3=-\gamma\lambda^\ald\der{\hat{\lambda}^\ald}~.
\end{equation}

The field equations \eqref{shCS1} and \eqref{shCS2} now take the form
\begin{eqnarray}
\bar{V}_\alpha \hat{\CA}_\beta-\bar{V}_\beta \hat{\CA}_\alpha
+[\hat{\CA}_\alpha,\hat{\CA}_\beta]&=&0~,\\
\bar{V}_3 \hat{\CA}_\alpha-\bar{V}_\alpha \hat{\CA}_3
+[\hat{\CA}_3,\hat{\CA}_\alpha]&=&0
\end{eqnarray}
and yield the same equations \eqref{SDYMeom} for the physical
fields.

\section{Some mathematical definitions}

{\bf Interior product.} For the interior product of a vector $V$
with a one-form $A$, we use the notation $V\lrcorner A:=\langle
V,A\rangle$. A second common notation for this product is $\di_V
A$.

\smallskip\noindent{\bf Holomorphic line bundles.} Given the Riemann sphere
$\CPP^1\cong S^2$ with standard patches $U_+$ and $U_-$ and
coordinates $\lambda_\pm$ on the corresponding patches and
$\lambda_\pm=1/\lambda_\mp$ on $U_+\cap U_-$, the {\em holomorphic
line bundle} $\CO(n)$ is defined by its transition function
$z_+=\lambda_+^n z_-$, where $z_\pm$ are complex coordinates on
fibres over $U_\pm$. For $n\geq 0$, global sections of the bundle
$\CO(n)$ are polynomials of degree $n$ in the coordinates
$\lambda_\pm$ and homogeneous polynomials of degree $n$ in
homogeneous coordinates (see also appendix A). The $\CO(n)$ line
bundle has first Chern number $n$. The complex conjugate bundle to
$\CO(n)$ is denoted by $\bar{\CO}(n)$. Its sections have
transition functions $\bar{\lambda}_+^n :
\bar{z}_+=\bar{\lambda}_+^n \bar{z}_-$.

\smallskip\noindent{\bf Spinor conventions.} All objects with
space-time indices are rewritten in spinor notation by
$x^{\alpha\ald}=\sigma_\mu^{\alpha\ald}x^\mu$ etc., where the
sigma-matrices are determined by the metric under consideration.
The homogeneous coordinates $\lambda_{\dot{1}}$ and
$\lambda_{\dot{2}}$ for a point in $\CPP^1$ are regarded as
components of a complex commuting spinors. Their indices are
raised and lowered with the antisymmetric $\eps$-tensors. We use
the convention
$\eps_{12}=\eps_{\dot{1}\dot{2}}=-\eps^{12}=-\eps^{\dot{1}\dot{2}}=1$,
implying
$\eps^{\alpha\beta}\eps_{\beta\gamma}=\delta^\alpha_\gamma$. The
complex conjugate is obtained by conjugating the components of the
spinor. A second anti-linear conjugation, denoted by $\hat{\cdot}$ is
performed for different types of spinors as
\begin{equation}
(\hat{\mu}_\alpha):=C (\bar{\mu}_\alpha)~,~~~
(\hat{\mu}^\alpha):=C(\bar{\mu}^\alpha)~,~~~
(\hat{\lambda}_\ald):=C(\bl_\ald)
~~~\mbox{and}~~~
(\hat{\lambda}^\ald):=C(\bl^\ald)~,
\end{equation}
where the $2\times2$-matrix $C$ is given by
\begin{equation}
C=\left(\begin{array}{cc} 0 & \eps\\ 1 & 0\end{array}\right)~.
\end{equation}
The conventions for Gra\ss mann variables are discussed in the
text around \eqref{tau1}-\eqref{tau5} and \eqref{Mtau}. These imply in
particular that
\begin{equation}
\overline{\der{\xi}}=\der{\bar{\xi}}~.
\end{equation}
Furthermore, we adopt the following convention for the conjugation
of products of Gra\ss mann variables and supernumbers in general:
\begin{equation}
\tau(\xi^1\xi^2)=\tau(\xi^1)\tau(\xi^2)~~~\mbox{and}~~~\tau(z^1
z^2)=\tau(z^1)\tau(z^2)~.
\end{equation}
With this choice, products of two real objects will be real. Note
that this is {\em not} the common convention used for
supersymmetry in Minkowski space, and here, we define
$\tau_M(\xi^1\xi^2)=\tau_M(\xi^2)\tau_M(\xi^1)$. A more detailed
discussion can be found in \cite{Cartier:2002zp}.

\smallskip\noindent{\bf Flag manifolds.} Complex flag manifolds
are a major tool in the context of twistors and the Penrose-Ward
correspondence. They can be considered as a generalization of
projective spaces and Gra\ss mann manifolds. An $r$-tuple of
vector spaces $(L_1,...,L_r)$ of dimensions $\dim_\FC L_i=d_i$
with $L_1\subset...\subset L_r\subset\FC^n$ and $0<d_0<...<d_r<n$
is called a {\em flag} in $\FC^n$. A {\em (complex) flag manifold}
is the (compact) space
\begin{equation}
F_{d_1...d_r}:=\{\mbox{all flags }(L_1,...,L_r)\mbox{ with
}\dim_\FC
L_i=d_i,~i=1,...,r\,\}~.
\end{equation}
Simple examples are $F_1=\CPP^{n-1}$ and $F_k=G_{k,n}(\FC)$.

{}To see how flag manifolds naturally arise, consider the
following reformulation of the (bosonic part of the) discussion
following \eqref{dblfibration}. We fix the full space to be
$\FC^4$. Then we can establish the following double fibration:
\begin{equation}\label{flgdblfibration1}
\begin{picture}(75,40)
\put(0.0,0.0){\makebox(0,0)[c]{$F_1$}}
%\put(32.0,0.1){\makebox(0,0)[c]{$\Leftrightarrow$}}
\put(64.0,0.0){\makebox(0,0)[c]{$F_2$}}
\put(32.0,33.0){\makebox(0,0)[c]{$F_{12}$}}
\put(7.0,18.0){\makebox(0,0)[c]{$\pi_2$}}
\put(55.0,18.0){\makebox(0,0)[c]{$\pi_1$}}
\put(25.0,25.0){\vector(-1,-1){18}}
\put(37.0,25.0){\vector(1,-1){18}}
\end{picture}
\end{equation}
Let $(L_1,L_2)$ be an element of $F_{12}$, i.e.\ $\dim_\FC L_1=1$,
$\dim_\FC L_2=2$ and $L_1\subset L_2$. Thus $F_{12}$ fibres over
$F_2$ with $\CPP^1$ as a typical fibre, which parametrizes the
freedom to choose a complex one-dimensional subspace in a complex
two-dimensional vector space. The projections are defined as
$\pi_2(L_1,L_2)=L_1$ and $\pi_1(L_1,L_2)=L_2$. The full connection
to \eqref{dblfibration} becomes obvious, when we note that
$F_1=\CPP^3=\CP^3\cup\CPP^1$ and that $F_2=G_{2,4}(\FC)$ is the
complexified and compactified version of $\FR^4$. The advantage of
the formulation in terms of flag manifolds is related to the fact,
that the projections are immediately clear: one has to shorten the
flags to suit the structure of the flags of the base space.

The compactified version of the ``dual" fibration
\eqref{superdblfibration5} is
\begin{equation}\label{flgdblfibration2}
\begin{picture}(80,40)
\put(0.0,0.0){\makebox(0,0)[c]{$F_3$}}
%\put(32.0,0.1){\makebox(0,0)[c]{$\Leftrightarrow$}}
\put(64.0,0.0){\makebox(0,0)[c]{$F_2$}}
\put(32.0,33.0){\makebox(0,0)[c]{$F_{23}$}}
\put(7.0,18.0){\makebox(0,0)[c]{$\pi_2$}}
\put(55.0,18.0){\makebox(0,0)[c]{$\pi_1$}}
\put(25.0,25.0){\vector(-1,-1){18}}
\put(37.0,25.0){\vector(1,-1){18}}
\end{picture}
\end{equation}
where $F_3$ is the space of hyperplanes in $\FC^4$. This space is
naturally dual to the space of lines, as every hyperplane is fixed
by a vector orthogonal to the elements of the hyperplane.
Therefore, we have $F_3=F_1^*=\CPP_*^3\supset \CP^3_*$.

Also the third double fibration \eqref{ambidblfibration2}, which
we used in the case of full $\CN{=}3$ SYM, is a restricted version
of the diagram
\begin{equation}\label{flgdblfibration3}
\begin{picture}(80,40)
\put(0.0,0.0){\makebox(0,0)[c]{$F_{13}$}}
%\put(32.0,0.1){\makebox(0,0)[c]{$\Leftrightarrow$}}
\put(64.0,0.0){\makebox(0,0)[c]{$F_2$}}
\put(32.0,33.0){\makebox(0,0)[c]{$F_{123}$}}
\put(7.0,18.0){\makebox(0,0)[c]{$\pi_2$}}
\put(55.0,18.0){\makebox(0,0)[c]{$\pi_1$}}
\put(25.0,25.0){\vector(-1,-1){18}}
\put(37.0,25.0){\vector(1,-1){18}}
\end{picture}
\end{equation}
where $F_2=G_{2,4}(\FC)$ is again the complexified and
compactified version of $\FR^4$. The flag manifold $F_{13}$ is
topologically the zero locus of a quadric in $\CPP^3\times\CPP^3_*$.
For further details and the super generalization, see
e.g.~\cite{Wells, Howe}.

\smallskip\noindent{\bf Supermanifolds and Calabi-Yau
supermanifolds.} The space $\FR^{r|s}$ is described by coordinates
$x^i$ and $\theta^j$ with $1\leq i\leq r$, $1\leq j\leq s$, where
the $\theta^j$ are real Gra\ss mann variables satisfying the
algebra $\{\theta^j,\theta^k\}=0$. The superspace $\FC^{r|s}$ is
defined analogously, with complex coordinates: $\bar{x}^i\neq
x^i$, $\bar{\theta}^j\neq \theta^j$. For our considerations, a
{\em supermanifold} is defined to be a topological space which is
locally diffeomorphic to $\FR^{r|s}$ or $\FC^{r|s}$.

A supermanifold contains a purely bosonic part (the ``body'')
which is parametrized in terms of bosonic coordinates. The body of
a supermanifold is a real or complex manifold by itself. The
$\RZ_2$-grading of the superspace used for parametrizing the
supermanifold induces a grading on the ring of functions on the
supermanifold. For objects like subspaces, forms etc. which come
with a dimension, a degree etc., we use the notation $(i|j)$,
where $i$ and $j$ denote the bosonic and fermionic part,
respectively.

We further introduce the parity-changing operator $\Pi$ which,
when acting on a fibre bundle, changes the parity of the fibre
coordinates. For example, $\Pi \CO(n)\rightarrow\CPP^1$ is
parametrized by complex variables $\lambda_\pm$ and Gra\ss
mann variables $\theta_\pm$ with
$\theta_+=\lambda_+^n\theta_-$ on $U_+\cap U_-$.

For a more extensive discussion of supermanifolds, see
\cite{Cartier:2002zp} and references therein.

Calabi-Yau manifolds are manifolds with vanishing first
Chern class which implies the existence of a globally
well-defined holomorphic volume form. For our purposes, we define
a {\em Calabi-Yau supermanifold} to be a supermanifold with a
globally defined holomorphic volume form. Note that the body of a
super CY is not a CY, in general.

In the purely bosonic case, the 3-fold
$\CO(m)\oplus\CO(n)\rightarrow \CPP^1$ with coordinates
$z^1_\pm,z^2_\pm,\lambda_\pm$ is a CY, if and only if $m+n=-2$,
and a volume form is then given by $\Omega^{3,0}_\pm=\pm\dd
z_\pm^1\wedge\dd z_\pm^2\wedge\dd \lambda_\pm$. In the super case,
the fermionic coordinates can also be assigned to some line
bundle, but because the Berezinian (i.e.\ the fermionic Jacobi
determinant) enters as an inverse in the integration, a fermionic
coordinate living in $\CO(n)$ will contribute $-n$ to the overall
first Chern number. Thus the bundle
\begin{equation}
\CP^{3|4}=
\CO(1)\oplus\CO(1)\oplus\Pi\CO(1)\oplus\Pi\CO(1)\oplus\Pi\CO(1)
\oplus\Pi\CO(1)\rightarrow \CPP^1
\end{equation}
is a CY supermanifold. Its holomorphic volume form is given by
$\hat\Omega^{3,0|4,0}_\pm=\pm\dd z_\pm^1\wedge\dd z_\pm^2\wedge\dd
\lambda_\pm\dd\theta^1_\pm\dd\theta^2_\pm $
$\dd\theta^3_\pm\dd\theta^4_\pm$, where $z^i_\pm$ and
$\theta^j_\pm$ are coordinates of the bosonic and fermionic line
bundles, respectively. The body of this supermanifold is
$\CO(1)\oplus\CO(1)\rightarrow \CPP^1$ and it is obviously not a
CY manifold.

\section{The twistor geometry in the Kleinian case $\eps=+1$}

As mentioned several times in the text, one should consider hCS
theory on domains $\hat{\CU}_\pm$ of the supertwistor space
$\CP^{3|\CN}$ for which $|\lambda_\pm|\neq1$ when working in the
Kleinian case, i.e.~when using the reality conditions obtained
from the involution\footnote{For $\tau_0$, the description is
similar and for that reason we focus on $\tau_1$. Note, however,
that the $\CPP^1$ embedded in the twistor spaces reduces to
different $S^1$s: for $\tau_1$, the constraint is
$\lambda_\pm=\bl^{-1}_\pm$ and for $\tau_0$ we have
$\lambda_\pm=\bl_\pm$.} $\tau_1$. In this and the following
appendix, we will discuss this aspect in more detail.

Let us start from the double fibration \eqref{superdblfibration},
\begin{equation}\label{C1}
\begin{picture}(50,40)
\put(0.0,0.0){\makebox(0,0)[c]{$\CP^{3|\CN}$}}
%\put(37.0,0.1){\makebox(0,0)[c]{$\Leftrightarrow$}}
\put(74.0,0.0){\makebox(0,0)[c]{$\CM_R^{4|2\CN}$}}
\put(42.0,37.0){\makebox(0,0)[c]{$\CF_R^{5|2\CN}$}}
\put(7.0,20.0){\makebox(0,0)[c]{$\pi_2$}}
\put(65.0,20.0){\makebox(0,0)[c]{$\pi_1$}}
\put(25.0,27.0){\vector(-1,-1){18}}
\put(47.0,27.0){\vector(1,-1){18}}
\end{picture}
\end{equation}
which describes the complex supertwistor correspondence for
$0\leq\CN\leq 4$. As before, we have complex coordinates
$(z_\pm^\alpha,\lambda_\pm,\eta_i^\pm)$ on the patches
$\hat{\CU}_\pm$ which cover $\CP^{3|\CN}$ and
$(x_R^{\alpha\ald},\lambda_\ald^\pm,\eta_i^\ald)$ on
$\CF^{5|2\CN}_R$. The projection $\pi_1$ is the trivial projection
$\pi_1(x_R^{\alpha\ald},\lambda_\ald^\pm,\eta_i^\ald)=(x_R^{\alpha\ald},\eta_i^\ald)$
and the projection $\pi_2$ is given by the formul\ae{}
\begin{equation}\label{C2}
z_\pm^\alpha=x_R^{\alpha\ald}\lambda_\ald^\pm~,~~~
\lambda_\ald^\pm=\lambda_\ald^\pm~~~\mbox{and}~~~
\eta_i^\pm=\eta_i^\ald\lambda_\ald^\pm~~~\mbox{with}
~~~(\lambda_\ald^+)=\left(\begin{array}{c} 1 \\ \lambda_+
\end{array}\right)~,~~~(\lambda_\ald^-)=\left(\begin{array}{c}
\lambda_- \\ 1
\end{array}\right)~.
\end{equation}
The action of the involution $\tau_1$ on the coordinates of
$\CP^{3|\CN}$ is given by formul\ae{} \eqref{three} together with
$\tau_1(\eta_i^\pm)=\bar{\eta}_i^\pm/\bl_\pm$. It yields the
reality conditions
\begin{equation}
z_\pm^2=\frac{\bar{z}_\pm^1}{\bl_\pm}~,~~~
\lambda_\pm=\frac{1}{\bl_\pm}~~~\mbox{and}~~~
\eta_i^\pm=\frac{\bar{\eta}^\pm_i}{\bl_\pm}
\end{equation}
on $\CP^{3|\CN}$, which imply
\begin{equation}\label{C4}
x_R^{2\dot{2}}=\bar{x}_R^{1\dot{1}}~,~~~
x_R^{2\dot{1}}=\bar{x}_R^{1\dot{2}}~~~\mbox{and}~~~
\eta_i^{\dot{2}}=\bar{\eta}_i^{\dot{1}}
\end{equation}
on $\CM_R^{4|2\CN}$ and $\CF^{5|2\CN}_R=\CM_R^{4|2\CN}\times
\CPP^1$.

The set of fixed points under this involution\footnote{Although
$\tau_1$ was defined on $\CP^{3|\CN}$, it induces an involution on
$\CF^{5|2\CN}_R$ which we will denote by the same symbol in the
following.} of the spaces contained in the double fibration
\eqref{C1} form real subsets $\CT^{3|\CN}\subset\CP^{3|\CN}$,
$\CR_R^{4|2\CN}\subset\CM_R^{4|2\CN}$ and $\CR_R^{4|2\CN}\times
S^1\subset\CF_R^{5|2\CN}$. Recall that the body $\CT^3$ of the
supermanifold\footnote{For $\CN=4$, $\CT^{3|\CN}$ has a globally
defined real volume form invariant under rescaling of homogeneous
coordinates.} $\CT^{3|\CN}$ is diffeomorphic to the space
$\RPS^3\backslash\RPS^1$ (cf. \eqref{realtwistor}) fibred over
$S^1\cong\RPS^1\subset\CPP^1$. Thus, we obtain the real version
\begin{equation}\label{C5}
\begin{picture}(50,50)
\put(0.0,0.0){\makebox(0,0)[c]{$\CT^{3|\CN}$}}
%\put(37.0,0.1){\makebox(0,0)[c]{$\Leftrightarrow$}}
\put(74.0,0.0){\makebox(0,0)[c]{$\CR_R^{4|2\CN}$}}
\put(38.0,37.0){\makebox(0,0)[c]{$\CR_R^{4|2\CN}\times S^1$}}
\put(7.0,20.0){\makebox(0,0)[c]{$\pi_2$}}
\put(65.0,20.0){\makebox(0,0)[c]{$\pi_1$}}
\put(25.0,27.0){\vector(-1,-1){18}}
\put(47.0,27.0){\vector(1,-1){18}}
\end{picture}
\end{equation}
of the double fibration \eqref{C1}. Here, $\pi_1$ is again the
trivial projection and $\pi_2$ is given by equations
\eqref{C2}-\eqref{C4} with $|\lambda_\pm|=1$.

The tangent spaces to the (real) $(2|\CN)$-dimensional leaves of
the fibration $\pi_2$ in \eqref{C5} are spanned by the vector
fields
\begin{equation}\label{C6}
v^+_\alpha:=\lambda_+^\ald\der{x_R^{\alpha\ald}}~~~\mbox{and}~~~
v_+^i:=\lambda^\ald_+\der{\eta_i^\ald}~,
\end{equation}
which satisfy the reality conditions
\begin{equation}\label{C7}
v^+_2=-\lambda_+ \bar{v}^+_1~~~\mbox{and}~~~v^i_+=-\lambda_+
\bar{v}_+^i~,
\end{equation}
where $|\lambda_+|=1$. Equivalently, one could also use the vector
fields
\begin{equation}\label{C6.2}
v^-_\alpha:=\lambda_-^\ald\der{x_R^{\alpha\ald}}=\lambda_-
v^+_\alpha~~~\mbox{and}~~~
v_-^i:=\lambda^\ald_-\der{\eta_i^\ald}=\lambda_-v_+^i~~~\mbox{with}~~\lambda_-=\frac{1}{\lambda_+}=\bl_+~.
\end{equation}
The vector fields \eqref{C6} and \eqref{C6.2} are the restrictions
of the vector fields $\bar{V}^\pm_\alpha$ and $\bar{\dpar}_\pm^i$
to $|\lambda_\pm|=1$.

Consider now the double fibration
\begin{equation}\label{C8}
\begin{picture}(50,50)
\put(0.0,0.0){\makebox(0,0)[c]{$\CP^{3|\CN}$}}
%\put(37.0,0.1){\makebox(0,0)[c]{$\Leftrightarrow$}}
\put(74.0,0.0){\makebox(0,0)[c]{$\CR_R^{4|2\CN}$}}
\put(38.0,37.0){\makebox(0,0)[c]{$\CR_R^{4|2\CN}\times \CPP^1$}}
\put(7.0,20.0){\makebox(0,0)[c]{$\pi_2$}}
\put(65.0,20.0){\makebox(0,0)[c]{$\pi_1$}}
\put(25.0,27.0){\vector(-1,-1){18}}
\put(47.0,27.0){\vector(1,-1){18}}
\end{picture}
\end{equation}
where the map $\pi_2$ is defined by the formul\ae{} \eqref{C2}
with $x_R^{\alpha\ald},\eta_i^\ald$ satisfying the reality
condition \eqref{C4} and complex $\lambda_\ald^\pm$. For Kleinian
signature $(--++)$, we have the local isomorphism
$\sSO(2,2)\cong\sSU(1,1)\times\sSU(1,1)$ and under the action of
the group $\sSU(1,1)$, the Riemann sphere $\CPP^1$ of projective
spinors decomposes into the disjoint union $\CPP^1=H_+^2\cup
S^1\cup H_-^2$ of three orbits. Here $H^2=H_+^2\cup H_-^2$ is the
two-sheeted hyperboloid with the boundary $S^1$ which is stable
under the involution $\tau_1:\lambda_\pm\mapsto\bl_\pm^{-1}$. We
have $\tau_1(S^1)=S^1$ and $\tau_1(H_\pm^2)=H_\mp^2$. This
decomposition can be carried over to the spaces in the double
fibration \eqref{C8}. We have
\begin{equation}\label{CC10}
\begin{array}{ccccccc} \CR_R^{4|2\CN}\times
\CPP^1&=&\CR_R^{4|2\CN}\times H_+^2&\cup&\CR_R^{4|2\CN}\times
S^1&\cup&\CR_R^{4|2\CN}\times
H_-^2\\
&&&&\downarrow&&\\
\downarrow&&\downarrow&&\raisebox{-4pt}{$\CT^{3|\CN}$}&&\downarrow\\
&&&&\cap&&\\
\CP^{3|\CN}&=&\CP^{3|\CN}_+&\cup&\CP_{0,\CN}&\cup&\CP^{3|\CN}_-~,
\end{array}
\end{equation}
where $\downarrow$ symbolizes the projection $\pi_2$. Here
$\CP_+^{3|\CN}$, $\CP_{0,\CN}$ and $\CP_-^{3|\CN}$ are
restrictions of the complex vector bundle
$\CP^{3|\CN}\rightarrow\CPP^1$ to $H_+^2$, $S^1$ and $H_-^2$,
respectively, and thus
$\CP_{0,\CN}:=\left.\CP^{3|\CN}\right|_{|\lambda_\pm|=1}$ is the
common boundary of the spaces $\CP^{3|\CN}_+$ and $\CP^{3|\CN}_-$.
Note that restricting the maps $\pi_1$ and $\pi_2$ in \eqref{C8}
to the space $\CR_R^{4|2\CN}\times S^1$, we obtain the real double
fibration \eqref{C5}.

The map $\pi_2$ in \eqref{C8} restricted to the space
$\CR_R^{4|2\CN}\times H^2$ is a smooth bijection (diffeomorphism)
of $\CR^{4|2\CN}_R\times H^2$ onto
$\tilde{\CP}^{3|\CN}:=\CP^{3|\CN}\backslash\CP_{0,\CN}=\CP^{3|\CN}_+\cup\CP^{3|\CN}_-$
which is defined by the formul\ae{} \eqref{C2} with
$|\lambda_\pm|\neq 1$ and $x_R^{\alpha\ald}$, $\eta_i^\ald$
subject to \eqref{C4}. Its inverse is given by
\begin{align}\nonumber
x_R^{1\dot{1}}=\frac{z^1_+-z_+^3\bar{z}^2_+}{1-z_+^3\bar{z}_+^3}
=\frac{\bz^2_--\bz_-^3z^1_-}{1-z_-^3\bar{z}_-^3}~,~~~
x_R^{2\dot{1}}=\frac{z^2_+-z_+^3\bar{z}^1_+}{1-z_+^3\bar{z}_+^3}
=\frac{\bz^1_--\bz_-^3z^2_-}{1-z_-^3\bar{z}_-^3}~,\\
\eta_i^{\dot{1}}=\frac{\eta_i^+-z^3_+\bar{\eta}_i^+}{1-z_+^3\bar{z}_+^3}
=\frac{\bar{\eta}_i^--\bz^3_-\eta_i^-}{1-z_-^3\bar{z}_-^3}~,~~~
\lambda_\pm=z_\pm^3\hspace*{2cm}\label{CC12}
\end{align}
and $x_R^{2\dot{2}}$, $x_R^{1\dot{2}}$ and $\eta^{\dot{2}}_i$
fixed by \eqref{C4}. Due to this diffeomorphism, the diagram
\eqref{C8} with the maps $\pi_1$ and $\pi_2$ restricted to
$\CR_R^{4|2\CN}\times H^2$ becomes a nonholomorphic fibration
\begin{equation}\label{C13}
\tilde{\CP}^{3|\CN}\rightarrow \CR^{4|2\CN}_R
\end{equation}
and on $\tilde{\CP}^{3|\CN}$, one can use either set of
coordinates $(z_\pm^\alpha,\lambda_\pm,\eta_i^\pm)$ and
$(x_R^{\alpha\ald},\lambda_\pm,\eta_i^\ald)$.

For the dual supertwistor space $\CP^{3|\CN}_*$, the discussion
follows along the same lines. One merely replaces the coordinates
$(z_\pm^\alpha,\lambda_\pm,\eta_i^\pm)$ of $\CP^{3|\CN}$ with the
coordinates $(w_\pm^\ald,\mu_\pm,\theta^i_\pm)$ of $\CP_*^{3|\CN}$
and the moduli $(x^{\alpha\ald}_R,\eta_i^\ald)\in\CM_R^{4|2\CN}$
with the moduli $(x^{\alpha\ald}_L,\theta^{\alpha
i})\in\CM_L^{4|2\CN}$. Considering then the set of fixed points of
the involution $\tau_1$ as done above leads to fibrations similar
to \eqref{C5}, \eqref{C8} and \eqref{C13}.

\section{Comments on hCS theory in the Kleinian case $\eps=+1$}

In the case of the real structure $\tau_1$, i.e.\ $\eps=+1$, which
yields Kleinian signature $(2,2)$ on $\FR^4$, we always discussed
hCS theory on $\tilde{\CP}^{3|\CN}$ in the text. This is due to a
peculiarity of the Penrose-Ward correspondence in this case which
we now discuss more explicitly.

Consider the real supertwistor space
$\CT^{3|\CN}\subset\CP^{3|\CN}$ and a real-analytic function
$f_{+-}^\tau:\CT^{3|\CN}\rightarrow\sGL(n,\FC)$ which can be
understood as an isomorphism
$f_{+-}^\tau:\CE_-^\tau\rightarrow\CE_+^\tau$ between two trivial
complex vector bundles $\CE^\tau_\pm\rightarrow\CT^{3|\CN}$. We
assume that $f_{+-}^\tau$ satisfies the reality condition
\begin{equation}\label{D1}
\left(f_{+-}^\tau\left(z_+^\alpha,\lambda_+,\eta_i^+\right)\right)^\dagger
=f_{+-}^\tau(z_+^\alpha,\lambda_+,\eta_i^+)~.
\end{equation}
Given such a function $f_{+-}^\tau$, one can extend it
holomorphically into a neighborhood $\hat{\CU}$ of $\CT^{3|\CN}$
in $\CP^{3|\CN}$, such that the extension $f_{+-}$ of
$f_{+-}^\tau$ satisfies the reality condition
\begin{equation}\label{DD3}
\left(f_{+-}\left(\tau_1(z_+^\alpha,\lambda_+,\eta_i^+)\right)\right)^\dagger
=f_{+-}(z_+^\alpha,\lambda_+,\eta_i^+)~,
\end{equation}
generalizing equation \eqref{D1}. The function $f_{+-}$ is
holomorphic on $\hat{\CU}=\hat{\CU}_+\cap\hat{\CU}_-$ and can be
identified with a transition function of a holomorphic vector
bundle $\CE$ over $\CP^{3|\CN}=\hat{\CU}_+\cup\hat{\CU}_-$ which
glues together two trivial bundles $\CE_+=\hat{\CU}_+\times\FC^n$
and $\CE_-=\hat{\CU}_-\times \FC^n$. Obviously, the two trivial
vector bundles $\CE^\tau_\pm\rightarrow \CT^{3|\CN}$ are
restrictions of the trivial bundles $\CE_\pm\rightarrow
\hat{\CU}_\pm$ to $\CT^{3|\CN}$.

In the twistor approach, it is assumed that the bundle $\CE$ is
holomorphically trivial when restricted to any curve
$\CPP^1_{x_R,\eta}\embd\CP^{3|\CN}$ and therefore there exists a
gauge in which the restriction of the transition function $f_{+-}$
to any $\CPP^1_{x_R,\eta}$ splits,
\begin{equation}\label{DD5}
f_{+-}=\psi_+^{-1}\psi_-~,
\end{equation}
into regular\footnote{Recall that by `regular', we mean smooth
with nonvanishing determinant.} matrix-valued functions $\psi_+$
and $\psi_-$ defined on $\hat{\CU}_+=\CP_+^{3|\CN}\cup\hat{\CU}$
and $\hat{\CU}_-=\CP_-^{3|\CN}\cup\hat{\CU}$ and holomorphic in
$\lambda_+\in H_+^2$ and $\lambda_-\in H_-^2$, respectively. Note
that the condition \eqref{DD3} is satisfied if
\begin{equation}\label{DDD4}
\psi_+^{-1}(\tau_1(x_R^{\alpha\ald},\lambda_+,\eta^\ald_i))=
\psi_-^\dagger(x_R^{\alpha\ald},\lambda_-,\eta_i^\ald)~.
\end{equation}
Restricting \eqref{DD5} to $S^1_{x_R,\eta}\embd\CPP^1_{x_R,\eta}$,
we obtain
\begin{equation}\label{DDD2}
f_{+-}^\tau=(\psi_+^\tau)^{-1}\psi_-^\tau~~~\mbox{with}~~~
(\psi_+^\tau)^{-1}=(\psi_-^\tau)^\dagger~,
\end{equation}
where $\psi_\pm^\tau$ are restrictions to $\CR_R^{4|2\CN}\times
S^1$ of the matrix-valued functions $\psi_\pm$ given by
\eqref{DD5} and \eqref{DDD4}. Thus the initial twistor data
consist of a real-analytic function\footnote{One could also
consider the extension $f_{+-}$ and the splitting \eqref{DDD2}
even if $f^\tau_{+-}$ is not analytic, but in this case the
solutions to the super SDYM equations can be singular. Such
solutions are not related with holomorphic bundles.} $f_{+-}^\tau$
on $\CT^{3|\CN}$ satisfying \eqref{D1} together with a splitting
\eqref{DDD2}, from which we construct a holomorphic vector bundle
$\CE$ over $\CP^{3|\CN}$ with a transition function $f_{+-}$ which
is a holomorphic extension of $f_{+-}^\tau$ to
$\hat{\CU}\supset\CT^{3|\CN}$. In other words, the space of real
twistor data is the moduli space of holomorphic vector bundles
$\CE\rightarrow\CP^{3|\CN}$ with transition functions satisfying
the reality conditions \eqref{DD3}.

In the purely real setting, one considers a real-analytic
$\sGL(n,\FC)$-valued function $f^\tau_{+-}$ on $\CT^{3|\CN}$
satisfying the hermiticity condition \eqref{D1} and the real
double fibration \eqref{C5}. Since the pull-back of $f_{+-}^\tau$
to $\CR^{4|2\CN}_R\times S^1$ has to be constant along the fibres
of $\pi_2$, we obtain the constraint equations
\begin{equation}\label{D3}
v^+_\alpha f^\tau_{+-}=0=v_+^i f^\tau_{+-}~,
\end{equation}
or equivalently
\begin{equation}\label{D3.2}
v^-_\alpha f^\tau_{+-}=0=v_-^i f^\tau_{+-}~,
\end{equation}
with the vector fields $v^\pm_\alpha$ and $v_\pm^i$ from
\eqref{C6} and \eqref{C6.2}. Using the splitting \eqref{DDD2} of
$f^\tau_{+-}$ on fibres $S^1_{x_R,\eta}$ of the projection $\pi_1$
in \eqref{C5} and substituting
$f^\tau_{+-}=(\psi^\tau_+)^{-1}\psi^\tau_-$ into \eqref{D3}, we
obtain the linear systems (cf. \eqref{slinsys1}-\eqref{slinsys4})
\begin{align}\nonumber
&(v^+_\alpha+\CA^+_\alpha)\psi^\tau_+=0~,&(v^-_\alpha+\CA^-_\alpha)\psi^\tau_-=0~,\\\label{D4}
&(v_+^i+\CA_+^i)\psi^\tau_+=0~,&(v_-^i+\CA_-^i)\psi^\tau_-=0~.
\end{align}
Here $\CA_\pm=(\CA_\alpha^\pm,\CA_\pm^i)$ are relative connections
on the bundles $\pi_2^*\CE^\tau_\pm$. {}From \eqref{D4}, one can
find $\psi^\tau_\pm$ for any given $\CA^\pm_\alpha$ and
$\CA_\pm^i$ and vice versa, i.e.\ find $\CA^\pm_\alpha$ and
$\CA_\pm^i$ for given $\psi^\tau_\pm$ by the formul\ae{}
\begin{align}\nonumber
&\CA_\alpha^+=\psi_+^\tau v_\alpha^+(\psi_+^\tau)^{-1}=\psi_-^\tau
v_\alpha^+(\psi_-^\tau)^{-1}~, &\CA_\alpha^-=\psi_+^\tau
v_\alpha^-(\psi_+^\tau)^{-1}=\psi_-^\tau
v_\alpha^-(\psi_-^\tau)^{-1}~,\\
\label{D8s} &\CA^i_+=\psi_+^\tau
v^i_+(\psi_+^\tau)^{-1}=\psi_-^\tau
v^i_+(\psi_-^\tau)^{-1}~,&\CA^i_-=\psi_+^\tau
v^i_-(\psi_+^\tau)^{-1}=\psi_-^\tau v^i_-(\psi_-^\tau)^{-1}~.
\end{align}
The compatibility conditions of the linear systems \eqref{D4} read
\begin{align}\nonumber
v^\pm_\alpha
\CA^\pm_\beta-v^\pm_\beta\CA^\pm_\alpha+[\CA^\pm_\alpha,\CA^\pm_\beta]&=0~,\\\nonumber
v^\pm_\alpha \CA_\pm^i-v_\pm^i\CA^\pm_\alpha+[\CA^\pm_\alpha,\CA_\pm^i]&=0~,\\
v_\pm^i
\CA_\pm^j-v_\pm^j\CA_\pm^i+[\CA_\pm^i,\CA_\pm^j]&=0~.\label{D5}
\end{align}
Geometrically, these equations imply flatness of the curvature of
the relative connections $\CA_\pm=(\CA^\pm_\alpha,\CA_\pm^i)$ on
the bundles $\pi_2^*\CE^\tau_\pm$ defined along the real
$(2|\CN)$-dimensional fibres of the projection $\pi_2$ in
\eqref{C5}.

Recall that $\psi^\tau_+$ and $\psi^\tau_-$ extend holomorphically
in $\lambda_+$ and $\lambda_-$ to $H_+^2$ and $H_-^2$,
respectively, and therefore we obtain from \eqref{D8s} that
$\CA^\pm_\alpha=\lambda_\pm^\ald\CA_{\alpha\ald}$ and
$\CA_\pm^i=\lambda_\pm^\ald\CA_\ald^i$, where $\CA_{\alpha\ald}$
and $\CA_\ald^i$ do not depend on $\lambda_\pm$. Then the
compatibility conditions \eqref{D5} of the linear systems
\eqref{D4} reduce to equations \eqref{compcon}. In section 5 it
was demonstrated that for $\eps=+1$, these equations are
equivalent to the field equations of $\CN$-extended SDYM theory on
$\FR^{2,2}$. Thus there are bijections between the moduli spaces
of solutions to equations \eqref{D5}, the field equations of
$\CN$-extended super SDYM theory on $\FR^{2,2}$ and the moduli
space of $\tau_1$-real holomorphic vector bundles $\CE$ over
$\CP^{3|\CN}$.

Consider now the extension of the linear systems \eqref{D4} to
open domains
$\hat{\CU}_\pm=\CP^{3|\CN}_\pm\cup\,\hat{\CU}\supset\CT^{3|\CN}$,
\begin{align}\nonumber
(\bar{V}_\alpha^\pm+\CA_\alpha^\pm)\psi_\pm&=0~,\\\nonumber
(\dparb_\pm^i+\CA_\pm^i)\psi_\pm&=0~,\\
\dpar_{\bl_\pm}\psi_\pm&=0~,\label{DDD10}
\end{align}
where $\bar{V}_\alpha^\pm$ and $\dparb_\pm^i$ are vector fields of
type $(0,1)$ on
$\hat{\CU}^s_\pm:=\hat{\CU}_\pm\backslash\CP_{0,\CN}$ as given in
\eqref{vectorfields1}-\eqref{vectorfields2}. These vector fields
annihilate $f_{+-}$ and from this fact and the splitting
\eqref{DD5}, one can also derive equations \eqref{DDD10}. Recall
that due to the existence of a diffeomorphism between the spaces
$\CR_R^{4|2\CN}\times H^2$ and $\tilde{\CP}^{3|\CN}$ which is
described in \eqref{CC10}-\eqref{CC12}, the double fibration
\eqref{C8} simplifies to the nonholomorphic fibration \eqref{C13}.
Moreover, since the restrictions of the bundle $\CE\rightarrow
\CP^{3|\CN}$ to the $\CPP^1_{x_R}$-fibres of the fibration
\eqref{C13} are trivial, there exist regular matrix-valued
functions $\hat{\psi}_\pm$ on $\hat{\CU}_\pm^s$ such that
\begin{equation}\label{DDD11}
f_{+-}=\hat{\psi}_+^{-1}\hat{\psi}_-
\end{equation}
on $\hat{\CU}^s=\hat{\CU}\backslash\CP_{0,\CN}$ and
\begin{equation}
\dparb_\pm^i\hat{\psi}_\pm=0~.
\end{equation}
The existence of this gauge was already implied in
\cite{Witten:2003nn}. Additionally, we impose the reality
condition
\begin{equation}\label{Drealcond}
\hat{\psi}_+^{-1}\left(x_R^{\alpha\ald},\frac{1}{\bl_+},\frac{\bar{\eta}^+_i}{\bl_+}\right)=
\hat{\psi}_-^\dagger(x_R^{\alpha\ald},\lambda_-,\eta_i^-)
\end{equation}
on $\hat{\psi}_\pm$. Although $\hat{\CU}^s$ consists of two
disconnected pieces, the functions $\hat{\psi}_\pm$ are not
independent on each piece because of the condition
\eqref{Drealcond}, which also guarantees \eqref{DD3} on
$\hat{\CU}^s$. The functions $\hat{\psi}_\pm$ and their inverses
are ill-defined on $\CP_{0,\CN}$ since the restriction of $\pi_2$
to $\CR_R^{4|2\CN}\times S^1$ is a noninvertible projection onto
$\CT^{3|\CN}$, see \eqref{CC10}. Equating \eqref{DD3} and
\eqref{DDD11}, one sees that the singularities of $\hat{\psi}_\pm$
on $\CP_{0,\CN}$ split off in a matrix-valued function
$\varphi^{-1}$, i.e.
\begin{equation}
\hat{\psi}_\pm=\varphi^{-1}\psi_\pm~,
\end{equation}
which disappears from
\begin{equation}\label{DDD15}
f_{+-}=\hat{\psi}_+^{-1}\hat{\psi}_-=(\psi_+^{-1}\varphi)(\varphi^{-1}\psi_-)=\psi_+^{-1}\psi_-~.
\end{equation}
Therefore $f_{+-}$ is a nonsingular holomorphic matrix-valued
function on all of $\hat{\CU}$.

{}From \eqref{DDD11}-\eqref{DDD15} it follows that on
$\tilde{\CP}^{3|\CN}$, we have a well-defined gauge transformation
\eqref{5.24}-\eqref{5.27} generated by $\varphi$ and one can
introduce gauge potentials $\hat{\CA}_+^{0,1}$ and
$\hat{\CA}_-^{0,1}$ which are defined on $\hat{\CU}_+^s$ and
$\hat{\CU}_-^s$, respectively, but not on $\CP_{0,\CN}$. By
construction,
$\hat{\CA}^{0,1}=(\hat{\CA}^{0,1}_+,\hat{\CA}^{0,1}_-)$ satisfies
the hCS equations \eqref{shCS1}-\eqref{shCS2} on
$\tilde{\CP}^{3|\CN}=\CP_+^{3|\CN}\cup\CP_-^{3|\CN}$ which are
equivalent to the $\CN$-extended super SDYM equations on
$\FR^{2,2}$. Conversely, having a solution $\hat{\CA}^{0,1}$ of
the hCS field equations on the space $\tilde{\CP}^{3|\CN}$, one
can find regular matrix-valued functions $\hat{\psi}_+$ on
$\hat{\CU}^s_+$ and $\hat{\psi}_-$ on $\hat{\CU}_-^s$ which
satisfy the reality condition \eqref{Drealcond}. These functions
define a further function
$f^s_{+-}=\hat{\psi}_+^{-1}\hat{\psi}_-:\hat{\CU}^s\rightarrow
\sGL(n,\FC)$ which can be completed to a holomorphic function
$f_{+-}:\hat{\CU}\rightarrow \sGL(n,\FC)$ due to \eqref{DDD15}.
The latter one can be identified with a transition function of a
holomorphic vector bundle $\CE$ over the supertwistor space
$\CP^{3|\CN}$. The restriction of $f_{+-}$ to $\CT^{3|\CN}$ is a
real-analytic function $f^\tau_{+-}$ which is {\em not
constrained} by any differential equation. Thus, in the case
$\eps=+1$ (and also for the real structure $\tau_0$), one can
either consider two trivial complex vector bundles $\CE^\tau_\pm$
defined over the space $\CT^{3|\CN}$ together with an isomorphism
$f_{+-}^\tau:\CE_-^\tau\rightarrow\CE_+^\tau$ or a single complex
vector bundle $\CE$ over the space $\CP^{3|\CN}$. However, the
appropriate hCS theory which has the same moduli space as the
moduli space of (equivalence classes of) these bundles is defined
on $\tilde{\CP}^{3|\CN}$. Moreover, real Chern-Simons theory on
$\CT^{3|\CN}$ has no moduli, since its solutions correspond to
flat bundles over $\CT^{3|\CN}$ with constant transition
functions\footnote{Note that these transition functions are in no
way related to the transition functions $f_{+-}$ of the bundles
$\CE$ over $\CP^{3|\CN}$ or to the functions $f_{+-}^\tau$ defined
on the whole of $\CT^{3|\CN}$.} defined on the intersections of
appropriate patches covering $\CT^{3|\CN}$.

{}To sum up, there is a bijection between the moduli spaces of
solutions to equations \eqref{D5} and to the hCS field equations
on the space $\tilde{\CP}^{3|\CN}$ since both moduli spaces are
bijective to the moduli space of holomorphic vector bundles over
$\CP^{3|\CN}$. In fact, whether one uses the real supertwistor
space $\CT^{3|\CN}$, or works with its complexification
$\CP^{3|\CN}$, is partly a matter of taste. However, the complex
approach is more geometrical and more natural from the point of
view of an action principle and the topological B-model. For
example, equations \eqref{D5} cannot be transformed by a gauge
transformation to a set of differential equations on $\CT^{3|\CN}$
as it was possible on $\tilde{\CP}^{3|\CN}$ in the complex case.
This is due to the fact that the transition function $f_{+-}$,
which was used as a link between the two sets of equations in the
complex case does not satisfy any differential equation after
restriction to $\CT^{3|\CN}$. {}From this we see that we cannot
expect any action principle on $\CT^{3|\CN}$ to yield equations
equivalent to \eqref{D5} as we had in the complex case. For these
reasons, we have chosen to use the complex approach throughout the
paper.

%####

\end{document}